\documentclass[fontsize=9pt,parskip=yes,onecolumn,abstract=true]{scrartcl}
\usepackage{graphicx}
\graphicspath{ {./figures/} }
\usepackage[
figurename=Figure,
font={sf,normalsize},
format=plain,
labelfont=bf,
]{caption}
\usepackage{float}
\usepackage{makecell} 
\usepackage{pbox} 
\usepackage{subcaption}
\usepackage{xcolor}
\usepackage{soul} 

\usepackage{indentfirst} 
\usepackage{amsmath}
\usepackage{amssymb}
\usepackage{geometry}
\usepackage[utf8]{inputenc}
\usepackage[section]{placeins} 
\usepackage{hyperref}

\usepackage{orcidlink}

\newcommand{\eg}{e.g.}
\newcommand{\ie}{i.e.}
\newcommand{\si}[1]{SI Section #1}

\newcommand{\hltmp}[1]{#1}

\providecommand{\keywords}[1]
{
  \small	
  \textbf{\textit{Keywords---}} #1
}

\usepackage[
natbib=true,
maxcitenames=1,
minbibnames=3,
maxbibnames=40,
style=authoryear
]{biblatex}
\usepackage{tabularx,ragged2e,booktabs}
\usepackage{authblk}
\usepackage{tcolorbox} 
\usepackage{algorithm}
\usepackage{algorithmic}

\date{}

\begin{document}

\title{A hybrid framework for compartmental models enabling simulation-based inference}

\author[a,b,*]{Domenic P.J. Germano \orcidlink{0000-0001-5893-4840}}
\author[a,c,*]{Alexander E. Zarebski  \orcidlink{0000-0003-1824-7653}}
\author[a]{Sophie Hautphenne      \orcidlink{0000-0002-8361-1901}}
\author[d]{Robert Moss            \orcidlink{0000-0002-4568-2012}}
\author[a]{Jennifer A. Flegg      \orcidlink{0000-0002-8809-726X}}
\author[e]{Mark B. Flegg          \orcidlink{0000-0002-4697-4789}}

\affil[a]{The School of Mathematics and Statistics, The University of Melbourne, Parkville, VIC, Australia}
\affil[b]{The School of Mathematics and Statistics, The University of Sydney, Camperdown, NSW, Australia}
\affil[c]{Pandemic Sciences Institute, University of Oxford, Oxford, United Kingdom}
\affil[d]{Melbourne School of Population and Global Health, The University of Melbourne, Parkville, Vic, Australia}
\affil[e]{School of Mathematics, Monash University, Clayton, VIC, Australia}
\setcounter{Maxaffil}{0}
\renewcommand\Affilfont{\itshape\small}

\maketitle

\def\thefootnote{*}\footnotetext{These authors contributed equally to this work}\def\thefootnote{\arabic{footnote}}

\keywords{Hybrid simulation | stochastic modelling | viral clearance | extinction | compartmental modelling}

\begin{abstract}
Multi-scale systems often exhibit a combination of stochastic and deterministic dynamics.
{In compartmental models, low occupancy compartments tend to exhibit stochastic dynamics while high occupancy compartments tend to follow deterministic dynamics.
Representing both dynamics with existing methods is challenging.}
Failing to account for stochasticity in small populations can produce ``atto-foxes'', for example in the Lotka-Volterra ordinary differential equation (ODE) model.
This limitation becomes problematic when studying the extinction of species or the clearance of infection, but it can be overcome by using discrete stochastic models, such as continuous time Markov chains (CTMCs).
Unfortunately, simulating CTMCs is {impractical} for many realistic models, where discrete events have very high frequencies.

In this work, we develop a novel mathematical framework to couple continuous ODEs and discrete CTMCs: ``Jump-Switch-Flow'' (JSF).
In this framework, compartments can reach extinct states (``absorbing states''), thereby resolving atto-fox-type problems.
JSF has the desired behaviours of exact CTMC simulation, but is substantially {computationally} faster than existing alternatives, {by at least one order of magnitude, and can even obtain constant scaling, irrespective of compartment occupancy}.

We demonstrate JSF's utility for simulation-based inference, particularly multi-scale problems, with several case-studies.
In a simulation study, we demonstrate how JSF can enable a more nuanced analysis of the efficacy of public health interventions.
We also carry out a novel analysis of longitudinal within-host data from SARS-CoV-2 infections to quantify the timing of viral clearance.
{In this work, we show how} JSF offers a novel approach to compartmental model simulation.
\end{abstract}

\section*{Introduction}
Dynamical systems are a powerful method for describing the world,
and have successfully been applied in many fields.
However, modelling populations which change in size across multiple scales remains a challenge \cite{fowler2021atto}.
Population sizes in biological processes often change over orders of magnitude, \eg~the spread of infectious disease \citep{anderson1991infectious};
the boom-and-bust of insect populations \citep{ludwig1978qualitative}; 
and the immune response to infections, such as HIV \citep{perelson2002modelling}, and influenza virus \citep{baccam2006kinetics}.

The dynamics of small populations can be heavily influenced by stochastic effects, while for larger populations these fluctuations often average out, justifying the use of continuum models. 
When the population under consideration does not change in size over orders of magnitude, there is usually a natural (and obvious) choice between a stochastic or deterministic model.
However, for multi-scale models, this decision remains a challenge, and 
developing modelling methods that can bridge these scales
has been a long-standing goal of applied mathematics 
\citep{cotter2016error,flegg2014analysis,isaacson2013convergent}.

Compartmental models describe how quantities (\eg~the number of molecule, cells, or people) change in a dynamical system.
Two popular ways to represent compartmental models are ordinary differential equations (ODEs) and continuous time Markov chains (CTMCs).
Many ODEs one encounters are actually \emph{ensemble averages} of CTMCs, although the mathematical justification of this is not always straightforward \citep{Kurtz_1970,Kurtz_1971,kurtz1972relationship}.

In ODEs, the state of the system changes continuously, which can lead to ``atto-fox problems'' where populations shrink to infeasibly small sizes, \ie~where a real population would likely have gone extinct \citep{fowler2021atto, mollison1991dependence, lobry2015migrations}. 
In CTMCs, the discrete state changes stochastically and there may be absorbing states \eg~extinction of a population.
As a result, CTMCs offer a more natural description of small populations sizes, however applying them to large populations can be computationally challenging. 
Moreover, a range of powerful techniques can be brought to bear on ODEs, enabling more thorough analysis. 
This raises a natural question about what to do when compartment occupancy changes across orders of magnitude. 

Exact and approximate algorithms to simulate CTMCs exist \cite{gillespie1976general,simoni2019stochastic}.
Exact methods (\eg~Doob-Gillespie, first/next reaction, and rejection based methods) are computationally expensive when state transitions occur at a high rate \citep{sanft2015constant}.
Approximate methods (\eg~Tau-leaping \citep{gillespie2001approximate,cao2006efficient} and the use of chemical Langevin equations \citep{rao2003stochastic,gillespie2000chemical,gibson2000efficient})
scale to high transition rates but can have unacceptable approximation error.
To overcome the limitations of classical approximate methods, over the past two decades, there has been a concerted effort to develop \emph{hybrid} stochastic-deterministic approaches \citep{simoni2019stochastic, kreger2021hybrid, bressloff2014path, rebuli2017hybrid}.

Some hybrid modelling approaches involve partitioning the transitions in a CTMC into fast and slow transitions.
Despite having significant implementation bookkeeping and overheads, these approaches can be sampled individually and subsequently synchronised in an efficient manner \cite{simoni2019stochastic}. 
Jump-diffusion differential equations partition the model compartments into \emph{fluid} and \emph{discrete} compartments, to construct \emph{hybrid switching jump diffusion} processes \cite{buckwar2011rungekutta,angius2015approximate}.
Other hybrid simulation techniques have been developed in the context of within-host viral infection \cite{kreger2021hybrid}. However, this implementation is not generalised, but tailored to a single system, and does not allow switching between stochastic and deterministic regimes.
In the context of determining the hybridisation between stochastic and deterministic couplings, path-integrals have been utilised, \cite{bressloff2014path}, which permit analysis of hybrid processes.
Recently, \citeauthor{kynaston2023regime} considered extended systems that represent each compartment with both a continuous and a discrete version and allow the for conversion between them \cite{kynaston2023regime}.

Essential for the real-world utility of many mathematical models is their ability to be calibrated to data.
For deterministic compartment models, standard parameter inference methods, such as maximum likelihood and posterior sampling via Markov chain Monte Carlo (MCMC), can readily be applied \citep{ionides2006inference}. 
For stochastic models these inference problems are usually much harder.
Simulation-based inference has emerged as a way to handle these problems,
\eg~the \emph{particle filter}~\citep{arulampalam2002tutorial,kitagawa1996monte} and \emph{approximate Bayesian computation}~\citep{Alahmadi2020ABC}.

{Calibrated multi-scale simulations of ODE (compartment) models is an important milestone towards accurate predictive models that address stochastic behaviour. In a recent paper, Trindade and Zygalakis develop a hybrid scheme for chemical kinetics over different scales with the goal of demonstrating the utility for parameter estimation ~\citep{trindade2024hybrid}. In this paper, a CTMC is used to model reaction events on small populations (below a threshold $I_1>0$) and for better efficiency Tau-leaping is used for reactions that only impact large populations (above a threshold $I_2>I_1$). Between these two thresholds, a ``blending function'' is used to create a linear transition between an accurate CTMC and efficient Tau-leaping treatment of the reactions. The hybrid adoption of CTMC and Tau-leaping regimes has the property of maintaining appropriate levels of noise at large populations but struggles as populations grow much larger where a Chemical Langevin or deterministic ODE description may be more computationally appropriate.}

In this paper, we present a simple and efficient hybrid simulation
method which we call \emph{Jump-Switch-Flow} (JSF). This method enables adaptive transitions between stochastic
and deterministic regimes across compartments, while ensuring conservation principles are maintained. 
In this approach, compartments can individually switch regimes, allowing for the compartments to be split into stochastic and deterministic subsets, while remaining coupled. {This regime switching presents unique challenges in matching at interfaces: ensuring that the deterministic regime represents the expected state under transition between regimes and that mass is conserved. Conservation is not always essential when it pertains to small population sizes, however in many cases sustained small populations have a significant impact on larger populations (\eg~enzymes in chemical processes and viruses within hosts.)} 
We demonstrate the utility and properties of the JSF method through simulation studies, and a case study in which we reanalyse existing longitudinal SARS-CoV-2 viral load datasets.

Two simulation studies are presented to explore the properties of JSF and demonstrate its use in an inference setting.
In the first simulation study, we compare the computational efficiency and accuracy of JSF with the exact Doob-Gillespie method and the Tau-leaping method. 
In the second simulation study, we demonstrate the types of insight available when using an approach that supports absorbing (extinction) states.

In the case study, we demonstrate The significance of being able to perform inference with models with absorbing states
(\ie~models in which one of the compartments can go extinct.)
In the case study we reanalyse longitudinal SARS-CoV-2 viral load data \citep{ke2022daily} using JSF to simulate a TEIRV  
 (Target cells -- Eclipsed cell -- Infectious cells -- Refractory cell -- Virions)
model and infer the state of host and virus across the infection.
Understanding how a viral infection is cleared has important ramifications for both treatment and prevention.
For example, the initial exposure may fail to initiate a systematic infection, and understanding the conditions under which this happens is important for infection prevention \cite{pearson2011stochastic}. Moreover, the infection may reach low levels, potentially escaping detection, but the virus may rebound and cause disease. Understanding when the virus has been cleared is important for optimising treatment duration.
Inferring viral clearance can be computationally challenging \citep{yan2016extinction}, and this case study demonstrates how JSF enables the tractable estimation of viral clearance.

\section*{Methods}
\subsection*{Jump-Switch-Flow mathematical framework}

Consider a compartmental model with $n$ compartments, {$\vec{V} = \left( v_1, ..., v_n\right)$}, where the state variable,  {\(v_i := v_{i}(t)\)}, represents the value of the \(i\)th compartment at time \(t\). 
For example,  {\(v_i\)} could be the number of people infected with a pathogen, or the copy number of a molecule in a cell.
The state variables  {\(v_i\)} may take values from different domains, depending upon the resolution needed, for example in an ODE,  { \(v_i\)} will have real values and in a CTMC  { \(v_i\)} might have integer values.

When the variables represent quantities, typically, discrete values are used to represent small populations, while larger populations are represented with a continuum.
To accommodate both scales, let the domain of {\(v_{i}\) be the set \(\{0,1,\ldots,\Omega_{i}\}\cup(\Omega_{i},\infty)\)}. 
The \textit{switching threshold parameter}, \(\Omega_i\in \mathbb{Z}_{\geq 0}\), is where the $i$th compartment transitions between discrete and continuous dynamics.
If a compartment  {\(v_{i} \in \{0,1,\ldots,\Omega_{i}\}\)}, we call it \emph{discrete} (or \emph{jumping}), and if { \(v_{i} \in (\Omega_{i},\infty)\)}, we call it \emph{continuous} (or \emph{flowing}).
While the switching threshold can be compartment specific, for ease of exposition, we will only consider a single threshold shared between all compartments, \ie~{\(\Omega_i = \Omega\)}. 

 {At each time $t$, the components of $\vec{V}$ can be partitioned into $\vec{V}_F$ and $\vec{V}_J$, where $\vec{V}_F$ contains the flowing variables $v_i > \Omega$, and $\vec{V}_J$ contains the jumping variables $v_i \leq \Omega$.
Therefore, at any moment in time $\dim{(\vec{V}_F)}$ of the $n$ compartments are flowing, and $\dim{(\vec{V}_J)}$ of the $n$ compartments are jumping, with $\dim{(\vec{V}_F)}  + \dim{(\vec{V}_J)} = n$. }

The dynamics of each compartment  {\(v_i\)} are described by a set of $m$ reactions  {$\mathcal{R} = \left\{{r}_k \right\}_{k=1}^m$}.
Each reaction  {${r}_k$} is defined by two properties: 
the rate (per unit time) at which it occurs, \(\lambda_{k}\), which (usually) is a function of the state $\vec{V}$;
and the reaction's effect on the state, \ie~the change \(\eta_{ik}\) to the size of compartment  {$v_i$ when reaction ${r}_k$ occurs. As a vector, $\vec{\lambda} \in \mathbb{R}^{m}$ is referred to as the \textit{propensity vector}.} As a matrix, $\eta\in \mathbb{Z}^{n,m}$ is referred to as the \emph{stoichiometric matrix}. 
For ODE models  {that only contain flowing variables}, these reactions occur continuously and are written in the form:
\begin{equation}\label{ODEmodels}
\frac{\mathrm{d}\vec{V}}{\mathrm{d}t} = \eta \vec{\lambda}(t,\vec{V}).
\end{equation} 
For CTMC models, reactions in the system $\mathcal{R}$ occur as discrete events.
In the {latter} case, each reaction ${r}_k$ has a separate propensity described by $\lambda_k(t,\vec{V})$. This propensity remains constant between reactions, but when a reaction  {${r}_k$} occurs, there is a change in $\vec{V}$ (as specified by the elements of  {$\eta_{\cdot k}$}), and therefore in $\vec{\lambda}(t,\vec{V})$.

As an example, consider the SIR model of epidemics, {which, in ordinary differential equations, has the form $\frac{dS}{dt}=-\frac{\beta S I}{S + I + R}$, $\frac{dI}{dt}=\frac{\beta S I}{S + I + R} - \gamma I$, and $\frac{dR}{dt} = \gamma I$.}
The state of this model is $\vec{V} = (S,I,R)^\intercal$, and there are two ``reactions'': infections ({${r}_1$}) and recoveries ({${r}_{2}$}).
For infections, the rate of reaction may be modelled by  {\(\lambda_1=\beta S I/(S+I+R)\)}, and the entries of the associated column of the stoichiometric matrix are $\eta_{1,1} = -1$, $\eta_{2,1} = 1$ and $\eta_{3,1}=0$.
For recoveries, the rate of reaction may be modelled by  {\(\lambda_2=\gamma I\)}, and the entries of the associated column of the stoichiometric matrix are, $\eta_{1,2}=0$, $\eta_{2,2} = -1$, and $\eta_{3,2} = 1$.

{We define $\mathcal{R}_J\subseteq \mathcal{R}$ as the subset of reactions treated as stochastic \textit{jump} events. The set $\mathcal{R}_J$ can be defined more precisely in two ways. While we use only the second in this manuscript, both are described for clarity.}
The first way to define  {$\mathcal{R}_J$} is as: {\(\mathcal{R}_J = \left\{{r}_k:\exists i \text{ s.t. } v_i\in\vec{V}_J  \text{ and }  \eta_{ik}\neq 0   \right\}\)}.
In this definition, a reaction  {${r}_k$} is included in  {$\mathcal{R}_J$} if and only if the reaction has some material effect on a jumping (discrete) compartment  {$v_i\in\vec{V}_J$}.
This is a minimum requirement; a reaction should not be permitted to evoke a continuous change in a discrete compartment.
However, where it makes sense to do so,  it is possible to allow reactions to make discrete changes to flowing (continuous) compartments.
The second way to define  {$\mathcal{R}_J$}, which we use throughout this manuscript, captures a larger set of reactions:  {\mbox{\(\mathcal{R}_J = \left\{{r}_k:\exists i \text{ s.t. } v_i\in\vec{V}_J \text{ and }  \left(\eta_{ik}\neq 0 \text{ or } \partial_{v_i}\lambda_k \neq 0\right)   \right\}\)}}.
In this definition, a reaction is included in  {$\mathcal{R}_J$} if either (1) it causes a material change in jumping (discrete) compartments \textit{or} (2) it is influenced/caused by a discrete compartment (\eg~as catalyst of some reaction.)

Reactions in  {$\mathcal{R}_J$} are simulated using stochastically sampled times, similar to CTMC models (and described below in the section on `Jump events'). 
It is important to note that unlike time homogeneous CTMC models, the propensities are \textit{not} constant because the state $\vec{V}$ (and therefore $\vec{\lambda}$) are continuously varying.
When any reaction  {${r}_k\in\mathcal{R}_J$} occurs, we say the system has \emph{jumped} and an instantaneous change of $\eta_{ik}$ for each compartment  {$v_i$} occurs (irrespective of whether $v_i\in\vec{V}_J$ or $v_i\in\vec{V}_F$, to ensure the conservation of mass). We therefore refer to reactions in  {$\mathcal{R}_J$} as \textit{jumps}.
The reactions in  {\(\mathcal{R}_F=\mathcal{R}\setminus \mathcal{R}_J\) represent the continuous change of value of the relevant compartments, all of which are deterministic (see section on `Flow events' above.) 
We therefore refer to reactions in $\mathcal{R}_F$ as \textit{flows}.}
 {At any moment in time $|\mathcal{R}_F|$ is the number of reactions which are flowing, and $|\mathcal{R}_J|$ is the number of reactions which are jumping, such that $|\mathcal{R}_F| + |\mathcal{R}_J| = m$.}

Finally, the hybrid model that we propose is capable of \emph{switching}.
Switching events occur when the size of a compartment $v_i$ crosses the threshold $\Omega$ and therefore changes membership between the flowing $\vec{V}_F$ and jumping $\vec{V}_J$ sets {(see sections on `Jump clock updates' and `Switching events').
Importantly, switching events may also impact which reactions are jumps, $\mathcal{R}_J$. As such, switches are paradigm-defining events which ought to occur infrequently in comparison to jumps (which may occur frequently) and flows (which occur continuously)}.

Due to the way that $\mathcal{R}$ is partitioned, it is possible to order the rows and columns of $\eta$ into the upper-triangular block form:
\begin{align}
\eta = \left(\begin{array}{c|c}
\eta_{FF} & \eta_{FJ} \\ \hline
0 & \eta_{JJ} \end{array}\right),
\end{align}
 {where $\eta_{FF} \in \mathbb{Z}^{\dim{(\vec{V}_F)} , \vert \mathcal{R}_F \vert }$, $\eta_{JJ} \in \mathbb{Z}^{\dim{(\vec{V}_J)}, \vert \mathcal{R}_J \vert }$ and ${\eta}_{FJ} \in \mathbb{Z}^{\dim{(\vec{V}_F)} , \vert \mathcal{R}_J \vert }$ refer to stoichiometric coefficients for changes in flowing compartments under flows, jumping compartments under jumps, and flowing compartments under jumps, respectively. Note that, the lower left corner is zeros because, by definition, flowing reactions cannot influence the state of jumping variables.}
Written as a system of equations analogous to (\ref{ODEmodels}), the hybrid JSF framework we propose formally takes the following form. For any time interval  {$t_{i}<t<t_{i+1}$ between switching events $i$ and $i+1$}:
\begin{align}
    \frac{\mathrm{d} \vec{V}_F}{\mathrm{d} t} &= \eta_{FF} \vec{\lambda}_{F}(t,\vec{V}) + {\eta}_{FJ} \vec{\Lambda}_{J}(t,\vec{V}), \label{JSFModel1}\\
    \vec{V}_J(t) &= \vec{V}_J(t_{i}) + \eta_{JJ} \int_{t_{i}}^t  \vec{\Lambda}_{J}(s,\vec{V}) \ \mathrm{d} s, \label{JSFModel2}
\end{align}
where  {$\vec{\lambda}_{F}\in\mathbb{R}^{\vert \mathcal{R}_F \vert }$} are the reaction rates of flows and  {$\vec{\Lambda}_{J}$} is a stochastic vector of  {$\vert \mathcal{R}_J \vert$} delta-function spike trains that are derived from the realisations of  {$\vert \mathcal{R}_J \vert$} different jumps sampled at rates which are dependent on the dynamic changes in the propensities  {$\vec{\lambda}_{J}\in\mathbb{R}^{\vert \mathcal{R}_J \vert}$} for these jumps  {(see `Jump events')}.

\begin{figure*}[ht!]
    \centering
    \includegraphics[width=0.9\textwidth]{figures/PLOS_Figs/Figure1.png}
     \caption{\label{fig:fig-1} JSF provides a way to capture both the continuous deterministic dynamics of large populations and the discrete stochastic dynamics of small populations. \textbf{A} A compartmental model representation of the Lotka-Volterra system \eqref{eq:lv}: prey ($v_1$, depicted as red rabbits which, in reaction $r_1$, reproduce at rate $\alpha$), predators ($v_2$, depicted as purple foxes which, in reaction $r_2$, die at rate $\gamma$), and their interaction (predation of rabbits but foxes, in reaction $r_3$, at a rate proportional to $\beta$). \textbf{B} The compartmental model can be formalised as a reaction network via a stoichiometry matrix $\eta$ which details each species role in the reactions. The reactants consumed $\eta^-$ and the products produced $\eta^+$ can be written explicitly with $\eta = \eta^+ - \eta^-$. \textbf{C} A hypothetical JSF trajectory demonstrating how the representation captures the continuous variation of some compartment-reaction pairs (flow) and the discrete stochastic changes (jumps) in others. Transitioning (switching) between a continuous and discrete state occurs at a threshold $\Omega$ which is indicated with a horizontal dashed line. \textbf{D} An example trajectory from the Lotka-Volterra model exhibiting the characteristic cyclical behaviour with additional stochasticity influencing the dynamics at low population sizes. \textbf{E} An example trajectory showing the possibility for the predator species to go extinct and the subsequent growth of the prey species. }
\end{figure*}

\subsection*{Example: Lotka-Volterra model}
The Lotka-Volterra model describes the populations of two species: prey, $v_{1}$, and predators, $v_{2}$. 
The prey reproduce at rate \(\alpha\), the predators die at rate \(\gamma\) and new predators are produced through predation at rate \(\beta\). 
When modelled with ODEs, this gives us the following system:
\begin{equation}\label{eq:lv}
\begin{aligned}
    \frac{d v_{1}}{dt} = \alpha v_{1} - \beta v_{1}v_{2},\\
    \frac{d v_{2}}{dt} = \beta v_{1}v_{2} - \gamma v_{2}.
\end{aligned}
\end{equation}
\noindent
The periodic solutions to \eqref{eq:lv} persist even when the predator species reaches infeasible population sizes: populations of order \(10^{-18}\) giving us ``atto-foxes''.

Figure~\ref{fig:fig-1}A shows a compartmental diagram of the Lotka-Volterra model with the reaction:  {${r}_1$, birth of $v_1$; ${r}_{2}$, death of $v_2$; and ${r}_{3}$, conversion of $v_1$ into $v_2$}.
Figure~\ref{fig:fig-1}B shows a representation of the reactions and their corresponding stoichiometric matrix.

Figure~\ref{fig:fig-1}C depicts a hypothetical trajectory of the Lotka-Volterra model when represented as a JSF process.
There are three switching events (at $t_1$, $t_2$ and $t_3$), and each configuration of discrete (jumping) and continuous (flowing) states are featured:
\begin{enumerate}
    \item Before $t_1$, the representation is  {$\vec{V}_F = v_1$ and $\vec{V}_J = v_2$ and the reactions are $\mathcal{R}_J = \left\{{r}_2, {r}_3 \right\}$} (hybrid regime).
    \item During $[t_1,t_2)$, the representation is  {$\vec{V}_F = (v_1,v_2)^\intercal$ and $\vec{V}_J = \varnothing$ and the reactions are $\mathcal{R}_J = \varnothing$} (ODE regime).
    \item During $[t_2,t_3)$, the representation is  {$\vec{V}_F = v_2$ and $\vec{V}_J = v_1$ and the reactions are $\mathcal{R}_J = \left\{{r}_1, {r}_3 \right\}$}  (hybrid regime).
    \item From $t_3$ onwards the representation is  {$\vec{V}_F = \varnothing$ and $\vec{V}_J = (v_1,v_2)^\intercal$ and the reactions are $\mathcal{R}_J = \mathcal{R}$} (CTMC regime).
\end{enumerate}

Figures~\ref{fig:fig-1}D and E show two instances of random trajectories sampled from the Lotka-Volterra process as represented with JSF with a switching threshold of $\Omega = 30$. 
In the trajectory in Figure~\ref{fig:fig-1}D the populations undergo several cycles while in Figure~\ref{fig:fig-1}E, with the inclusion of discrete stochastic behaviour, the predator species goes extinct, allowing the prey to grow exponentially.

{\subsection*{Mathematical detail}
Here we provide the mathematical details of JSF: the evolution and computation of values for flowing compartments (see `Flow events'); the evolution and computation of values for jumping compartments (see `Jump events'); the sampling of jump times (see `Jump clock updates'); and the switching of compartments between jumping and flowing regimes (see `Switching events').
Further details, including pseudo-code listings of the algorithms used to sample from the JSF process, are given in \si{1}.}

{\subsubsection*{Flow events}\label{FlowEventsUpdates}
Equation~\eqref{JSFModel1} contains the differential equations describing the evolution of the flowing compartments.
If we consider an interval of time between jumps, $\left[t_0, t_1\right]$, then $\vec{V}_J(t)$ remains constant and therefore, $\vec{\Lambda}_{J}(t,\vec{V}) = \vec{0}$ across this interval. Therefore, the differential equations for the flowing compartments are:
\begin{equation}\label{JSF-Flowing}
    \frac{\mathrm{d} \vec{V}_F}{\mathrm{d} t} = \eta_{FF} \vec{\lambda}_{F}(t,\vec{V}).
\end{equation}
This is a standard dynamical system of ODEs, which we numerically integrate forward in time over discrete time steps of size $\Delta t$. While an arbitrary order method may be used, we use a simple Forward Euler method: 
\begin{align}
    \label{JSF-Flowing-update}
    \vec{V}_F(t+\Delta t)=& \vec{V}_F(t) + \Delta \vec{V}_F(t)\\
    =& \vec{V}_F(t) + \Delta t \, \eta_{FF} \vec{\lambda}_{F}(t,\vec{V}) + O(\Delta t^2).
\end{align}
Figures \ref{fig:EventFig}\textbf{A} and \ref{fig:EventFig}\textbf{B} depict examples of a flowing compartment and a jumping compartment, respectively. Flow updates are given at intervals of discrete time step $\Delta t$ as indicated. {In the Jump-Switch-Flow algorithm, flowing time steps are bounded above by a finite, but small, $\Delta t$. Flowing time steps have this maximum duration if no jumps are found on this time interval (that is, if (\ref{JSF-Flowing}) remains valid). If a jump occurs at $\Delta \tau < \Delta t$ from any given time (see Section `Jump events') then a flowing time step of $\Delta \tau$ is used instead of $\Delta t$ in Equation~\eqref{JSF-Flowing-update}. In this case, a jump is instantaneously applied after the flow event and the next interval of time is associated with new flow rates $\vec{\lambda}_F$ (which are therefore updated at the exact moment of time of the jump event). We note therefore that error associated with the ODE solver is determined by the size of the finite capped time interval size $\Delta t$ according to classical ODE error analysis. In this case since we use Forward Euler, the error scale like $O(\Delta t^2)$.} Using Equation~\eqref{JSF-Flowing-update}, over each time step (flowing interval), flow events only directly affect flowing compartments as can be seen in Figure \ref{fig:EventFig}\textbf{A}. Jumping compartments like that shown in Figure \ref{fig:EventFig}\textbf{B} are unaffected directly by flowing events {because the moments of a jumping event are pre-determined according to approximations in the trajectories of the flowing compartments which have the same accuracy as the ODE solver (for example, we assume a linear changes in the flowing compartments as consistent with Forward Euler method used in Equation (\ref{JSF-Flowing-update}) -- see Section `Jump events' for details}. As indicated by discontinuities in both Figure \ref{fig:EventFig}\textbf{A} and Figure \ref{fig:EventFig}\textbf{B}, jump events may be applied to flowing and jumping compartments alike in order to conserve mass in a manner which we shall now describe.} 

\begin{figure}[h!]
    \centering
    \includegraphics[width=0.99\linewidth]{figures/PLOS_Figs/Figure2.png}
    \caption{{
    The JSF process can be numerically integrated. \textbf{A} and \textbf{B}: Dynamic changes in flowing compartments (\textbf{A}) and jumping compartments (\textbf{B}). Flows are computed using the time step $\Delta t$. If a jump occurs within a flow update, the flows are adjusted up until the corresponding time $\Delta \tau$ as a time step instead of $\Delta t$ (indicated by a discontinuity in both \textbf{A} and \textbf{B}). Jumps occur when a jump clock reaches zero.
    \textbf{C}: The jump clock $J_k$ for reaction $k$ counts down until event $k$ occurs when $J_k=0$, \ie~between $t$ when $J_k>0$ and $t+\Delta t$ when $J_k<0$. The time of the jump is found by interpolating for $t+\Delta \tau$ using Equation~\eqref{jumptime}, where $\Delta \tau < \Delta t$. Once the jump has occurred and jumps implemented according to the respective stoichiometric updates, $J_k$ is reinitialised by sampling \(J_k = -\log(u_k)\) where \(u_k\sim\text{Unif}(0,1)\), and then counts down with successive time steps $\Delta t$ until the next jump $k$ is realised.
    \textbf{D} A switching event of type 1 occurs when a discrete compartment jumps above the switching threshold, $\Omega$. This is shown here after the jump time step $\Delta \tau_2$.
    \textbf{E} A switching event of type 2 occurs when a flowing compartment reaches the switching threshold, $\Omega$. Here, the time at which the switch occurs is found using continuation.
    \textbf{F} A switching event of type 3 occurs when a flowing compartment is caused to go below the switching threshold, $\Omega$, due to a jump in another compartment. Here the jump has caused the flowing compartment to end at the dotted circle (which we denote as $\hat{v}_i$). This compartment is then re-labelled as jumping. Since jumping compartments require integer states, the state is re-initialised to take the value $\lceil \hat{v}_i \rceil$ with probability $v_i - \lfloor \hat{v}_i \rfloor$, and otherwise $\lfloor \hat{v}_i \rfloor$ such that the expected value of the state is at the dotted circle (in the example presented here, the variable was rounded down).
    For all switching events, membership of $V_F$ and $V_J$ is recomputed after switching.}}
    \label{fig:EventFig}
\end{figure}

{\subsubsection*{Jump events} \label{JSFJumpClockUpdates}
The reactions in $\mathcal{R}_J$ produce discontinuous jumps in the state vector $\vec{V}$.
In Equations \eqref{JSFModel1} and \eqref{JSFModel2}, each element in the vector $\vec{\Lambda}_{J}$ corresponds to a reaction in $\mathcal{R}_J$. Consider, for example, $r_k \in \mathcal{R}_J$. 
The $k^{\text{th}}$ element of $\vec{\Lambda}_{J}$ is:
\begin{align}
\Lambda_{J}^{(k)} = \sum_{i} \delta\left(t-t_{i}^{(k)} \right),
\end{align}
where $\delta$ is the Dirac delta function and $t_{i}^{(k)}$ is the time at which the reaction $r_k$ occurs for the $i$th time. In this way, the term $\vec{\Lambda}_{J}$ in Equations \eqref{JSFModel1} and \eqref{JSFModel2} manifests as discrete jumps in both $\vec{V}_F$ and $\vec{V}_J$ at each jump. For the sake of the simulation, the computation of the jump times $t_{i}^{(k)}$ for each instance $i$ of each reaction $r_k\in \mathcal{R}_J$ is all that is required. 
The stoichiometric matrix describes the change to each compartment when that jump event occurs.}

{
The rate of reaction $r_k$, $\lambda_k(t,\vec{V})$, may be state dependent so can change due to the continuous change caused by flows. 
Figures \ref{fig:EventFig}\textbf{A} and \ref{fig:EventFig}\textbf{B} depict an example of flowing and jumping compartment (respectively) experiencing a jump event. In this illustration, we note that if $\lambda_k$ depends on $v_2$ the event time $t_i^{k}$ is not sampled from an exponential distribution which requires a constant propensity/rate between events. 
Formally, the jump times form an inhomogeneous Poisson process. There are multiple ways to sample the jump times $t_{i}^{(k)}$, see~\cite{klein1984time} for a detailed discussion.
We use a variant of the Next Reaction Method \cite{gibson2000efficient} (which is an optimised variant of the Doob-Gillespie method \cite{gillespie1976general}) to sample jump times. We first note that the propensity for a jump is dependent only on the instantaneous state $\vec{V}$, and therefore at time $t_0$ if there has been $i-1$ jumps associated with reaction ${r}_k$, it has no bearing on the distribution of the time $t_i^{(k)}$. Therefore, we shall simply denote $t_i^{(k)} = t_k$ as the \textit{next} jump time for reaction ${r}_k$.
The cumulative probability function from which $t_k$ is sampled depends on the current time, $t_0$, and the evolution of the state variables in time, $\vec{V}(t)$.
The cumulative distribution for the time of the next event associated with $r_k$ is:
\begin{align}
    \text{CDF}(t;k) = 1 - \exp\left\{- \int_{t_{0}}^{t} \lambda_{k}(s,\vec{V}(s)) ds \right\}. \label{eq:cdf}
\end{align}
To sample $t_k$, inverse transform sampling is used \cite{klein1984time}, \ie~sample \(u_{k}\sim\text{Unif(0,1)}\) and then solve $\text{CDF}(t_k;k) = u_k$ for $t_k$.
To account for the varying state, we define a new function, $J_k(t)$ the \emph{jump clock} for reaction ${r}_k$, {using Equation (\ref{eq:cdf}), and solving for when $\text{CDF}(t_k;k) = u_k$ (inverse transform sampling)}.
The jump clock acts as a timer, identifying the time $t_k$ for when $r_k$ next occurs, when $J_k(t_k)=0$. The function $J_k$ is defined by:
\begin{equation}\label{eq:jump-timer}
   J_k(t_k) = -\log(u_{k}) - \int_{t_{0}}^{t_k} \lambda_{k}(s,\vec{V}(s)) \ \mathrm{d}s,
\end{equation}
noting that $u_k$ and $1-u_k$ have the same distribution.
In general, we cannot solve directly for $t_k$, so we solve for it numerically by tracking the value of $J_k(t)$ as $\vec{V}$ evolves through flows, jumps and switches. }

{
For each reaction ${r}_k$, at some initial time, for example $t_{i-1}^{(k)}$, being the continuous time of the $(i-1)^{\text{th}}$ jump associated with reaction $r_k$, we sample $u_k$ and initialise $t_0 = t_{i-1}^{(k)}$. The initial value of $J_k(t)$ is therefore equal to the positive number $\log(u_{k}^{-1})$ (since \(u_{k}\sim\text{Unif(0,1)}\)). As time progresses, $J_k(t)$ decreases according to (\ref{eq:jump-timer}) since $\lambda_k\geq 0$. The value of $J_k(t)$ decreases to zero over time and when $J_k(t)=0$, a jump associated with ${r}_k$ is triggered (hence the name ``jump clock''). Once a jump clock reaches 0 and a jump is triggered, the clock is reset by sampling a new random number, $u_k\sim\text{Unif}(0,1)$. See Figure \ref{fig:EventFig}\textbf{C} for a schematic illustration of how the jump clock is updated.}


{
To update the jump clock, we require numerical integration of $\lambda_k(t,\vec{V}(t))$ forward in time. Fortunately, we also have piece-wise polynomial approximations for $\vec{V}_F(t)$ as a result of our numerical treatment of the continuous flows (see Subsection \ref{FlowEventsUpdates} `Flow events'), combined with piece-wise constant values for $\vec{V}_J(t)$ which only change when jumps occur. 
We discuss the numerical integration of the jump clock below.
}


{\subsubsection*{Jump clock updates} \label{JSFJumpClockUpdates}
For a given jump reaction ${r}_k$, a jump clock is initialised at time $t_0$ with $u_k\sim\text{Unif}(0,1)$ giving $J_k = \log(u_k^{-1})$.
From time $t$ to $t+\Delta t$, with $t>t_0$, Equation~\eqref{eq:jump-timer} tells us the clock ticks down from $J_k$ to $J_k - \Delta J_k$, where $\Delta J_k$ is
\begin{align}
\Delta J_k =  \int_{0}^{\Delta t} \lambda_{k}(t+s,\vec{V}(t+s)) \ \mathrm{d}s.
\end{align}
In general this integral is intractable, but we can approximate it as follows: consider the Taylor expansion of $\lambda_{k}(t+s,\vec{V}(t+s))$ about $s = 0$, which gives
\begin{align}
\Delta J_k &= \int_{0}^{\Delta t}  \lambda_{k}(t,\vec{V})  + \dfrac{\partial \lambda_k}{\partial \vec{V}} \dfrac{\mathrm{d}\vec{V}^{\intercal}}{\mathrm{d}t}  \, s + O(s^2)  \ \mathrm{d}s = \int_{0}^{\Delta t}  \alpha  + \beta s + O(s^2)  \ \mathrm{d}s,
\end{align}
where we write  $\alpha = \lambda_{k}(t,\vec{V})$, and $\beta =  \dfrac{\partial \lambda_k}{\partial \vec{V}} \dfrac{\mathrm{d}\vec{V}^{\intercal}}{\mathrm{d}t} $. 
Since (i) our numerical evaluation of $\vec{V}$ is piece-wise linear across the time interval, by virtue of the fact that we use a Forward Euler approximation to solving Equation~\eqref{JSF-Flowing}, and (ii) $\Delta t$ is small, $\Delta J_k$ is approximated to the precision of our algorithm by taking
\begin{align}
\Delta J_k &\approx  \frac{\Delta t}{2} \left(2\alpha  + \beta \Delta t \right), \label{delJ}
\end{align}
{noting that since $\vec{V}$ is piece-wise linear across the time interval, the convergence of the integration of the Jump-clock also behaves as $~ \Delta t$.}
Importantly, we know that between jumps $\dfrac{\mathrm{d}\vec{V}_F}{\mathrm{d}t}$ is given by Equation~\eqref{JSF-Flowing} whilst $\dfrac{\mathrm{d}\vec{V}_J}{\mathrm{d}t}=0$. Thus, $\beta = \dfrac{\partial \lambda_k}{\partial \vec{V}_F} \left( \eta_{FF} \vec{\lambda}_{F}(t,\vec{V}) \right)^{\intercal} $ (evaluated at $t$).
To calculate the updated jump clock, we compute $J_k(t + \Delta t) = J_k(t) - \Delta J_k$ as the provisional value. We then have two distinct cases: (i) $J_k(t) - \Delta J_k > 0$, then no jump occurred during the interval $(t, t+\Delta t)$ and we have the jump clock $J_k(t+\Delta t) := J_k(t) - \Delta J_k$: (ii) $J_k(t) - \Delta J_k < 0$, then a jump occurred (i.e. a ${r}_k$ reaction) during the interval $(t,t+\Delta t)$, that we need to account for in the updated jump clock.}
{
In the case (ii) where a jump occurs within the interval $(t,t+\Delta t)$, let $t + \Delta \tau$, where $0<\Delta \tau <\Delta t$, denote the time at which this jump occurs.
We can find $\Delta \tau$ by interpolation and solving the following for $\Delta\tau$:
\begin{align}
    2 \Delta  J_k - \Delta \tau(2\alpha + \beta\Delta \tau) = 0,
\end{align}
where $\Delta  J_k$ is the residual of the jump clock from $t$ to $t + \Delta \tau$.
$\Delta \tau$ is then given by
\begin{equation}\label{jumptime}
    \Delta \tau = 
    \begin{cases}
    \frac{\sqrt{\alpha^2 + 2\beta \Delta  J_k} - \alpha}{\beta}, \quad &\beta \neq 0,\\
    \frac{\alpha}{\Delta J_k}, \quad &\beta = 0.
    \end{cases}
\end{equation}
If there is a jump, rather than using $\Delta t$ to forward compute the flowing compartments in Equation~\eqref{JSF-Flowing-update} we instead use $\Delta \tau$ to take part of a flow step and then implement the jump after the flow to get the state at time $t+\Delta \tau$. After this, we reinitialise the jump clock $J_k$ as described above. This procedure is illustrated in Figure \ref{fig:EventFig}\textbf{C}.}

{\subsubsection*{Switching events}
The way in which we handle switching is one of the main contributions of this manuscript.
Switching events occur when variables transition between $\vec{V}_J$ and $\vec{V}_F$, i.e. between discrete or continuous values.
As a consequence, the contents of $\mathcal{R}_J$ and $\mathcal{R}_F$ may change.
There are three types of switching events.
The first involves a compartment moving from $\vec{V}_J$ to $\vec{V}_F$.
This transition is straightforward as a new equation is added to Equation~\eqref{JSF-Flowing}, and the state $\vec{V}_F$ is initialised at the switching time, with the new flowing compartment at $\Omega$. 
Figure \ref{fig:EventFig}\textbf{D} depicts an example of a type 1 switching event.
The second type involves a flowing compartment moving from $\vec{V}_F$ to $\vec{V}_J$.
Here, the flowing compartment reaches the switching threshold, $\Omega$, and becomes discrete and no longer follows the flowing dynamics (of Equation~\eqref{JSF-Flowing}).
The time at which this occurs is found by continuation, and computation is resumed.
Figure \ref{fig:EventFig}\textbf{E} depicts an example of a type 2 switching event.}

{
The third type of switching event involves a compartment jumping from $\vec{V}_F$ to $\vec{V}_J$.
In this case, a flowing compartment jumps down to the switching threshold $\Omega$, and becomes discrete.
{To highlight how the third type of switch event is necessary, we propose considering a simple model where two species, $X$ and $Y$, interact to produce a third species, $Z$, with $X$ also undergoing a  death dominated birth-death process. Suppose that the switching threshold is set as $\Omega = 1000$, and that $X(0) = 1001$, $Y(0) = 10$ and $Z(0) = 0$. Due to the death reaction of $X$, suppose that within the first (Flow) step, $X$ experiences some decay $\varepsilon < 1$. Therefore, $X(t_1) = 1001 - \varepsilon > \Omega = 1000$ is in a flowing (continuous) regime. Now let's suppose that in the next instance, $X$ and $Y$ combine to produce one $Z$. At this instance, $X$ must experience some integer loss (which we can assume to be 1, but in general may be any positive value). In this case, producing $Z$ will result in $X(t_2) < \Omega$ but non integer.}
{Therefore, to account for type three switch events, }let $v_i$ be the compartment switching from $\vec{V}_F$ to $\vec{V}_J$ due to a jump. 
In general, these types of jumps result in $v_i$ being non-integer, i.e. $v_i \notin \{0,1,\ldots,\Omega\}\cup(\Omega,\infty)$.
To ensure the values of \(v_{i}\) stay in \(\{0,1,\ldots,\Omega\}\cup(\Omega,\infty)\), we add another constraint to the process, initially proposed by \citeauthor{rebuli2017hybrid} \cite{rebuli2017hybrid}.
The idea is to randomly round the compartment value up or down to an integer in a way that conserves the average behaviour of the process.
Let $\hat{v}_i$ be the value of the flowing compartment $v_i$ after jumping down across the threshold $\Omega$ but before being rounded into \(\{0,1,\ldots,\Omega\}\cup(\Omega,\infty)\). We apply the following rule to round $v_i$ after the switch. We take  \(v_i = \lceil \hat{v}_i \rceil\) with probability \( \hat{v}_i-\lfloor \hat{v}_i \rfloor  \), and otherwise we round down by setting \(v_i = \lfloor \hat{v}_i \rfloor\). 
This ensures the expected value of the variable after the switch is \(\hat{v}_i\) as described under the flowing paradigm from which this compartment has come and that the variable remains in the domain \(\{0,1,\ldots,\Omega\}\cup(\Omega,\infty)\). Figure \ref{fig:EventFig}\textbf{F} depicts an example of a type 3 switching event.}


\section*{Results}

\subsection*{Effects of interventions to reduce transmission}

Simulation based inference (\eg~the particle filter) relies on being
able to efficiently simulate from the generative process. To
demonstrate how JSF can be used in this setting, we first present a
simulation study of forecasting the elimination of a infectious
pathogen.

\subsubsection*{SIRS model with demography}

The SIRS model with demography is an extension of the classic
susceptible-infectious-removed (SIR) model. In the SIR model
\emph{susceptible} individuals may be infected by contact with
\emph{infectious} individuals; infected individuals eventually cease
to be infectious and transition to the \emph{removed} compartment. 
The SIRS model with demography extends the SIR model by allowing removed
individuals to transition back to being susceptible to infection,
for individuals to give birth to new (susceptible) individuals, and to allow for death of individuals
(at equal rates). The SIRS model with demography can be written as the following ODE system, with susceptible (S), infectious (I), and recovered (R) individuals:
\begin{equation}\label{eq:sirs-odes}
    \begin{split}
        \frac{dS}{dt} &=  - \beta \frac{IS}{N} + \omega R + \kappa N - \mu S,\\
        \frac{dI}{dt} &= \beta \frac{IS}{N} - \gamma I - \mu I,\\
        \frac{dR}{dt} &= \gamma I - \omega R - \mu R,
    \end{split}
\end{equation}
where $N(t) = S(t) + I(t) + R(t)$ is the total population size, $\beta$ the infection rate, $\gamma$ the recovery rate, $\omega$ the immunity waning rate, $\kappa$ the birth rate, and $\mu$ the death rate.
Figure~\ref{fig:fig-2}A shows a compartmental diagram of the SIRS model with demography.

Figure~\ref{fig:fig-2}B shows our simulated (via the Doob-Gillespie algorithm) time series of daily noisy measurements of the prevalence of infection.
The true parameters used in the simulation are shown in Figure~\ref{fig:fig-2}C; these
values are broadly consistent with existing estimates for influenza or SARS-CoV-2.
We assume that these measurements are drawn from a negative binomial distribution
where the expected value is equal to the true prevalence of infection and there is a constant (known) dispersion parameter, $k=100$.

\subsubsection*{Estimating elimination probabilities}

\begin{figure*}[ht!]
    \centering
     \includegraphics[width=0.975\linewidth]{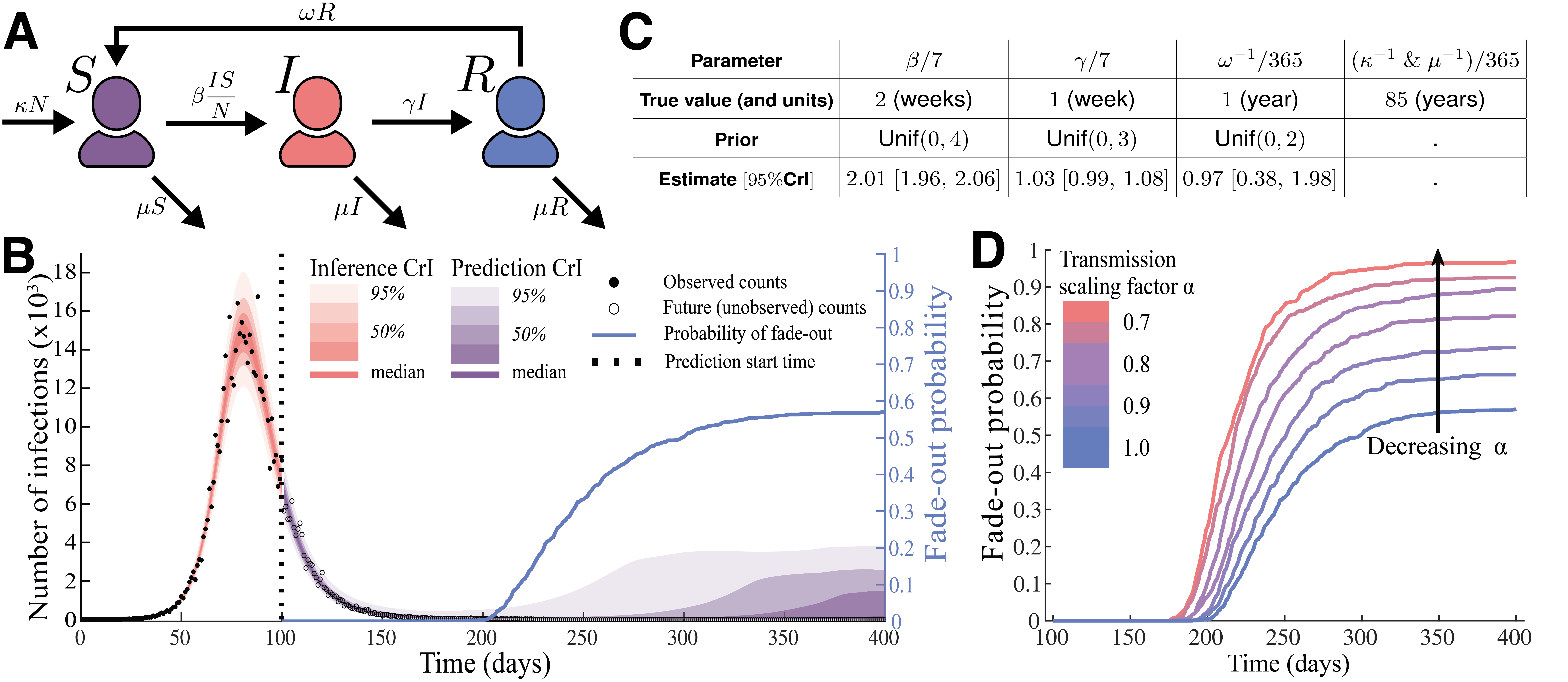}
     \caption{\label{fig:fig-2} JSF enables us to forecast when a pathogen will be eliminated from a population, \ie~when we can claim that an epidemic has ``ended'', and quantify the likely impact of varying levels of intervention. \textbf{A} A compartmental model representation of the SIRS with demography. \textbf{B} A simulated time series of noisy measurements of prevalence along with inferred prevalence trajectories and forecasts of future prevalence with associated uncertainty bands. The dashed vertical line indicates the date up until which we have data. The solid blue line indicates an estimate of the probability that the pathogen has been eliminated from the population (as opposed to persisting at very low levels). \textbf{C} The true parameters used in the simulation along with their posterior estimates and the associated priors used. Each of the marginal posterior distributions peaks tightly about the true value. \textbf{D} The probability of pathogen elimination increases with the strength of the intervention against transmission. Near certainty of elimination is achieved with a $30\%$ reduction of transmission. }
\end{figure*}

Combining the SIRS model with a particle filter \citep{Moss2024,kitagawa1996monte}
we used the first 100 days of the time series to forecast the remaining 150 days. 
Figure~\ref{fig:fig-2}B shows the estimated true prevalence across the first 100 days and forecasts the prevalence for the subsequent 150 days. Included in the forecast is a daily estimate of the probability that the pathogen has been eliminated (in the simulation the pathogen was eliminated on day 250).
It is important to note that there is an endemic equilibrium in this model, so it is reasonable for the probability to plateau as it does.

To demonstrate the utility of the posterior distribution (\ie~the capacity of the particle filter to forecast the epidemic after the 100 days of observed data), we considered the impact of a possible intervention. This intervention, introduced on day 100, reduces the force of infection (and hence the effective reproduction number) by a factor of $\alpha$. 
Figure~\ref{fig:fig-2}D shows how the probability of eliminating the pathogen increases substantially as we decrease $\alpha$ from $1$ (no intervention) to $0.7$ (a $30\%$ reduction in transmission).
The fade-out probability is calculated as the proportion of sampled (hybrid stochastic-deterministic) trajectories that go extinct (via $I = 0$) of the total number of sampled trajectories.

The full details of the configuration of the particle filter, and the marginal posterior distributions are given in \si{2E}.

\subsection*{Computational properties}

To further investigate the computational properties of JSF and to compare it with exact (Doob-Gillespie) simulation and (approximate) tau-leaping simulation we used repeated simulations from the SIRS model with demography.
We partitioned the simulations into three classes: early stochastic extinction, fade-out after a single epidemic, and sustained transmission (\si{2}.)
As seen in Table \hltmp{SI 3}, JSF and exact simulation generate very similar proportions of samples in each of these classes.
Two aspects of this system of practical importance are the timing and magnitude of peak infections, and the total number of infections.
As can be seen in Figure \hltmp{SI 7} the distribution of peak timing and total number of infections are in good agreement between JSF and exact simulation.
For the magnitude of peak infections, there is substantially less variability in the JSF samples than the exact samples, but the proportional difference is small.

With regards to computational efficiency, Figure \hltmp{SI 10B} shows the average time required per simulation using JSF, exact simulation (Doob-Gillespie), tau-leaping, {and the efficient Tau-hybrid method \cite{matthew2023gillespy2}} for the SIRS model for populations of different sizes.
For populations of size above about $10^{5}$ (\ie~approximately the size of a small city), JSF is substantially  {computationally} faster than either exact simulation or tau-leaping, {there being a difference of at least one order of magnitude in the amount of time needed, and (if a suitable Switching threshold is specified) a constant scaling, irrespective of compartment occupancy, may be obtained. To simplify the interpretation of the results in Figure \hltmp{SI 10B}, each of the methods were implemented in the same interpreted language so that the observed differences can be attributed to the algorithm rather than the platform on which it is implemented.}
{We have also included a comprehensive simulation case study of the simple birth-death process (see Section SI 2 of the supplementary information), which exhibits JSf computational efficiency for both birth dominated  and death dominated regimes, with comparable results to the exact stochastic process. We also demonstrate JSF's accuracy for particularly difficult to solve stochastic simulations which exhibit multiple time scales (see Section SI 3 of the supplementary information), where we see that JSF produces summary statistics and trajectories comparable to the exact stochastic process, out performing current hybrid methods.}

\subsection*{SARS-CoV-2 virus clearance informed by longitudinal data}

\begin{figure*}[t]
    \centering
     \includegraphics[width=0.975\linewidth]{figures/PLOS_Figs/Figure4.png}
     \caption{\label{fig:fig-3} The TEIRV model describes the within-host dynamics of host cells and virus during infection. The model accounts for delays in virus production post cellular infection and the capability of target cells to enter a refractory state as a defence against infection. \textbf{A} A compartmental diagram of the TEIRV model showing the interactions between: \textit{Target} cells ($T$), \textit{Refractory} cells ($R$), cells in the \textit{Eclipsed} phase of infection ($E$) and \textit{Infectious} cells ($I$); and the \textit{(Total) Virions} ($V$). Total Virion particles are utilised as to not distinguish between intra- and extra-cellular Virions. Solid arrows represent the exchange of mass (cells or virions) through compartments, while dashed lines represent the influence of a compartment on a reaction's rate of occurrence. \textbf{B} The representation of this model as a reaction network with a stoichiometric matrix. This formalises the dependency on different variables indicated by the dashed arrows in the compartmental diagram. \textbf{C} An example trajectory showing the exponential growth and decline in the amount of free virus across the duration of the infection, this process terminates in the virus, eclispse and infected cell populations reaching zero (shortly before time $12$). Reaching zero which is possible in this hybrid continuous/discrete stochastic model but does not occur in a purely ODE based representation.}
\end{figure*}

Understanding viral clearance is important when deciding how long treatment must be maintained.
This case study demonstrates how JSF enables inference of virus clearance, which has been computationally challenging to date \cite{yan2016extinction,farrukee2018characterization}.
In particular, we use a mathematical model --- the TEIRV model described below --- {to study viral clearance using longitudinal data from 6 individuals infected  with SARS-CoV-2 \cite{ke2022daily}}. 

\subsubsection*{TEIRV model of within-host viral dynamics}

The TEIRV model is an extension of the classic target-infected-virus
(TIV) model \cite{perelson2002modelling}. In the TIV model \emph{target}
cells may be \emph{infected} by the virus; before dying, infected
cells produce \emph{virus}; and the virus can degrade or infect
remaining target cells. The TEIRV extends the TIV model through the
inclusion of an \emph{eclipsed} compartment, to model the delay
between infection of a target cell and the subsequent production of
virus, and a \emph{refractory} compartment, to model heightened
antiviral defences of target cells, \eg~through the effects of
interferons \cite{ke2022daily,blancomelo2020imbalanced}. To simplify
the use of this model a quasi-steady-state approximation for
interferon production is employed. This approximation allows us to
avoid the need to explicitly model the amount of interferon present.

Figure~\ref{fig:fig-3}A shows a compartmental diagram of the TEIRV model.
Target cells becoming infected by virions at rate $\beta$, which then enter the eclipsed phase. 
These eclipsed cells then become infectious at rate $k$. 
Infectious cells are cleared from the population at rate $\delta$, and produce virions at rate $\pi$.
Virions are themselves cleared at rate $c$. We note that $V$ is the total number of virion particles, and therefore this model does not distinguish between intra- and extra-cellular virions. As a result, following the infection of a target cell ($T$), no virion particles are lost, in this model.
The infectious cells recruit interferon, which cause the target cells to become refractory (and hence protected against infection) at rate $\Phi$. 
The refractory cells return to a naive state as target cells at rate $\rho$.
These assumptions are represented with the following ODE system:
\begin{equation}\label{eq:teirv-odes}
    \begin{split}
        \frac{dT}{dt} &= - \beta V T - \Phi I T + \rho R,\\
        \frac{dE}{dt} &= \beta V T - k E,\\
        \frac{dI}{dt} &= k E - \delta I,\\
        \frac{dV}{dt} &= \pi I - c V,\\
        \frac{dR}{dt} &= \Phi I T - \rho R.
    \end{split}
\end{equation}

The stoichiometric matrix corresponding to these assumptions is shown in Figure~\ref{fig:fig-3}B. 
Figure~\ref{fig:fig-3}C, which shows a trajectory sampled from the TEIRV model when represented with JSF; the classic boom-bust dynamics of the viral populations can be seen in the exponential growth and decline of the $V$ compartment.
Unlike solutions of the ODE model in~\eqref{eq:teirv-odes}, we see that by time $t=14$, the populations of eclipsed and infected cells, and the virus have gone extinct, indicating a definitive end to the infection.

The basic reproduction number, $\mathcal{R}_{0}$ is a fundamental quantity of epidemiological models. 
For the TEIRV model, $\mathcal{R}_{0}$ can be calculated from the next-generation matrix \cite{diekmann2010construction}:
\begin{equation}\label{eq:r0}
\mathcal{R}_0 = \frac{\pi \beta T(0)}{\delta c}.
\end{equation}

\subsubsection*{Observation model linking virus to time series}

The longitudinal data contains cycle number (CN) values from nasal samples. 
The CN values are inversely proportional to the number of virions present. 
In our analysis we used an existing (empirical) model to link the cycle numbers to the (logarithmic) viral genome load \cite{ke2022daily}: $\log_{10}V = 11.35 - 0.25 \text{CN}$.
We assume the observed values are drawn from a normal distribution with mean $\log_{10}V$ and unit standard deviation, and that the values are truncated at the detection limit of $-0.65$.

\subsubsection*{Estimating virus reproduction and clearance}

Recall that the particle filter is capable of combining a mechanistic model of the within-host dynamics, such as the TEIRV, and an observation model, such as the viral load measurements, and will return a (Bayesian) posterior sample of both the parameters of the process and the trajectory of virus and cell populations through time.
We assume that the infection begins with a single exposed target cell (in a population of $8\times 10^{7}$ target cells) \cite{ke2022daily}.
This gives us an initial condition for the process: $T(0) = 8 \times 10^7$, $E(0) = 1$, $I(0) = 0$, $R(0) = 0$. 
We leave the initial viral load, $V(0)$, as a parameter to be fit to the data.

We selected six patient time series \cite{ke2022daily}, choosing ones that contained a full 14 data points, for both consistency and simplicity.
Two of the model parameters were fixed as in previous analysis \cite{ke2022daily}: $c=10$, $k=4$.
See \si{3} for full details of the particle filter and JSF configuration.

\begin{figure*}[ht!]
    \centering
     \includegraphics[width=0.975\linewidth]{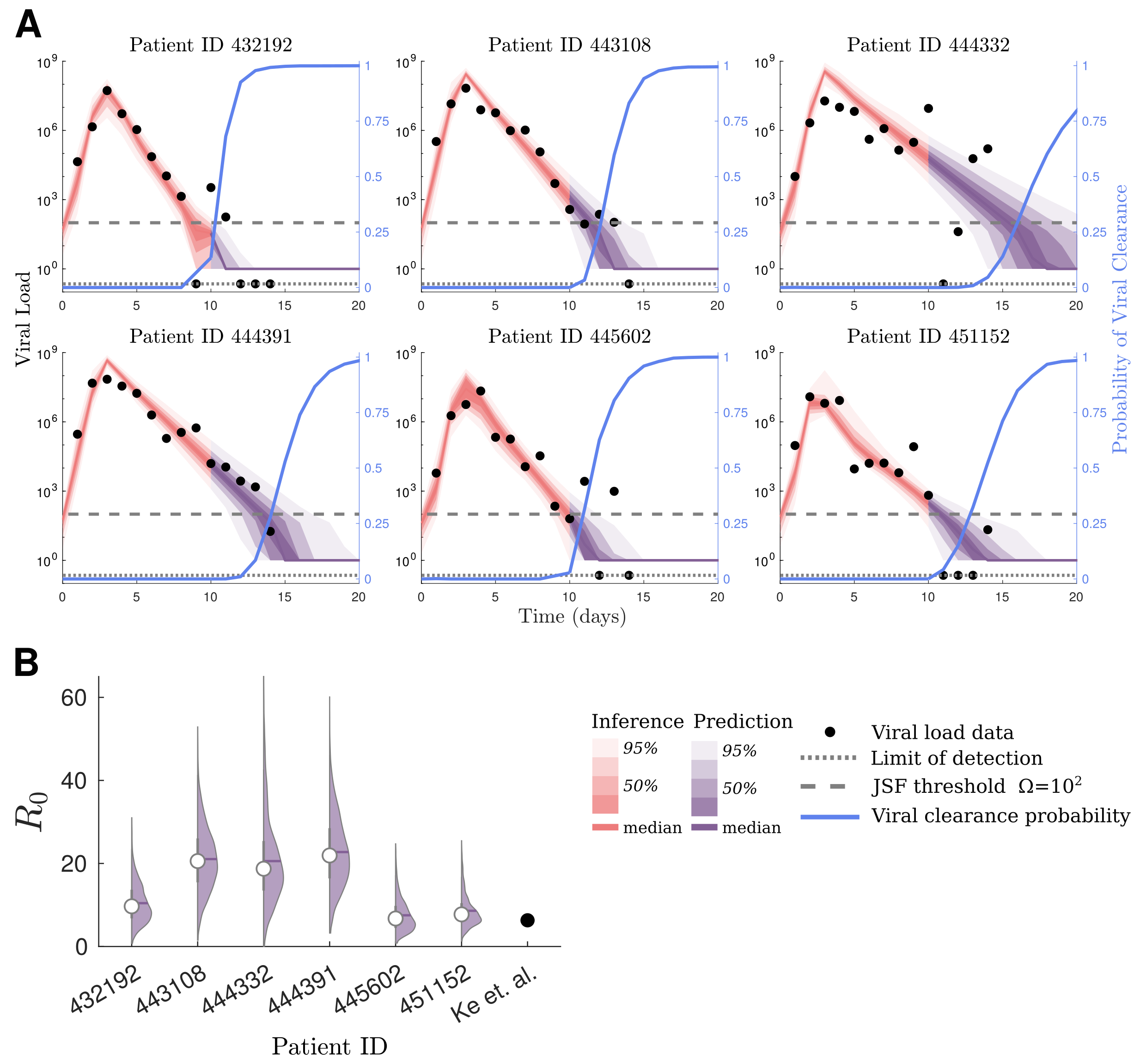}
     \caption{\label{fig:fig-4} JSF enables the inference of the viral load through time and the probability that the virus has been cleared from longitudinal data of SARS-CoV-2 viral load measurements. \textbf{A} Model fits to viral load data based on nasal swabs of six patients from \cite{ke2022daily} using a refractory cell within-host model, using our Jump-Switch-Flow method. Nasal viral load data is shown as solid dots, and the limit of detection is shown as a dotted line. We show the switching threshold, $\Omega = 10^2$ by a dashed line. The red and purple bands show mean (0\%), 25\%, 50\%, 75\% and 95\% credible intervals (from dark to light) for the inference and prediction, respectively. We also estimate the probability of viral clearance at a given point in time, based on the previous data to that point, in blue (light). \textbf{B} The posterior distribution of viral $R_0$ for each patient, along with the median (white circle), These patient specific estimates are contrasted with the single point estimate from previous analysis \cite{ke2022daily}.}
\end{figure*}

Figure~\ref{fig:fig-4}A shows our model fits to the first 10 days of viral load data for the six selected patients. After day 10, using the estimated parameters, we generate and predict the distribution of subsequent viral load until day 20.
The estimated viral peak coincides with the data for each of the patients and the predicted viral trajectories closely match subsequent observations.  

{We estimate the probability of viral clearance over time (right blue axis of in the panels of Figure~\ref{fig:fig-4}A). Viral clearance is defined by the point at which the virion, eclipsed, and infected cell populations all reach extinction. We note that this is a conservative measure if one were only interested in when a patient's \textit{infectiousness} becomes negligible. However, for proof of concept of inferring such quantities, we have chosen to estimate the probability of complete viral clearance.}

There is clear heterogeneity in the amount of time required to clear the virus (Figure~\ref{fig:fig-4}A).
We estimate that with high certainty patients 423192, 443108 and 445602 clear the virus within 20 days. 
In contrast, for patients 444332, 444391 and 451152 we infer that they have not cleared the virus within that time frame. 
Within those who did clear the virus, we observe that different patients require different amounts of time to clear the virus. 
Specifically, patient 432192 (\ref{fig:fig-4}A) is inferred to have cleared the virus first. Patients 443108 and 445602 both require an estimated 16 days to obtain viral clearance. 

{The posterior distributions of the parameters (Figure \hltmp{SI 11}) and priors (Table \hltmp{SI 5}), are given in \si{5}}. 
Because of potential identifiability issues with the rate parameters, we instead compare the estimated reproduction number, $R_0$, to results from previous analysis \citep{ke2022daily}.
Figure~\ref{fig:fig-4}B shows the posterior distributions of $R_0$, as well as the point estimates from \cite{ke2022daily}.
Since the previous analysis, \citep{ke2022daily}, reports identical estimates for $\pi$, $\beta$ and $\delta$ across these patients, we only include a single point estimate for their work.
Our estimate is consistent with previous results \citep{ke2022daily} for patients 432192, 445602 and 451152.
However, patients 443108, 444332 and 444391 all have higher $R_0$ estimates in our results. 
Our $R_0$ estimates are consistent with similar within-host viral infection analyses of other respiratory pathogens \citep{baccam2006kinetics,hernandez2020host, gubbins2024quantifying}.

\section*{Discussion} \label{discussion}
We have presented a simple hybrid simulation method for compartmental models: \emph{Jump-Switch-Flow} (JSF). This method facilitates efficient simulation of multi-scale models. Via simulation studies and an analysis of longitudinal data from SARS-CoV-2 infections we demonstrated the desirable computational properties of this method and the types of novel analyses which it enables.
JSF allows compartments to dynamically change between stochastic and deterministic behaviour.
Combining JSF with a simulation based inference method, \eg~a particle filter, enables inference for multi-scale models, and in situations where absorbing states (such as a population going extinct) are important.

Simulations shows JSF produces trajectories largely consistent with gold standard exact simulation techniques, \eg~Doob-Gillespie algorithm (see for example Figures \hltmp{SI 1}--\hltmp{SI 4} and \hltmp{SI 6} -- \hltmp{SI 9}). {We demonstrate how JSF is highly accurate at producing summary statistics comparable to gold standard exact simulation techniques (see Figures SI 1 -- SI 4), with a significant improvement in accuracy for problems with multiple time scales when compared to currently available hybrid methods (see Figure SI 3 and SI 4)}
However, JSF is also much faster for realistic population sizes (as can be seen in the comparison of computational speed in Figure \hltmp{SI 10B}). 
We demonstrate how our approach can be incorporated into a particle filter \citep{Moss2024}, to perform parameter and state estimation and prediction (see Figure~\ref{fig:fig-2}). 

{A strength of JSF is the capacity to infer when a process has reached an absorbing state, for example, an epidemic fading-out, or a virus being cleared. The ability to infer when an absorbing state has been reached is important for calibrating intervention measures \citep{parag2020exact} and understanding immune response \citep{yan2016extinction}. ODE based models rarely have absorbing states that can be reached in finite time, which fundamentally limits their capacity to describe extinction and clearance processes.}

We analysed viral load for SARS-CoV-2 infections using the JSF within a particle filter with a TEIRV, refractory cell model \citep{ke2022daily}.
In the subset of the data considered, we find consistent parameter values between patients (\si{3}). 
In a novel analysis, we estimated the probability of viral clearance through time, finding substantial heterogeneity in the time until viral clearance (see Figure~\ref{fig:fig-4}A). 
This is important as an accurate quantification of when the virus has been cleared is crucial for determining appropriate treatment regimes.

A key advantage of JSF is the ability to combine the stochasticity of discrete small populations with convenient deterministic models for large populations.
The point at which a compartment will transition between these descriptions is specified by the threshold parameters $\Omega_{i}$ (one for each compartment).
Setting these parameters to low values speeds up computation, while setting them higher captures more of the stochasticity in the process.
As demonstrated in the simulation studies, an appropriate value can be determined through some preliminary simulations (see SI 4 for an example of preliminary simulation analysis for the SIRS model example).
An important consideration in selecting this parameter is ensuring that the absorbing states can still be reached (\eg~compartment extinction), and that stable and steady states can be perturbed by random fluctuations. 
{Potential considerations when specifying the values $\Omega_i$ may also include preliminary simulations, increasing $\Omega_i$ and observing how the computational complexity scales (see Figure SI 10).}
Furthermore, developing theory to better understand when a CTMC is well approximated by an ODE is at the core to determine how the switching threshold should be chosen. {Understanding into the numerical properties of the Jump-Switch-Flow process, and how convergence of the method behaves for arbitrary choices of Flow event solvers will also prove to be an exciting area of further research.}
However, these lies outside the scope of this work.

Extensions include the incorporation of an intermittent Stochastic Differential Equation approximation between the CTMC and ODE regimes, and extension to spatial processes which exhibit both stochastic and deterministic behaviours.

The modelling framework presented in this paper has the potential to change the way compartmental models are developed and calibrated, moving us towards more accurate and more efficient hybrid methods. These models will help to form the basis of informed decision making, based on realistic and accurate descriptions of the system. 
This has broad applicability, from ecological models, chemical systems and single cell models, to infectious diseases and within-host models.

\section*{Acknowledgement}
We thank Ada Yan for helpful discussions regarding within-host modelling.
J.A.F.'s research is supported by the Australian Research Council (FT210100034, CE230100001) and the National Health and Medical Research Council (APP2019093). S.H.'s research is supported by the Australian Research Council (DP200101281).

\section*{Data Availability}
The manuscript has associated data in a data repository available at \url{https://github.com/DGermano8/JSFGermano2024}. Original data related to SARS-CoV-2 viral load measurements can also be found at \cite{ke2022daily}.

\section*{Code Availability}
All code to reproduce our results is available at \url{https://github.com/DGermano8/JSFGermano2024}.
A Python package implementing the Jump-Switch-Flow method is available at \url{https://DGermano8.github.io/JSF}.

\section*{Author Contributions}
J.A.F and M.B.F conceptualised research;
D.P.J.G, A.E.Z, S.H., R.M, J.A.F and M.B.F designed research;
D.P.J.G and A.E.Z performed research;
D.P.J.G, A.E.Z and R.M contributed software tools;
D.P.J.G and A.E.Z analyzed data;
J.A.F and M.B.F supervised research;
D.P.J.G and A.E.Z wrote the paper;
D.P.J.G, A.E.Z, S.H., R.M, J.A.F and M.B.F reviewed the paper;
J.A.F acquired funding.

\printbibliography[heading=bibintoc]

@article{kreger2021hybrid,
  title={A hybrid stochastic-deterministic approach to explore multiple infection and evolution in HIV},
  author={Kreger, J. and Komarova, N.L. and Wodarz, D.},
  journal={PLOS Computational Biology},
  volume={17},
  number={12},
  pages={e1009713},
  year={2021},
  publisher={Public Library of Science San Francisco, CA USA}
}

@article{bressloff2014path,
  title={Path integrals and large deviations in stochastic hybrid systems},
  author={Bressloff, P.C. and Newby, J.M.},
  journal={Physical Review E},
  volume={89},
  number={4},
  pages={042701},
  year={2014},
  publisher={APS}
}

@article{matthew2023gillespy2,
  title={GillesPy2: a biochemical modeling framework for simulation driven biological discovery},
  author={Matthew, S. and Carter, F. and Cooper, J. and Dippel, M. and Green, E. and Hodges, S. and Kidwell, M. and Nickerson, D. and Rumsey, B. and Reeve, J. and Petzold, L.R., and Sanft, K.R., and Drawert, B.},
  journal={Letters in biomathematics},
  volume={10},
  number={1},
  pages={87},
  year={2023}
}

@article{Alahmadi2020ABC,
author = {Alahmadi, A.A.  and Flegg, J.A.  and Cochrane, D.G.  and Drovandi, C.C.  and Keith, J.M. },
title = {A comparison of approximate versus exact techniques for Bayesian parameter inference in nonlinear ordinary differential equation models},
journal = {Royal Society Open Science},
volume = {7},
number = {3},
pages = {191315},
year = {2020},
doi = {10.1098/rsos.191315},
}

@article{trindade2024hybrid,
  author =	 {Trindade, T.T. and Zygalakis, K.C. },
  title =	 {A hybrid tau-leap for simulating chemical kinetics
                  with applications to parameter estimation},
  journal =	 {Royal Society Open Science},
  volume =	 11,
  number =	 12,
  pages =	 240157,
  year =	 2024,
  doi =		 {10.1098/rsos.240157}
}

@article{kynaston2023regime,
  author =	 {Kynaston, J.C. and Yates, C.A. and
                  Hekkink, A.V.F. and Guiver, C.},
  title =	 {{T}he regime-conversion method: a hybrid technique
                  for simulating well-mixed chemical reaction
                  networks},
  journal =	 {Frontiers in Applied Mathematics and Statistics},
  volume =	 9,
  year =	 2023,
  doi =		 {10.3389/fams.2023.1107441}
}

@book{anderson1991infectious,
  title={Infectious diseases of humans: dynamics and control},
  author={Anderson, R.M. and May, R.M.},
  year={1991},
  publisher={Oxford University Press}
}

@article{ludwig1978qualitative,
  title={Qualitative analysis of insect outbreak systems: the spruce budworm and forest},
  author={Ludwig, D. and Jones, D.D. and Holling, C.S.},
  journal={Journal of Animal Ecology},
  volume={47},
  number={1},
  pages={315--332},
  year={1978}
}

@article{yan2016extinction,
  author =	 "Yan, A.W.C. and Cao, P. and McCaw, J.M.",
  title =	 {{O}n the extinction probability in models of
                  within-host infection: the role of latency and
                  immunity},
  journal =	 "Journal of Mathematical Biology",
  year =	 2016,
  day =		 01,
  volume =	 73,
  number =	 4,
  pages =	 "787--813",
  doi =		 "10.1007/s00285-015-0961-5"
}

@article{pearson2011stochastic,
  doi =		 {10.1371/journal.pcbi.1001058},
  author =	 {Pearson, J.E. AND Krapivsky, P. AND Perelson, A.S.},
  journal =	 {PLOS Computational Biology},
  publisher =	 {Public Library of Science},
  title =	 {{S}tochastic {T}heory of {E}arly {V}iral
                  {I}nfection: {C}ontinuous versus {B}urst
                  {P}roduction of Virions},
  year =	 2011,
  volume =	 7,
  pages =	 {1--17},
  number =	 2
}

@article{blancomelo2020imbalanced,
  author =	 {Blanco-Melo, D. and Nilsson-Payant, B.E.
                  and Liu, W.C. and Uhl, S. and Hoagland,
                  D. and M{\o}ller, R. and Jordan, T.X.
                  and Oishi, K. and Panis, M. and Sachs,
                  D. and Wang, T.T. and Schwartz, R.E. and
                  Lim, J.K. and Albrecht, R.A. and tenOever,
                  B.R.},
  title =	 {{I}mbalanced {H}ost {R}esponse to {SARS}-{C}o{V}-2
                  {D}rives {D}evelopment of {COVID}-19},
  journal =	 "Cell",
  year =	 2020,
  day =		 28,
  publisher =	 "Elsevier",
  volume =	 181,
  number =	 5,
  pages =	 "1036--1045.e9",
  doi =		 "10.1016/j.cell.2020.04.026"
}

@article{perelson2002modelling,
  author =	 "Perelson, A.S.",
  title =	 {{M}odelling viral and immune system dynamics},
  journal =	 "Nature Reviews Immunology",
  year =	 2002,
  day =		 01,
  volume =	 2,
  number =	 1,
  pages =	 "28--36",
  doi =		 "10.1038/nri700"
}

@article{gubbins2024quantifying,
  title={Quantifying the relationship between within-host dynamics and transmission for viral diseases of livestock},
  author={Gubbins, S.},
  journal={Journal of the Royal Society Interface},
  volume={21},
  number={211},
  pages={20230445},
  year={2024},
  publisher={The Royal Society},
doi = {10.1098/rsif.2023.0445}
}

@article{hernandez2020host,
title = {{I}n-host {M}athematical {M}odelling of {COVID}-19 in {H}umans},
journal = {Annual Reviews in Control},
volume = {50},
pages = {448--456},
year = {2020},
doi = {10.1016/j.arcontrol.2020.09.006},
author = {E.A. Hernandez-Vargas and J.X. Velasco-Hernandez},
}

@article{Moss2024, 
doi = {10.21105/joss.06276}, 
url = {10.21105/joss.06276}, 
year = {2024}, publisher = {The Open Journal}, 
volume = {9}, 
number = {96}, 
pages = {6276}, 
author = {R. Moss}, 
title = {pypfilt: a particle filter for Python}, 
journal = {Journal of Open Source Software} 
}

@article{gillespie2001approximate,
  title={Approximate accelerated stochastic simulation of chemically reacting systems},
  author={Gillespie, D.T.},
  journal={The Journal of chemical physics},
  volume={115},
  number={4},
  pages={1716--1733},
  year={2001},
  publisher={American Institute of Physics}
}

@article{fowler2021atto,
  author =	 "Fowler, A.C.",
  title =	 {{A}tto-{F}oxes and {O}ther {M}inutiae},
  journal =	 "Bulletin of Mathematical Biology",
  year =	 2021,
  day =		 31,
  volume =	 83,
  number =	 10,
  pages =	 104,
  doi =		 "10.1007/s11538-021-00936-x"
}

@article{rebuli2017hybrid,
  title={Hybrid Markov chain models of S--I--R disease dynamics},
  author={Rebuli, N.P. and Bean, N.G. and Ross, J.V.},
  journal={Journal of Mathematical Biology},
  volume={75},
  pages={521--541},
  year={2017},
  publisher={Springer}
}

@article{Kurtz_1970, 
title={{S}olutions of ordinary differential equations as limits of pure jump {M}arkov processes}, 
volume={7},
DOI={10.2307/3212147}, 
number={1}, 
journal={Journal of Applied Probability}, 
    author = {Kurtz, T.G.},
year={1970}, 
pages={49--58}
}

@article{Kurtz_1971, 
title={{L}imit theorems for sequences of jump {M}arkov processes approximating ordinary differential processes}, 
volume={8}, 
DOI={10.2307/3211904}, 
number={2}, 
journal={Journal of Applied Probability}, 
    author = {Kurtz, T.G.},
year={1971}, 
pages={344--356}
}

@article{kurtz1972relationship,
    author = {Kurtz, T.G.},
    title = "{The Relationship between Stochastic and Deterministic Models for Chemical Reactions}",
    journal = {The Journal of Chemical Physics},
    volume = {57},
    number = {7},
    pages = {2976--2978},
    year = {1972},
    doi = {10.1063/1.1678692},
}

@article{gillespie1976general,
title = {A general method for numerically simulating the stochastic time evolution of coupled chemical reactions},
journal = {Journal of Computational Physics},
volume = {22},
number = {4},
pages = {403--434},
year = {1976},
doi = {10.1016/0021-9991(76)90041-3},
author = {D.T. Gillespie}
}

@article{ionides2006inference,
  author={Ionides, E.L. and Bret{\'o}, C. and King, A.A.},
title = {Inference for nonlinear dynamical systems},
journal = {Proceedings of the National Academy of Sciences},
volume = {103},
number = {49},
pages = {18438--18443},
year = {2006},
doi = {10.1073/pnas.0603181103},
}

@article{diekmann2010construction,
  author =	 {Diekmann, O. and Heesterbeek, J.A.P. and Roberts,
                  M.G. },
  title =	 {{T}he construction of next-generation matrices for
                  compartmental epidemic models},
  journal =	 {Journal of The Royal Society Interface},
  volume =	 7,
  number =	 47,
  pages =	 {873--885},
  year =	 2010,
  doi =		 {10.1098/rsif.2009.0386}
}

@article{sanft2015constant,
  author = {Sanft, K.R. and Othmer, H.G.},
    title = "{Constant-complexity stochastic simulation algorithm with optimal binning}",
    journal = {The Journal of Chemical Physics},
    volume = {143},
    number = {7},
    pages = {074108},
    year = {2015},
    doi = {10.1063/1.4928635},
}

@article{gillespie2000chemical,
      author = {Gillespie, D.T.},
    title = "{The chemical Langevin equation}",
    journal = {The Journal of Chemical Physics},
    volume = {113},
    number = {1},
    pages = {297--306},
    year = {2000},
    doi = {10.1063/1.481811},
}

@article{rao2003stochastic,
  author = {Rao, C.V. and Arkin, A.P.},
    title = "{Stochastic chemical kinetics and the quasi-steady-state assumption: Application to the Gillespie algorithm}",
    journal = {The Journal of Chemical Physics},
    volume = {118},
    number = {11},
    pages = {4999--5010},
    year = {2003},
    doi = {10.1063/1.1545446},
}

@article{cao2006efficient,
    author = {Cao, Y. and Gillespie, D.T. and Petzold, L.R.},
    title = "{Efficient step size selection for the tau-leaping simulation method}",
    journal = {The Journal of Chemical Physics},
    volume = {124},
    number = {4},
    pages = {044109},
    year = {2006},
    doi = {10.1063/1.2159468},
}

@article{lobry2015migrations,
  title={{M}igrations in the {R}osenzweig-{M}ac{A}rthur model and the ``atto-fox'' problem},
  author={Lobry, C. and Sari, T.},
  journal={Revue Africaine de la Recherche en Informatique et Math{\'e}matiques Appliqu{\'e}es},
  volume={20},
  pages={95--125},
  year={2015}
}

@article{mollison1991dependence,
  title = {Dependence of epidemic and population velocities on basic parameters},
journal = {Mathematical Biosciences},
volume = {107},
number = {2},
pages = {255--287},
year = {1991},
doi = {10.1016/0025-5564(91)90009-8},
author = {D. Mollison},
}

@article{cotter2016error,
author = {Cotter, S.L. and Erban, R.},
title = {Error Analysis of Diffusion Approximation Methods for Multiscale Systems in Reaction Kinetics},
journal = {SIAM Journal on Scientific Computing},
volume = {38},
number = {1},
pages = {B144--B163},
year = {2016},
doi = {10.1137/14100052X},
}

@article{isaacson2013convergent,
  author = {Isaacson, S.A.},
    title = "{A convergent reaction-diffusion master equation}",
    journal = {The Journal of Chemical Physics},
    volume = {139},
    number = {5},
    pages = {054101},
    year = {2013},
    doi = {10.1063/1.4816377},
}

@article{flegg2014analysis,
author = {Flegg, M.B. and Chapman, S.J. and Zheng, L. and Erban, R.},
title = {Analysis of the Two-Regime Method on Square Meshes},
journal = {SIAM Journal on Scientific Computing},
volume = {36},
number = {3},
pages = {B561--B588},
year = {2014},
doi = {10.1137/130915844},
}

@article{ke2022daily,
  author =	 "Ke, R. and Martinez, P.P. and Smith, R.L. and Gibson, L.L. and Mirza, A. and Conte, M. and Gallagher, N. and Luo, CH.
                  and Jarrett, J. and Zhou, R. and Conte,
                  A. and Liu, T. and Farjo, M. and
                  Walden, K.K.O. and Rendon, G. and
                  Fields, C.J. and Wang, L. and
                  Fredrickson, R. and Edmonson, D.C. and
                  Baughman, M.E. and Chiu, K.K. and Choi,
                  H. and Scardina, K.R. and Bradley, S.
                  and Gloss, S.L. and Reinhart, C. and
                  Yedetore, J. and Quicksall, J. and
                  Owens, A.N. and Broach, J. and Barton, B.
                  and Lazar, P. and Heetderks, W.J. and
                  Robinson, M.L. and Mostafa, H.H. and
                  Manabe, Y.C. and Pekosz, A. and McManus,
                  D.D. and Brooke, C.B.",
  title =	 {{D}aily longitudinal sampling of {SARS}-{C}o{V}-2
                  infection reveals substantial heterogeneity in
                  infectiousness},
  journal =	 "Nature Microbiology",
  year =	 2022,
  day =		 01,
  volume =	 7,
  number =	 5,
  pages =	 "640--652",
  doi =		 "10.1038/s41564-022-01105-z"
}

@article{gibson2000efficient,
author = {Gibson, M.A. and Bruck, J.},
title = {Efficient Exact Stochastic Simulation of Chemical Systems with Many Species and Many Channels},
journal = {The Journal of Physical Chemistry A},
volume = {104},
number = {9},
pages = {1876--1889},
year = {2000},
doi = {10.1021/jp993732q},
}

@article{buckwar2011rungekutta,
  title =	 {Runge–{K}utta methods for jump-diffusion differential
                  equations},
  journal =	 {Journal of Computational and Applied Mathematics},
  volume =	 236,
  number =	 6,
  pages =	 {1155--1182},
  year =	 2011,
  doi =		 {10.1016/j.cam.2011.08.001},
  author =	 {E. Buckwar and M.G. Riedler}
}

@article{klein1984time,
  author =	 {Klein, R.W. and Roberts, S.D.},
  title =	 {{A} time-varying {P}oisson arrival process
                  generator},
  journal =	 {Simulation},
  year =	 1984,
  volume =	 43,
  number =	 4,
  pages =	 {193--195},
  doi =		 {10.1177/003754978404300406}
}

@article{simoni2019stochastic,
  author =	 {Simoni, G. and Reali, F. and Priami,
                  C. and Marchetti, L.},
  title =	 {{S}tochastic simulation algorithms for computational
                  systems biology: {E}xact, approximate, and hybrid
                  methods},
  journal =	 {WIREs Systems Biology and Medicine},
  volume =	 11,
  number =	 6,
  pages =	 {e1459},
  doi =		 {10.1002/wsbm.1459},
  year =	 2019
}

@article{angius2015approximate,
  title =	 {{A}pproximate analysis of biological systems by
                  hybrid switching jump diffusion},
  journal =	 {Theoretical Computer Science},
  volume =	 587,
  pages =	 {49--72},
  year =	 2015,
  doi =		 {10.1016/j.tcs.2015.03.015},
  author =	 {A. Angius and G. Balbo and M.
                  Beccuti and E. Bibbona and A. Horvath and
                  R. Sirovich}
}

@article{arulampalam2002tutorial,
  author={Arulampalam, M.S. and Maskell, S. and Gordon, N. and Clapp, T.},
  journal={IEEE Transactions on Signal Processing}, 
  title={A tutorial on particle filters for online nonlinear/non-{G}aussian {B}ayesian tracking}, 
  year={2002},
  volume={50},
  number={2},
  pages={174--188},
  doi={10.1109/78.978374}}

@article{kitagawa1996monte,
  author = {G. Kitagawa},
title = {Monte {C}arlo {F}ilter and {S}moother for {N}on-{G}aussian {N}onlinear {S}tate {S}pace {M}odels},
journal = {Journal of Computational and Graphical Statistics},
volume = {5},
number = {1},
pages = {1--25},
year = {1996},
publisher = {Taylor \& Francis},
doi = {10.1080/10618600.1996.10474692},
}

@article{farrukee2018characterization,
  author =	 {R. Farrukee and A.E. Zarebski and J.M. McCaw and
                  J.D. Bloom and P.C. Reading and A.C. Hurt},
  title =	 {{C}haracterization of {I}nfluenza {B} {V}irus
                  {V}ariants with {R}educed {N}euraminidase
                  {I}nhibitor {S}usceptibility},
  journal =	 {Antimicrobial Agents and Chemotherapy},
  volume =	 62,
  number =	 11,
  pages =	 {10.1128/aac.01081-18},
  year =	 2018,
  doi =		 {10.1128/aac.01081-18}
}

@article{baccam2006kinetics,
  author =	 {P. Baccam and C. Beauchemin and
                  C.A. Macken and F.G. Hayden and A.S. Perelson},
  title =	 {{K}inetics of {I}nfluenza {A} {V}irus {I}nfection in
                  Humans},
  journal =	 {Journal of Virology},
  volume =	 80,
  number =	 15,
  pages =	 {7590--7599},
  year =	 2006,
  doi =		 {10.1128/jvi.01623-05}
}

@article{parag2020exact,
  author =       {Parag, Kris V. AND Donnelly, Christl A. AND Jha, Rahul AND
                  Thompson, Robin N.},
  journal =      {PLOS Computational Biology},
  publisher =    {Public Library of Science},
  title =        {An exact method for quantifying the reliability of
                  end-of-epidemic declarations in real time},
  year =         2020,
  month =        11,
  volume =       16,
  pages =        {1--21},
  number =       11,
  doi =          {10.1371/journal.pcbi.1008478}
}

\makeatletter
\renewcommand{\theequation}{SI\arabic{equation}}
\renewcommand{\thefigure}{SI\arabic{figure}}
\renewcommand{\thesection}{SI\arabic{section}}
\renewcommand{\thetable}{SI\arabic{table}}
\renewcommand{\thealgorithm}{SI\arabic{algorithm}}
\setcounter{algorithm}{0}
\setcounter{equation}{0}
\setcounter{figure}{0}
\setcounter{table}{0}
\setcounter{section}{0}

\title{Supplementary Information: A hybrid framework for compartmental models enabling simulation-based inference}

\date{}
\maketitle
\def\thefootnote{*}\footnotetext{These authors contributed equally to this work}\def\thefootnote{\arabic{footnote}}

\section{Jump-Switch-Flow: implementation details}

To draw exact samples from the JSF process requires solving the differential equations for the flowing variables $\vec{V}_F$. 
For nonlinear systems these differential equations are usually intractable, so they are numerically integrated.
Here we describe an algorithm for approximately sampling trajectories, assuming we have a numerical solver for the differential equations (when viewed as an initial value problem (IVP)), to approximate the solutions.

\subsection{Jump-clock}
We implement the JSF algorithm with a slightly different jump clock as defined by $J_k$ in Section \ref{JSFJumpClockUpdates}. In particular, equate $u_k= \text{CDF}(t_k;k)$ but do not take the log of this expression before defining $J_k$ in Equation (\ref{eq:jump-timer}). As a result, in our codes we consider the jump clock $\tilde{J}_k$:
\begin{align} \label{jump-clock-implementation}
     \tilde{J}_{k}(t) = u_{k} - \left( 1 - \exp\left\{- \int_{t_{0}}^{t} \lambda_{k}(\vec{V}(s)) \ \mathrm{d}s  \right\} \right),
\end{align}
where $u_k\sim\text{Unif}(0,1)$ and the time of the previous jump event $t_0$.
Computing the times for the jump events follows a similar process as above. We compute the jump clock on the regular ODE mesh used to solve $\vec{V}_F$ in Equation \eqref{JSF-Flowing}. 
From Equation~\eqref{jump-clock-implementation} we can write \(\tilde{J}_k(t+\Delta t )\) in terms
of $\tilde{J}_k(t)$ and an integral of the reaction rate:

\begin{equation}\label{eq:jump-clock-exp}
  \tilde{J}_{k}(t+\Delta t)=\tilde{J}_k(t)+\left(\exp\left\{-\int_{t}^{t+\Delta t}\lambda_{k}(\vec{V}(s)) \ \mathrm{d}s \right\}-1\right)\left(\tilde{J}_k(t)+1-u_k\right).
\end{equation}

For steps where $\tilde{J}_k(t+\Delta t)>0$ we can do a single step of this
process to get the updated jump clock values. However, if $\tilde{J}_k(t+\Delta t)<0$ a jump has occurred at some time $t + \Delta \tau \in \left( t, t+\Delta t\right)$, and we need to solve $\tilde{J}_k(t+ \Delta \tau)=0$ to find when the jump occurred.
Therefore, we require a method to find $\Delta \tau$, where $0 < \Delta \tau < \Delta t$. To do this, we start by letting the right hand side
of Equation~\eqref{eq:jump-clock-exp} equal zero and rearrange to
get:
\begin{align} \label{eq:lambda_integral}
    \int_{t}^{t+\Delta \tau}\lambda_{k}(\vec{V}(s)) \ \mathrm{d}s = \ln\left\{\frac{\tilde{J}_k(t)+1-u_k}{1-u_k}\right\}.
\end{align}

If $\lambda_k$ is a positive constant then:
\begin{align}
    \Delta \tau = \frac{1}{\lambda_k} \ln\left\{\frac{\tilde{J}_k(t)+1-u_k}{1-u_k}\right\}.
\end{align}
If \(\lambda_k\) is not constant, we approximate $\lambda_k$, which we call $\widehat{\lambda}_k$, using the integration of $\vec{V}(t)$, as per Equation \eqref{JSF-Flowing}, since we know $\vec{V}(t)$ and $\vec{V}(t+\Delta t)$. Therefore, we also know \(\lambda_k\) at \(t\) and \(t+\Delta t\). We then write $\widehat{\lambda}_k$, at $t + \Delta \tau$  between times times $t$ and $t+\Delta t$ via linear interpolation:
\begin{equation}\label{eq:lambda-approx}
  \widehat{\lambda}_k(t+\Delta \tau)\approx\left(1 - \frac{\Delta \tau}{\Delta t}\right)\lambda_k(\vec{V}(t))+\frac{\Delta \tau}{\Delta t}\lambda_k(\vec{V}(t+\Delta t)).
\end{equation}

We can now substitute this approximation into Equation (\ref{eq:lambda_integral}), solve it, and rearrange for $\Delta \tau$. This gives a quadratic in $\Delta \tau$ with the following solution:

\begin{equation} \label{eq:firing-time}
\Delta \tau = \frac{\sqrt{\left(\Delta t \widehat{\lambda}_k(t)\right)^{2} + 2\alpha\Delta t\Delta\lambda_k}-\Delta t \widehat{\lambda}_k(t) }{\Delta\lambda_k},
\end{equation}
\noindent
where we have introduced the shorthand
\(\Delta\lambda_k=\lambda_{k}(\vec{V}(t+\Delta t))-\lambda_{k}(\vec{V}(t))\).

\subsection{Pseudocode details}
As with a typical IVP, we require the following data:
\begin{itemize}
    \item the initial condition of the system, $\vec{V}(0)$; 
    \item the final time $T_{\max}>0$;
    \item the stoichiometric Reactant, $\eta^{-}$, and Product, $\eta^{+}$, matrices, from which we derive the Exchange matrix, \mbox{$\eta = \eta^{+}-\eta^{-}$};
    \item a set of the associated propensities of the reactions, $\mathcal{R}$;
    \item a time step size for the numerical integrator, $\Delta t$;
    \item a vector of switching thresholds, $\mathbf{\Omega}$;
\end{itemize}
To sample the Jump-Switch-Flow process, we have identified both an \textit{exact} sampler, that samples the process exactly, and an \textit{operator-splitting} sampler, that uses some simplifying approximations that makes sampling substantially faster.

\subsubsection{Exact Jump-Switch-Flow sampler}
Here, we describe an \textit{exact} algorithm for sampling trajectories from Jump-Switch-Flow.
The algorithm to sample \emph{exactly} from Jump-Switch-Flow is given in Algorithm~\ref{alg:JSF-algorithm-E}.
First, initialise the jump clocks (line 3). 
On each iteration through the ODE loop, we first partition the reactions based on inclusion in $\mathcal{S}$ (line 6). 
We use the IVP solver to compute $\Delta \vec{V}_F(t)$ based on the flowing reactions (line 7). 
For each step of the ODE loop, we update the jump clocks (line 9). 
On lines 10--12, we determine if a jump event has occurred based on the sign of the the Jump-clock. 
If a jump event has occurred, compute the corresponding event time.
On line 13 we then rewind the state (and jump clocks) to when the jump occurred.
Resample the appropriate jump clock (line 14), and update the state to account for the jump (line 15). 
Repeat until the time has reached the maximum desired time.

\begin{algorithm}[h!]
\caption{An algorithmic description of the Exact Jump-Switch-Flow sampler.\label{alg:JSF-algorithm-E}}
\begin{algorithmic}[1]
\REQUIRE $\vec{V}(0)$, $\Delta t >0$ and $T_{\max}>0$
\STATE Initialise model state
\STATE $t \leftarrow 0$
\STATE Initialise jump clocks $u_i\sim\text{Unif}(0,1)$ for $i=1,\ldots,N$
\STATE {\color{gray} ($\triangleright$  Start ODE loop)}
\WHILE{$t < T_{\max}$}
    \STATE Compute Jumping reactions $\mathcal{S}(t)$
    \STATE Compute flow event, $\Delta \vec{V}_F(t)$
  \STATE {\color{gray} ($\triangleright$  Perform jump loop for ODE mesh step)}
      \STATE Update jump clocks, $\tilde{J}_k$, to $t+\Delta t$ {\, \color{gray} ($\triangleright$ As per Equation (\ref{delJ}))}
      \IF{any $\tilde{J}_k \leq 0$ {\color{gray} ($\triangleright$ Reaction Occurred)}}
          \STATE Identify first fired Reaction, $j$
          \STATE Compute time of jump event, $\Delta \tau$ {\, \color{gray} ($\triangleright$ As per Equation (\ref{jumptime}))}
          \STATE Reverse jump clocks to time $t + \Delta \tau$
          \STATE Resample $u_j\sim \text{Unif}(0,1)$
          \STATE Update both: $\vec{V}_J(t+\Delta \tau) = \vec{V}_J(t) + \mathbf{\eta}_j$ \textbf{and} $\vec{V}_F(t+\Delta \tau) = \vec{V}_F(t) + \Delta \tau \, \Delta \vec{V}_F(t) + \mathbf{\eta}_j$
          \STATE $t \leftarrow t + \Delta \tau$
      \ELSIF{ $\tilde{J}_k > 0, \forall k$ }
        \STATE Update: $\vec{V}_F(t+\Delta t) = \vec{V}_F(t) + \Delta t \, \Delta \vec{V}_F(t) $
  \STATE $t \leftarrow t + \Delta t$
      \ENDIF
\ENDWHILE
\end{algorithmic}
\end{algorithm}

\subsubsection{Operator Splitting Jump-Switch-Flow sampler}
Here, we describe the algorithm for the \textit{operator-splitting} algorithm, for sampling trajectories from Jump-Switch-Flow, which utilises some simplifying approximations to improve sampling performance.

The Operator Splitting Jump-Switch-Flow sampler is described in Algorithm \ref{alg:JSF-algorithm-OS}. Unlike the Exact JSF sampler, when a jump event occurs, we continue to integrate the system with the same flow event, $\Delta \vec{V}_F(t)$, instead of resampling it as before. This, in turn, ensures that the ODE is not continually being resampled. While this adds complexity at the implementation stage, significant computational time saving is achieved. 

We first initialise the Jump-Switch-Flow process by initialising all the jump clocks (line 3). 
On each iteration through the ODE loop, we first partition the reactions based on inclusion in $\mathcal{S}$ (line 6). 
We use the IVP solver to compute $\Delta \vec{V}_F(t)$ based on the flowing reactions (line 7). 
We now iterate through this current time-step, $\Delta t$, checking if any jump events occur (lines 10 -- 24), named the \textit{jump loop}. 
First, we track how much ``relative time'' ($\delta t$) passes through this jump loop (line 9). We then update the jump clocks to the end of the jump loop, accounting for any relative time that has occurred (line 11). 
As with the Exact JSF sampler, on lines 12 -- 14, we determine if a jump event has occurred using the Jump-clock. 
This gives us which Reaction has occurred, and at what time, $t+\Delta \tau$. On line 15 we then reverse the system to the time the jump event occurs, noting that we must reverse the excess between when the jump event occurred and the remainder of the time-step. We then resample the associated jump clock that just fired (line 16), and update the whole state (i.e. both flowing and Jumping variables) to the current time (line 17). We update the current time to time $t+\Delta \tau$ (line 18) and track the relative time in the jump loop (line 19). We then continue through the jump look. If no jump events have occurred, we leave the loop (line 21). We then simply update the whole state according to the flow event calculated, accounting for any time spent inside the jump loop (line 24).

\begin{algorithm}[h!]
\caption{An algorithmic description of the Operator Splitting Jump-Switch-Flow sampler .}\label{alg:JSF-algorithm-OS}
\begin{algorithmic}[1]
  \REQUIRE $\vec{V}(0)$, $\Delta t >0$ and $T_{\max}>0$
  \STATE Initialise model state
  \STATE $t \leftarrow 0$
  \STATE Initialise jump clocks $u_i\sim\text{Unif}(0,1)$ for $i=1,\ldots,N$
  \STATE {\color{gray} ($\triangleright$  Start ODE loop)}
  \WHILE{$t < T_{\max}$}
        \STATE Compute jumping reactions $\mathcal{S}(t)$
        \STATE Compute flow event, $\Delta \vec{V}_F(t)$
      \STATE {\color{gray} ($\triangleright$  Perform jump loop for ODE mesh step)}
      \STATE Set $\delta t \leftarrow 0$
      \WHILE{$\Delta t > \delta t $}
          \STATE Update jump clocks, $\tilde{J}_k$, to $t+(\Delta t-\delta t)$ 
          \IF{any $\tilde{J}_k \leq 0$ {\color{gray} ($\triangleright$ Reaction Occurred)}}
              \STATE Identify first fired Reaction, $j$
              \STATE Compute time of jump event, $t+\Delta \tau$ 
              \STATE Reverse jump clocks to time $t + \Delta \tau$
              \STATE Resample $u_j\sim \text{Unif}(0,1)$
              \STATE Update both: $\vec{V}_J(t+\Delta \tau) = \vec{V}_J(t) + \mathbf{\eta}_j$ \textbf{and} $\vec{V}_F(t+\Delta \tau) = \vec{V}_F(t) + \Delta \tau \, \Delta \vec{V}_F(t) + \mathbf{\eta}_j$
              \STATE $t \leftarrow t + \Delta \tau$
              \STATE $\delta t \leftarrow \delta t + \Delta \tau$
          \ELSIF{ $\tilde{J}_k > 0, \forall k$ }
            \STATE Leave while loop
          \ENDIF
      \ENDWHILE
      \STATE Update: $\vec{V}_F(t+(\Delta t - \delta t)) = \vec{V}_F(t) + (\Delta t - \delta t) \, \Delta \vec{V}_F(t) $
      \STATE $t \leftarrow t + (\Delta t - \delta t)$
  \ENDWHILE
\end{algorithmic}
\end{algorithm}

\section{{Simulation Study: Birth-death processes} \label{sec:BD-processes}}
{Fundamentally, one can conceptualise any compartmental model as a collection of interacting birth-death processes. For this reason, we will consider two very simple such processes: 1) a birth dominated birth-death process, and 2) a death dominated birth-death process. In this section, we will perform a simulation study to highlight the advantages of our Jump-Switch-Flow method, and compare the results to the gold standard, Doob-Gillespie Exact method, and the highly computationally efficient Tau-hybrid method, which is provided by GillesPy2, written in both C++ and Python.}

{The exact model we will utilise is a single species, $X$, experiencing birth with rate $\alpha$, which results in $X \rightarrow 2X$, and death with rate $\beta$, which results in $X \rightarrow \varnothing$.}

\subsubsection{{A birth dominated birth-death process}}
\begin{figure}[H]
    \centering
    \includegraphics[width=0.98\linewidth]{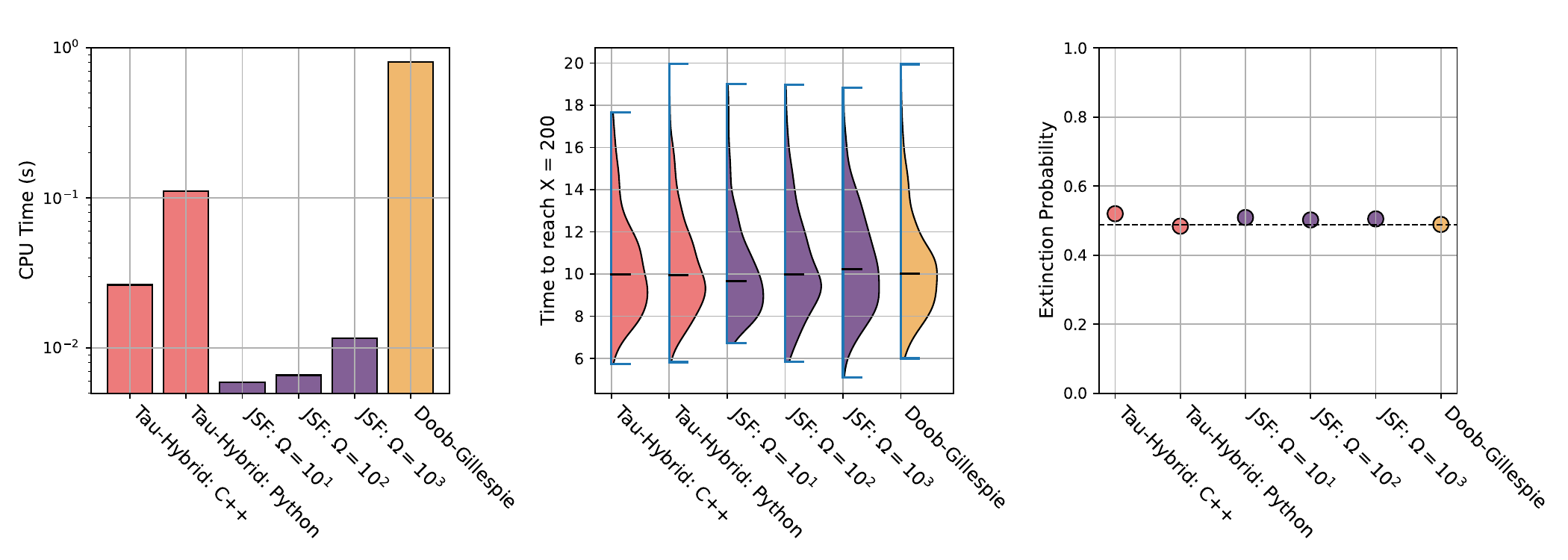}
    \caption{{Summary statistics for the birth dominated birth-death process for the Tau-hybrid (red) JSF for various switching thresholds, $\Omega$, (purple) and the Doob-Gillespie (yellow) methods. Left: CPU time required. We observe that as $\Omega$ increases, the computational time increases, but remains the computationally least expensive. Middle: Distributions in time to reach $X=200$. We observe that JSF results in comparable distributions to the exact process, with diminishing returns for $\Omega > 10^2$. Right: Extinction probability. We observe that all point estimates are comparable to that of the exact method.}}
    \label{fig:birth-drive}
\end{figure}

{For the birth dominated process, we require $\alpha > \beta$. Therefore, we specify $\alpha = 1$, and $\beta = 0.5$. We also choose the initial species population $X(0) = 1$, and a simulation time of $t = 20$. The quantities of interest to test the accuracy and efficiency of the methods are (i) the CPU time for a particular simulation instance, (ii) the simulation time required to reach a particular value, $X = 200$, and (iii) the accuracy of the method to experience initial population extinction. Figure \ref{fig:birth-drive} shows results for each of the three quantities of interest for the Tau-hybrid method (red), Doo-Gillespie (yellow) and three different Jump-Switch-Flow examples (purple) with varying switching thresholds of $\Omega = 10^1$, $\Omega = 10^2$, and $\Omega = 10^3$, respectively. Here, 1000 unique trajectories are considered for each method.}

{Considering the CPU time, we can observe as we increase the switching threshold, JSF requires more CPU time, since more of the simulation is in a discrete, Jumping regime. We also observe that the CPU time is, at worst, around two orders of magnitude less, compared to that of Doob-Gillespie. In comparison, we observe that the C++ implementation of Tau-hybrid method requires at best three times the computational time to simulate the equivalent process as JSF, while the Python implementation is at least one order of magnitude slower, indicating that JSF is much more efficient.}
{To assess the accuracy of each method, we now compare the (simulation) time for the species to reach 200. Here, both Tau-hybrid and JSF with switching thresholds of $\Omega = 10^2$, and $\Omega = 10^3$ result in distributions and median value comparable to that of Doob-Gillespie. The only exception is that of JSF with a switching threshold of $\Omega = 10^1$.}
{Lastly, we observe that all methods result in an initial extinction probability comparable to that of Doob-Gillespie.}

{These results suggest that JSF is capable of recovering the key quantities of the exact stochastic method, while being computationally cheaper than current hybrid methods and the exact method, for the birth dominated birth-death process.}

\subsubsection{{A death dominated birth-death process}}
\begin{figure}
    \centering
    \includegraphics[width=0.98\linewidth]{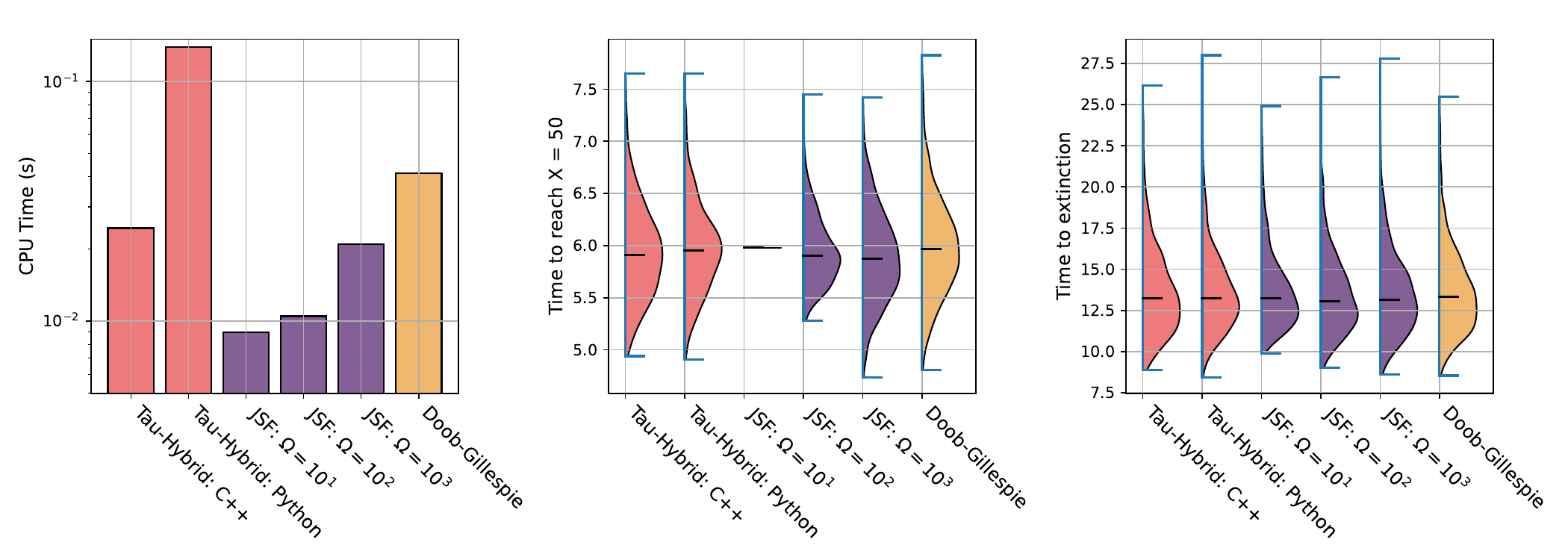}
    \caption{{Summary statistics for the death dominated birth-death process for the Tau-hybrid (red) JSF for various switching thresholds, $\Omega$, (purple) and the Doob-Gillespie (yellow) methods. Left: CPU time required. We again observe that as $\Omega$ increases, the computational time increases, but still remains the computationally least expensive. Middle: Distributions in time to reach $X=50$. We observe that a switching threshold of  $\Omega = 10^3$ is required for JSF to result in comparable distributions to the exact process, Right: Distribution in the time to extinction. We observe that for $\Omega > 10^1$, JSf results in comparable distributions to that of the exact method.}}
    \label{fig:death-drive}
\end{figure}

{For the birth dominated process, we require $\alpha < \beta$. Therefore, we specify $\alpha = 0.5$, and $\beta = 1$. We choose the initial species population $X(0) = 100$, and a simulation time of $t = 30$.}
{As before, the quantities of interest to test the accuracy and efficiency of the methods are (i) the CPU time for a particular simulation instance, (ii) the simulation time required to reach a particular value, $X = 50$, and (iii) the simulation time require for the species to go extinct. Figure \ref{fig:death-drive} shows results for each of the three quantities of interest for the Tau-hybrid method (red), Doo-Gillespie (yellow) and three different Jump-Switch-Flow examples (purple) with varying switching thresholds of $\Omega = 10^1$, $\Omega = 10^2$, and $\Omega = 10^3$, respectively. Again, 1000 unique trajectories are considered for each method.}
{Unlike in the previous example, here we observe that all JSF examples require more CPU time to complete the equivalent computation, despite this being a decaying death process. However, we still observe that all JSF methods require less CPU time when compared to the C++ implementation of the Tau-hybrid method, this time consuming at least 1.5 times less computation time, and still at least one order of magnitude quicker when compared to the Python implementation. We also note that JSF requires at most half the computational time of that of the exact Doob-Gillespie method.}
{We now consider the time for the species to reach a population of size $X = 50$. In this example, we see that JSF with $\Omega=10^1$ does not capture the distribution of the exact method, but only reports a single value. This is due to trajectories from this example all being deterministic at this value. If we consider JSF with $\Omega=10^2$, we see that while the median is comparable to the exact method, the distribution is not as broad. This is because the method only becomes stochastic at $X = 100$, resulting in variance in the trajectory prior to this value being lost. As we increase the switching threshold to $\Omega=10^3$, we observe that distribution in the Time to reach $X = 50$ become comparable to that of thee exact method. This provides a clear guide for how the switching threshold may be chosen, given some prior knowledge of the system. Lastly, we observe that extinction time for the JSF with $\Omega=10^2$ and $\Omega=10^3$ resulting in a comparable distribution to that of the Doob-Gillespie method. Again, we also observe that JSF with $\Omega=10^1$ results in a truncated distribution to that of the exact method.}

{These results also suggest that JSF is capable of recovering the key quantities of the exact stochastic method, while being computationally cheaper than current hybrid methods, for the death dominated birth-death process. Through this example, we also show how JSF may exhibit sensitivity in the value of the switching threshold chosen, and how to circumvent these sensitivities.}

\section{{Simulation study: Multiple time-scales in autocatalysis}}
{To emphasise the utility and accuracy of our Jump-Switch-Flow method, we will consider a simple autocatalysis example, with multiple time-scales. Consider two species, $X$, and $Y$, which react together via the following reactions: $X + Y \rightarrow 2 X + Y$, with rate $\alpha$, $X \rightarrow 2 X$, with rate $\beta$, $X \rightarrow \varnothing$, with rate $\gamma$, and $Y \rightarrow \varnothing$ with rate $\delta$. To demonstrate the ability of JSF to simulate the trajectories with multiple time-scales, but importantly, simulate them \textit{accurately}, we choose the following rate parameters:
\begin{enumerate}
    \item a moderate growth rate of $X$: $\beta=10$,
    \item a large death rate of $Y$, to ensure we cannot assume a constant $Y$: $\delta=100$,
    \item a large death rate of $X$, meaning $X$ quickly vanishes: $\gamma = 50$,
    \item and lastly a small autocatalysis rate, resulting in $X$ varying drastically in scale: $\alpha = 1$.
\end{enumerate}}
{Despite being a relatively simple model, with an analytic deterministic description, simulating stochastic trajectories requires careful consideration of the partitioning between the stochastic reactions to ensure an accurate representation, as there are reactions that are occurring a multiple different timescales. For example, autocatalysis occurs initially relatively frequently, due to the large number of species $Y$, despite the small rate $\alpha$. However, following the rapid decline of $Y$, $X$ will experience moderate decay. Therefore, we expect to observe two different timescales in the species $X$: a rapid increase, followed by gradual decline.}

\begin{figure}[H]
    \centering
    \includegraphics[width=0.98\linewidth]{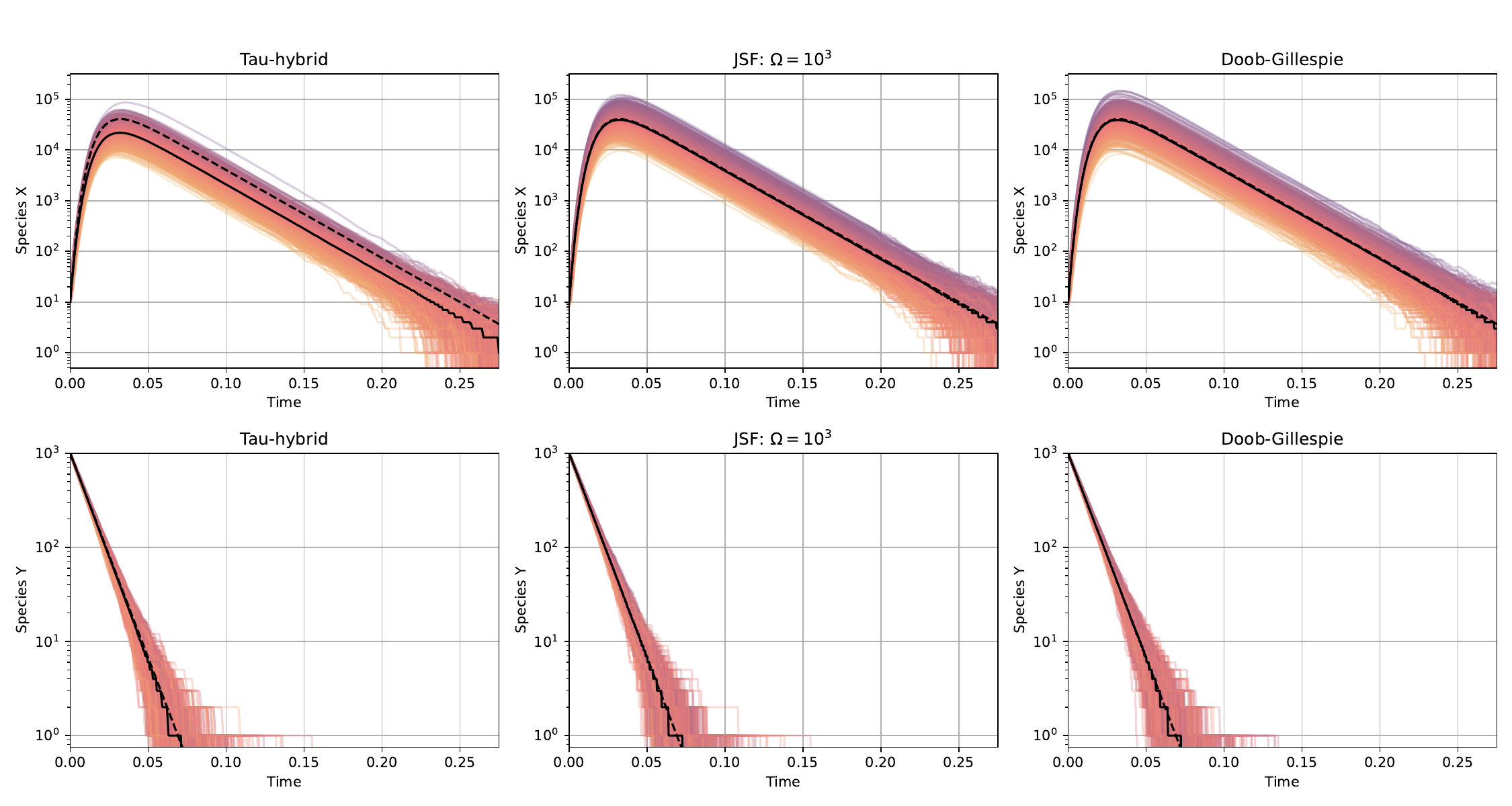}
    \caption{{1000 trajectories for the multiscale autocatalyst example for the Tau-hybrid method (left), the JSF method (middle) and the exact Doob-Gillespie method (right). Top shows species $X$, and bottom shows species $Y$. We also show the median species value (solid black line) and the analytic ODE solution (dashed black line). We see that the median trajectory of Tau-hybrid is less than the analytic solution, while both Doob-Gillespie and JSF match the analytic solution.}}
    \label{fig:AutocatalystTrajectores}
\end{figure}

{We assign the initial species sizes $X(0) = 10$, and $Y(0) = 10^2$, and sample 1000 trajectories using the exact Doob-Gillesie method, GillesPy2's C++ hybrid method (Tau-hybrid) and JSF with a switching threshold of $\Omega=10^2$. Figure \ref{fig:AutocatalystTrajectores} shows the 1000 trajectories for Tau-hybrid, JSF (with $\Omega = 10^2$), and Doob-Gillespie. We also present the analytic ODE solutions (dashed black) and the median value at a point (solid black).}

\begin{figure}[H]
    \centering
    \includegraphics[width=0.98\linewidth]{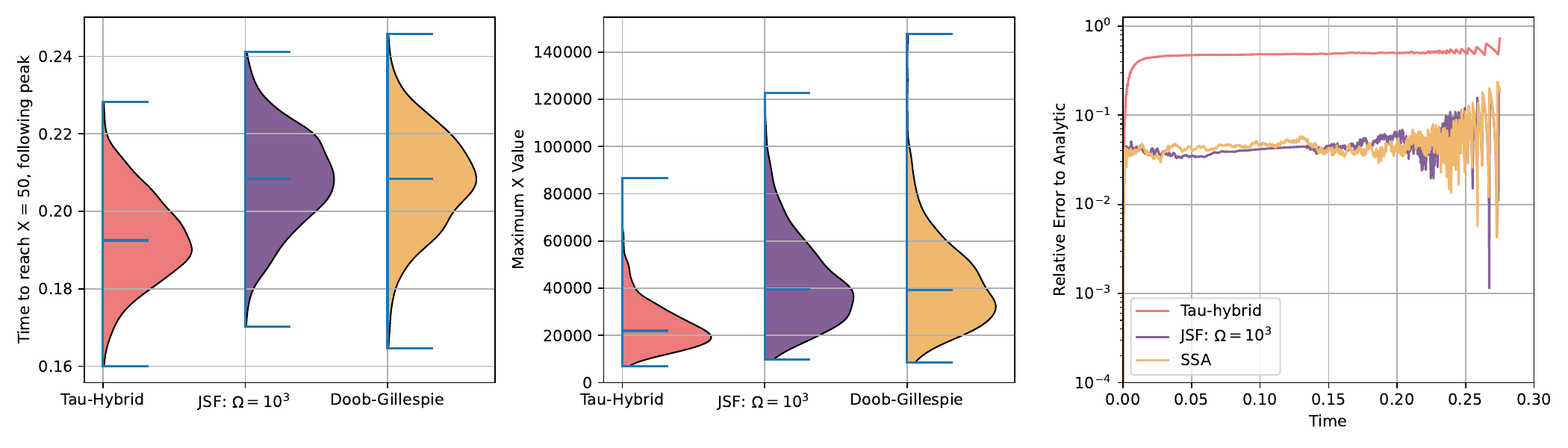}
    \caption{{Summary statistics for the multiscale autocatalyst example for the Tau-hybrid (red), JSF with $\Omega=10^3$ (purple) and Doob-Gillespie (yellow). Left: time to reach $X=50$ following initial peak. We see that the distributions of Doob-Gillespie and JSF are comparable, while tau-hybrid is skewed downward with a different median value. Middle: Maximum $X$ value. Again, we observe JSF and Doob-Gillespie resulting in matching distributions, while Tau-hybrid is skewed downwards. Right: relative error to analytic solution. JSF and Doob-Gillespie display similarly small error, an order of magnitude less than Tau-hybrid.}}
    \label{fig:AutocatalystSummaryData}
\end{figure}

{Here, we can see that all $Y$ species solutions match the analytic solution, since species $Y$ undergoes constant decay only. However, we clearly see that, while the median trajectories of JSF and Doob-Gillespie match that of the analytic description, the Tau-hybrid method median is less than the analytic solution. We can further quantify these discrepancies by looking at the distributions of the time for species $X$ to reach a particular value, the distributions of the maximum $X$ value, and also the relative point error between the median of the particular approach, and the analytic solution. We present these quantities in Figure \ref{fig:AutocatalystSummaryData}.}

{Figure \ref{fig:AutocatalystSummaryData} shows the distributions for time to reach $X = 50$ following the peak, where we see that JSF and Doob-Gillespie produce comparable results. However we also observe that the Tau-hybrid method under-performs, and results in earlier hitting times. We observe a similar trend in in the maximum $X$ value, with Tau-hybrid producing a skewed downwards distribution, and JSF and Doob-Gillespie producing comparable distributions. Lastly, the relative point error of JSF and Doob-Gillespie are comparable, with Tau-hybrid an order of magnitude greater.}

{From the results presented here, we see that our Jump-Switch-Flow method is capable of producing highly accurate hybrid trajectories, even with highly multiscale systems, as the one presented here. We also observe the summary statistics produced by JSF being comparable to those of the exact Doob-Gillespie method, and outperforming the highly efficient and sophisticated Tau-hybrid method.}

\section{Simulation study: SIRS with demography}\label{ssec:methods-sim-study}

We first present how the SIRS model with demography can be expressed with the Jump-Switch-Flow method. For the Jump-Switch-Flow method to be a useful method, it must reproduce behaviour that is representative of the exact process (i.e. the CTMC). That is, given some summary statistic, distributions produced via sampling the JSF process should be comparable to those obtained via sampling the CTMC.
Moreover, we require that it does so with acceptable computational efficiency. We show, through simulation experiments, how the Jump-Switch-Flow method compares to the Doob-Gillespie method \citep{gillespie1976general} for accuracy, and how our approach can exhibit sensitivity with respect to model inputs. We present how our Jump-Switch-Flow method compares to both the Doob-Gillespie and Tau-Leaping methods. 

\begin{figure}[H]
    \centering
    \begin{subfigure}[b]{0.32\textwidth}
     \centering
     \includegraphics[width=0.95\linewidth]{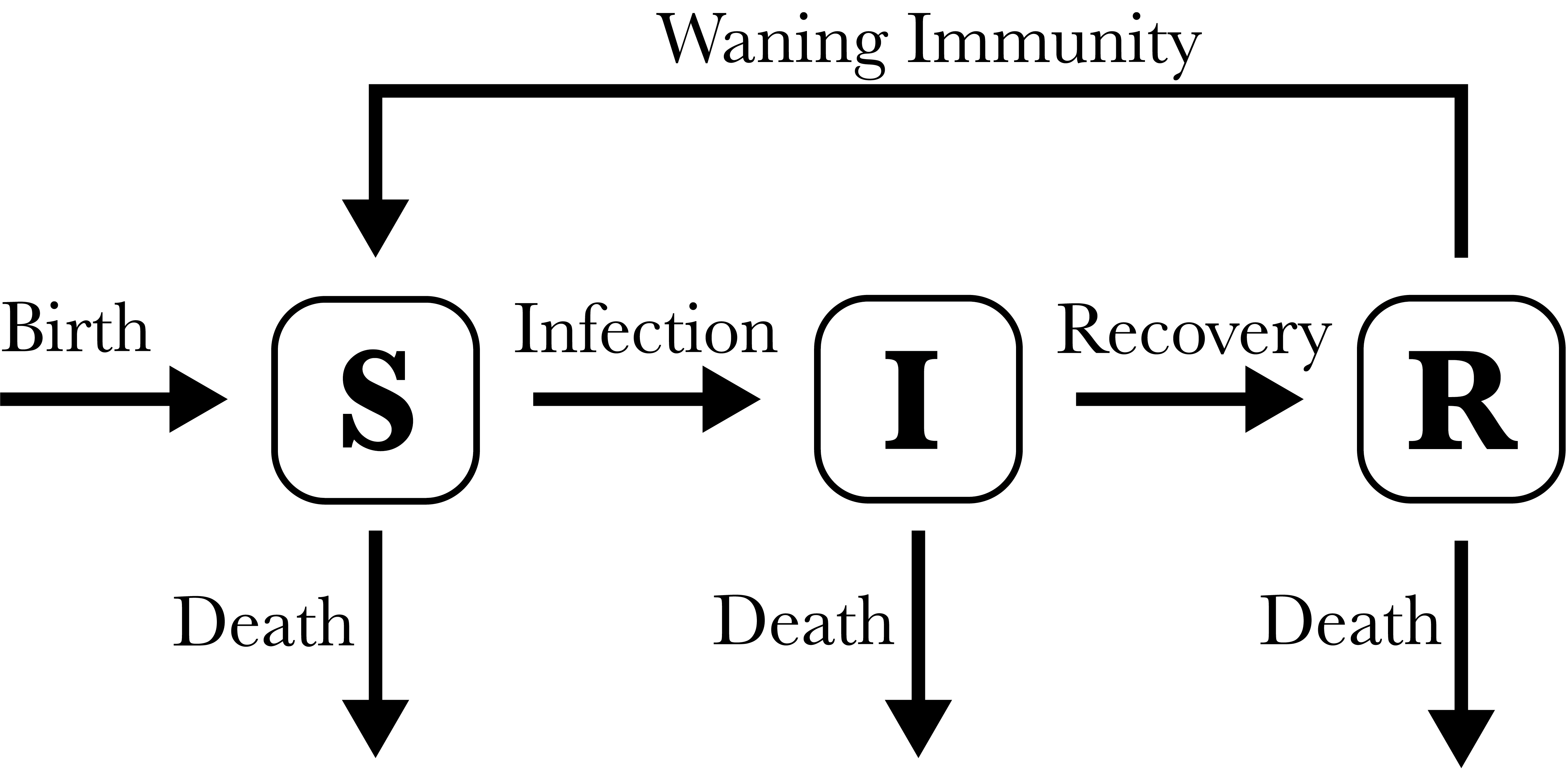}
     \caption{\label{fig:SIRS_Compartment_Description}}
     \end{subfigure}
     \begin{subfigure}[b]{0.32\textwidth}
     \centering
     \includegraphics[width=0.95\linewidth]{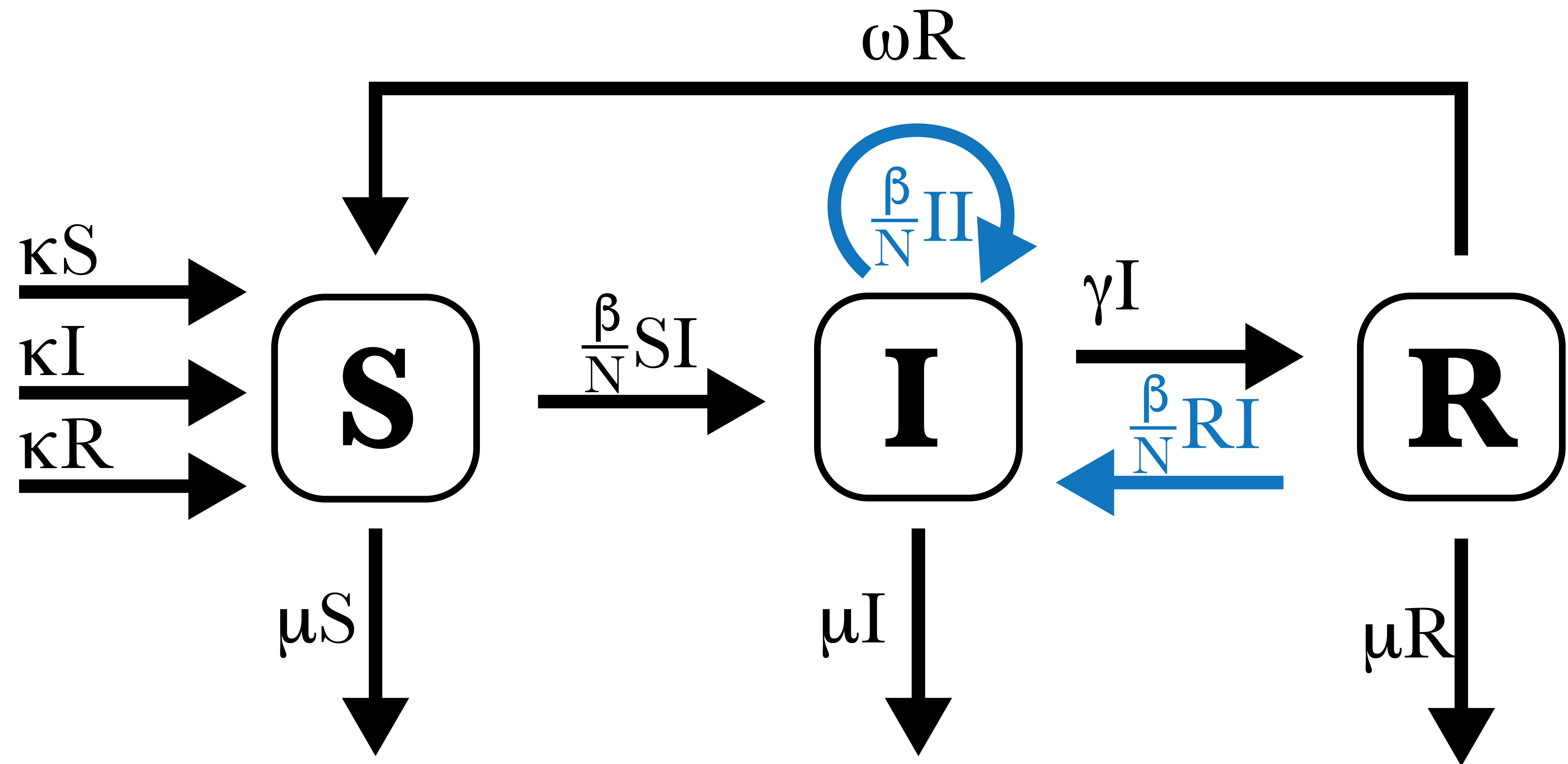}
     \caption{\label{fig:SIRS_Full}}
     \end{subfigure}
     \begin{subfigure}[b]{0.32\textwidth}
     \centering
     \includegraphics[width=0.95\linewidth]{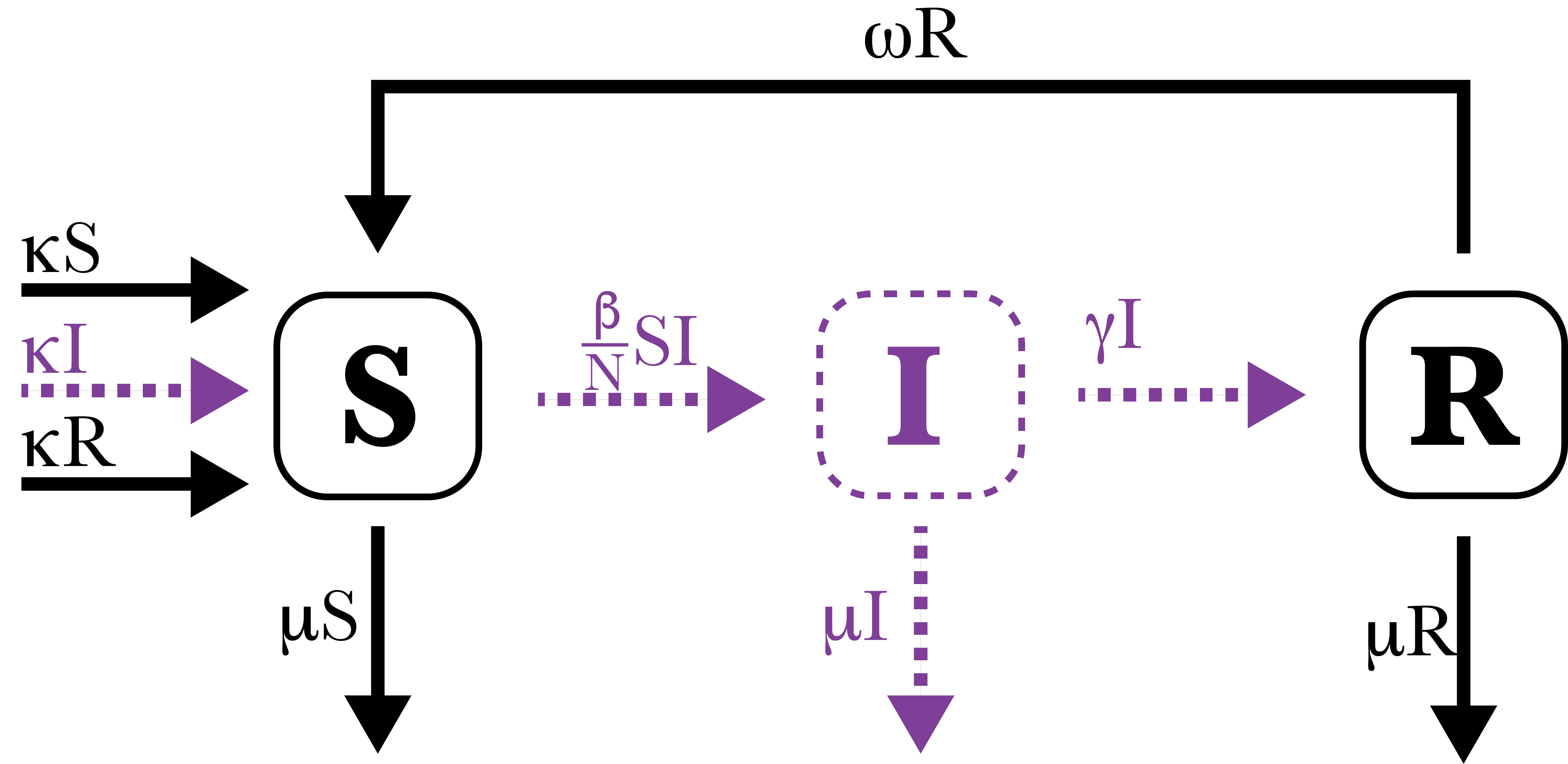}
     \caption{\label{fig:SIRS_Switching}}
     \end{subfigure}

     \captionsetup{subrefformat=parens}
     \caption{\label{fig:SIRS_compartmental} \subref{fig:SIRS_Compartment_Description} The labelled compartmental SIRS model with demography. \subref{fig:SIRS_Full} The SIRS model with demography where the arrows describe the interactions between the reactants and products. Here, blue arrows do not result in any flow through the system, and can be ignored. \subref{fig:SIRS_Switching} The model as a Jump-Switch-Flow system with continuous (flowing) S and R (black), and discrete (jumping) I (purple). Since I is jumping, the reactions involving I are modelled as discrete, jumping reactions.}
    \end{figure}

Figure \ref{fig:SIRS_Compartment_Description} shows the main reactions of the SIRS model with demography: individuals fall into one of three compartments: susceptible ($S$), infectious ($I$) and recovered ($R$). Here, arrows represent how individuals move and progress through compartments. Specifically, individuals are born into the susceptible compartment. Births occur from individuals in the $S$, $I$ and $R$ compartments, and therefore depend explicitly on the populations (discrete or continuous) in each of these compartments. Individuals can die within each of the compartments. Finally, individuals may progress through compartments via susceptible individuals experiencing infection, infected individuals experiencing recovery, and recovered individuals experiencing waning immunity.

In order to represent the SIRS model with demography as a Jump-Switch-Flow process, we need to associate rates, reactants and products to each of the arrows (see Figure \ref{fig:SIRS_Full}). Note that the births arrow is split into three arrows depending upon the compartment of the parent. Moreover, one of the key assumptions of any SIR-type model is that the population is homogeneously mixed, which enables us to describe how individuals interact with one another. This enables us to say that susceptible individuals interact with infectious individuals with a rate inversely proportional to the whole population. However, there are likewise infectious individuals interacting with other infectious individuals, and also with recovered individuals. However, these last two interactions (as shown in blue in Figure \ref{fig:SIRS_Full}) do not result in any exchange of individuals between compartments.

Figure \ref{fig:SIRS_Switching} shows the  SIRS model with demography as a Jump-Switch-Flow process. In this example figure, the $S$ and $R$ compartments are continuous (flowing), while the $I$ compartment is discrete (jumping). To decide which of the arrows/reactions are jumping, i.e. in $\mathcal{S}$, and which are flowing, i.e. in $\mathcal{S}'$, we consider the reactants and products of the arrows: if an arrow has any discrete reactant or product, then that arrow is also jumping, otherwise it is flowing. Table \ref{tab:sird-stoich} shows the reactants and products of the SIRS model with demography.

\begin{table}[!ht]
    \centering
    \begin{tabular}{lc|ccc|ccc|ccc}
   & \multicolumn{1}{l|}{} & \multicolumn{3}{c|}{\textbf{\small Reactants, $\eta^{-}$ }} & \multicolumn{3}{c|}{\textbf{\small Products, $\eta^{+}$ }} & \multicolumn{3}{c|}{\textbf{ \small $\eta$}} \\
    \multicolumn{1}{c|}{\textbf{\small Reaction}} & \textbf{\small Rate}  & $S$ & $I$ & $R$ & $S$ & $I$ & $R$  & $S$  & $I$ & \multicolumn{1}{c|}{$R$} \\ \hline
    \multicolumn{1}{l|}{Birth by $S$} & $\kappa$ & 1 & $\cdot$ & $\cdot$ & 2 & $\cdot$ & $\cdot$ & 1 & $\cdot$ &\multicolumn{1}{c|}{$\cdot$} \\
    \multicolumn{1}{l|}{$^*$Birth by $I$} & $\kappa$ & $\cdot$ & 1 & $\cdot$ & 1 & 1 & $\cdot$ & 1 & $\cdot$ & \multicolumn{1}{c|}{$\cdot$} \\
    \multicolumn{1}{l|}{Birth by $R$} & $\kappa$ & $\cdot$ & $\cdot$ & 1 & 1 & $\cdot$ & 1 & 1 & $\cdot$ & \multicolumn{1}{c|}{$\cdot$} \\
    \multicolumn{1}{l|}{$^*$Infection of $S$} & $\beta_{S}/N$ & 1 & 1 & $\cdot$ & $\cdot$ & 2 & $\cdot$ & -1 & 1 & \multicolumn{1}{c|}{$\cdot$} \\
    \multicolumn{1}{l|}{\textcolor{blue}{$^*$Infection of $I$}}& \textcolor{blue}{$\beta_{I}/N$} & \textcolor{blue}{$\cdot$} & \textcolor{blue}{2} & \textcolor{blue}{$\cdot$} & \textcolor{blue}{$\cdot$} & \textcolor{blue}{2} & \textcolor{blue}{$\cdot$} & \textcolor{blue}{$\cdot$} & \textcolor{blue}{$\cdot$} & \multicolumn{1}{c|}{\textcolor{blue}{$\cdot$}} \\
    \multicolumn{1}{l|}{ \textcolor{blue}{$^*$Infection of $R$}} & \textcolor{blue}{$\beta_{R}/N$} & \textcolor{blue}{$\cdot$} & \textcolor{blue}{1} & \textcolor{blue}{1} & \textcolor{blue}{$\cdot$} & \textcolor{blue}{1} & \textcolor{blue}{1} & \textcolor{blue}{$\cdot$} & \textcolor{blue}{$\cdot$} & \multicolumn{1}{c|}{\textcolor{blue}{$\cdot$}} \\
    \multicolumn{1}{l|}{$^*$Recovery of $I$} & $\gamma$ & $\cdot$ & 1 & $\cdot$ & $\cdot$ & $\cdot$ & 1 & $\cdot$ & -1 & \multicolumn{1}{c|}{1} \\\multicolumn{1}{l|}{Waning of $R$} & $\omega$ & $\cdot$ & $\cdot$ & 1 & 1 & $\cdot$ & $\cdot$ & 1 & $\cdot$ & \multicolumn{1}{c|}{-1}\\
    \multicolumn{1}{l|}{Death of $S$} & $\mu$ & 1 & $\cdot$ & $\cdot$ & $\cdot$ & $\cdot$ & $\cdot$ & -1 & $\cdot$ & \multicolumn{1}{c|}{$\cdot$}  \\
    \multicolumn{1}{l|}{$^*$Death of $I$} & $\mu$ & $\cdot$ & 1 & $\cdot$ & $\cdot$ & $\cdot$ & $\cdot$ & $\cdot$ & -1 & \multicolumn{1}{c|}{$\cdot$} \\
    \multicolumn{1}{l|}{Death of $R$} & $\mu$ & $\cdot$ & $\cdot$ & 1 & $\cdot$ & $\cdot$ & $\cdot$ & $\cdot$ & $\cdot$ & \multicolumn{1}{c|}{-1} \\
    \end{tabular}
    \caption{\label{tab:sird-stoich}The stoichiometric matrices for the SIRS model with demography. The (*) indicates reactions that are jumping when \(I\) is discrete and \(S\) and \(R\) are continuous. The rows with blue coloured entries do not alter the exchange through the system, and so need not be represented.}
\end{table}
  
For the remainder of this section, we suppose parameter values for the SIRS model with demography that would be typical for a newly introduced, yearly seasonal communicable disease, as presented in Table \ref{tab:SIRS_Params}. With these parameter values, on average, we expect an infected individual to infect 2 other people per week (during the initial outbreak), and be infectious for 1 week, with immunity to the disease lasting for 1 year. For the population turnover, we assume individuals live for, on average, 85 years.
Initially there are two infectious individuals (i.e. $I(0)=2$), no recovered individuals (i.e. $R(0)=0$), and the remainder of the population is susceptible (i.e. $S(0)=N(0)-I(0)$).

\begin{table}[!ht]
    \centering
    \begin{tabular}{|c|c|c|c|c|c|}
        \hline
        Parameter & $\beta$ & $\gamma$ & $\omega$ & $\kappa$ & $\mu$ \\
        \hline
        Rate description & Infection & Recovery & Immunity Waning & Birth  & Death\\
        \hline
        Value & $2/7$  & $1/7$ & $1/365$ & $1/(85\times365)$ & $1/(85\times365)$\\
        \hline
    \end{tabular}
    \caption{Parameter values used for SIRS model with demography.}
    \label{tab:SIRS_Params}
\end{table}

\begin{figure}[H]
    \centering
    \begin{subfigure}[t]{0.25\textwidth}
    \centering {Extinction}
    \end{subfigure}
    \begin{subfigure}[t]{0.25\textwidth}
    \centering {Fade-out}
    \end{subfigure}
    \begin{subfigure}[t]{0.25\textwidth}
    \centering {Endemic}
    \end{subfigure}

     \begin{subfigure}[b]{0.25\textwidth}
     \centering
     \includegraphics[width=1.0\linewidth]{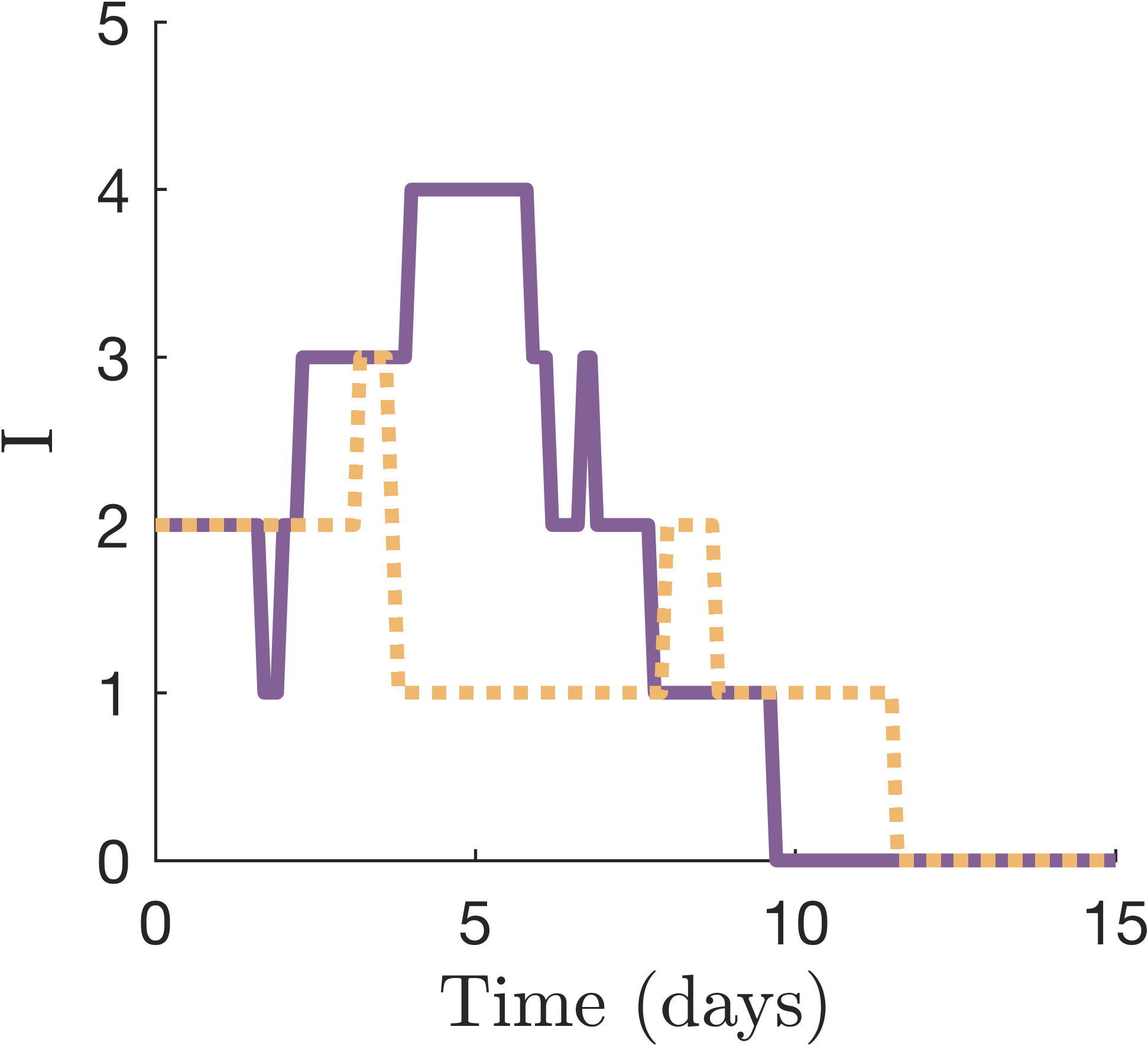}
     \caption{\label{fig:SIRS_Ext_I}}
     \end{subfigure}
     \begin{subfigure}[b]{0.25\textwidth}
     \centering
     \includegraphics[width=1.0\linewidth]{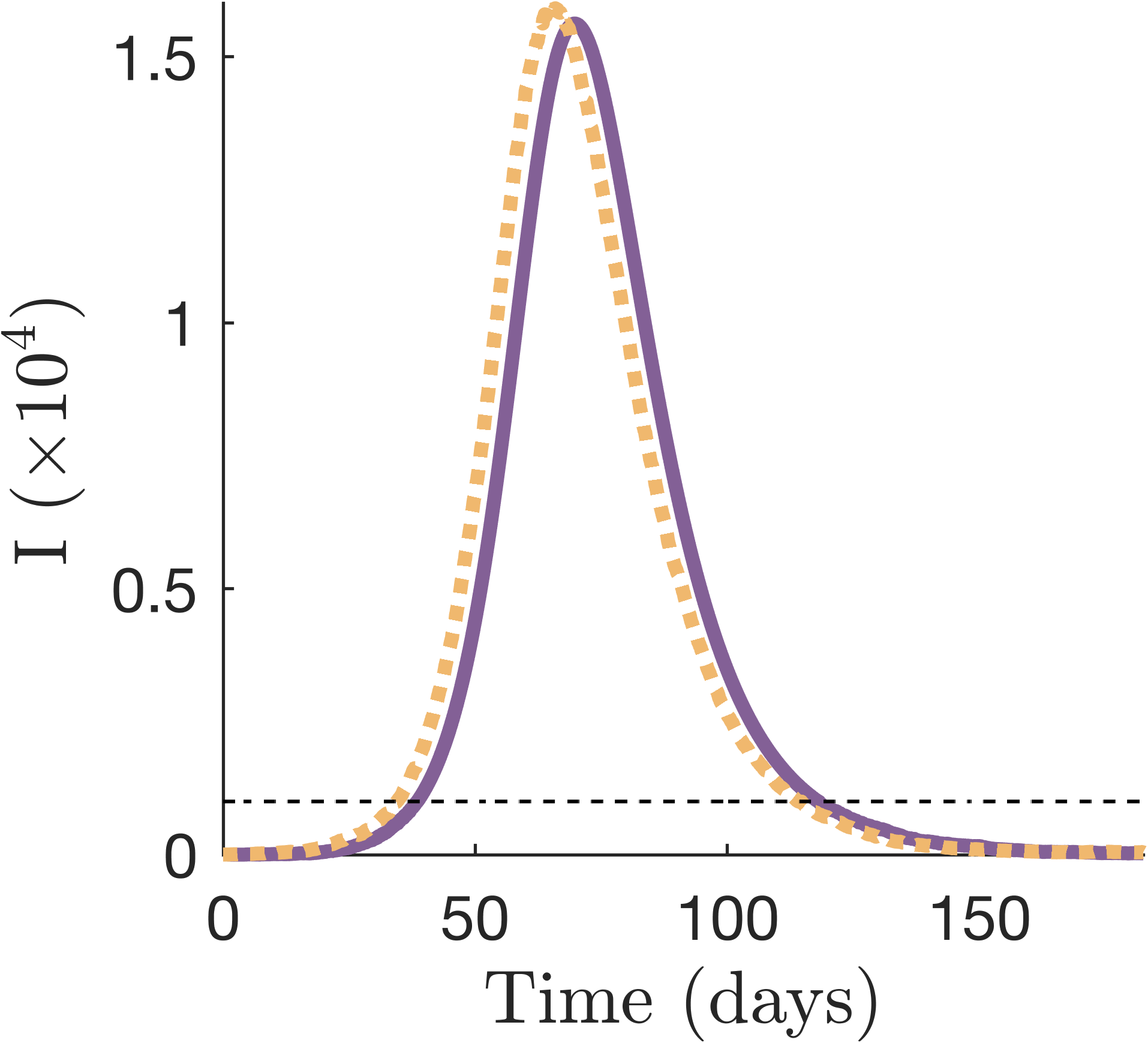}
     \caption{\label{fig:SIRS_Fad_I}}
     \end{subfigure}
     \begin{subfigure}[b]{0.25\textwidth}
     \centering
     \includegraphics[width=1.0\linewidth]{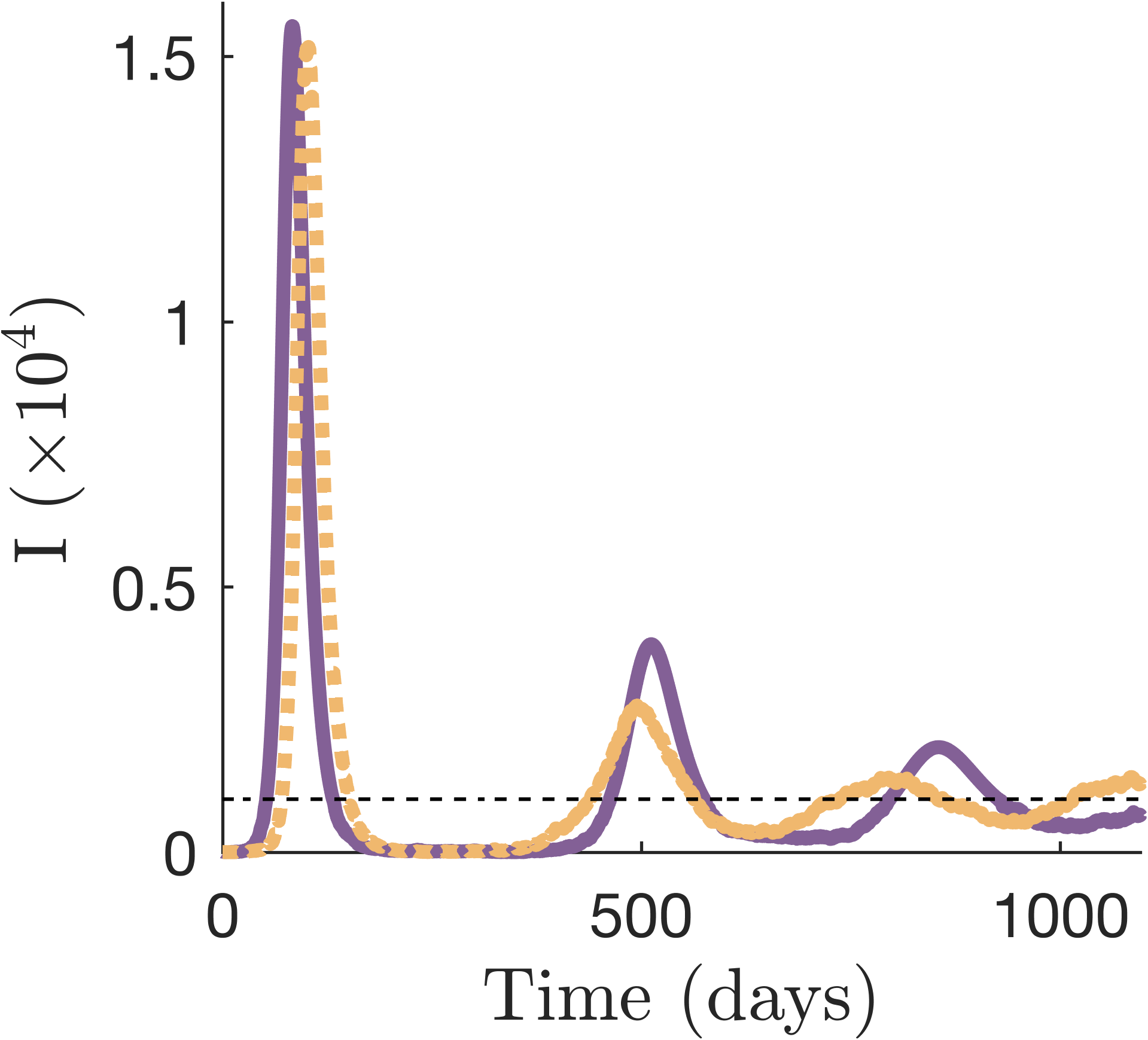}
     \caption{\label{fig:SIRS_End_I}}
     \end{subfigure}

     \begin{subfigure}[b]{0.25\textwidth}
     \centering
     \includegraphics[width=1.0\linewidth]{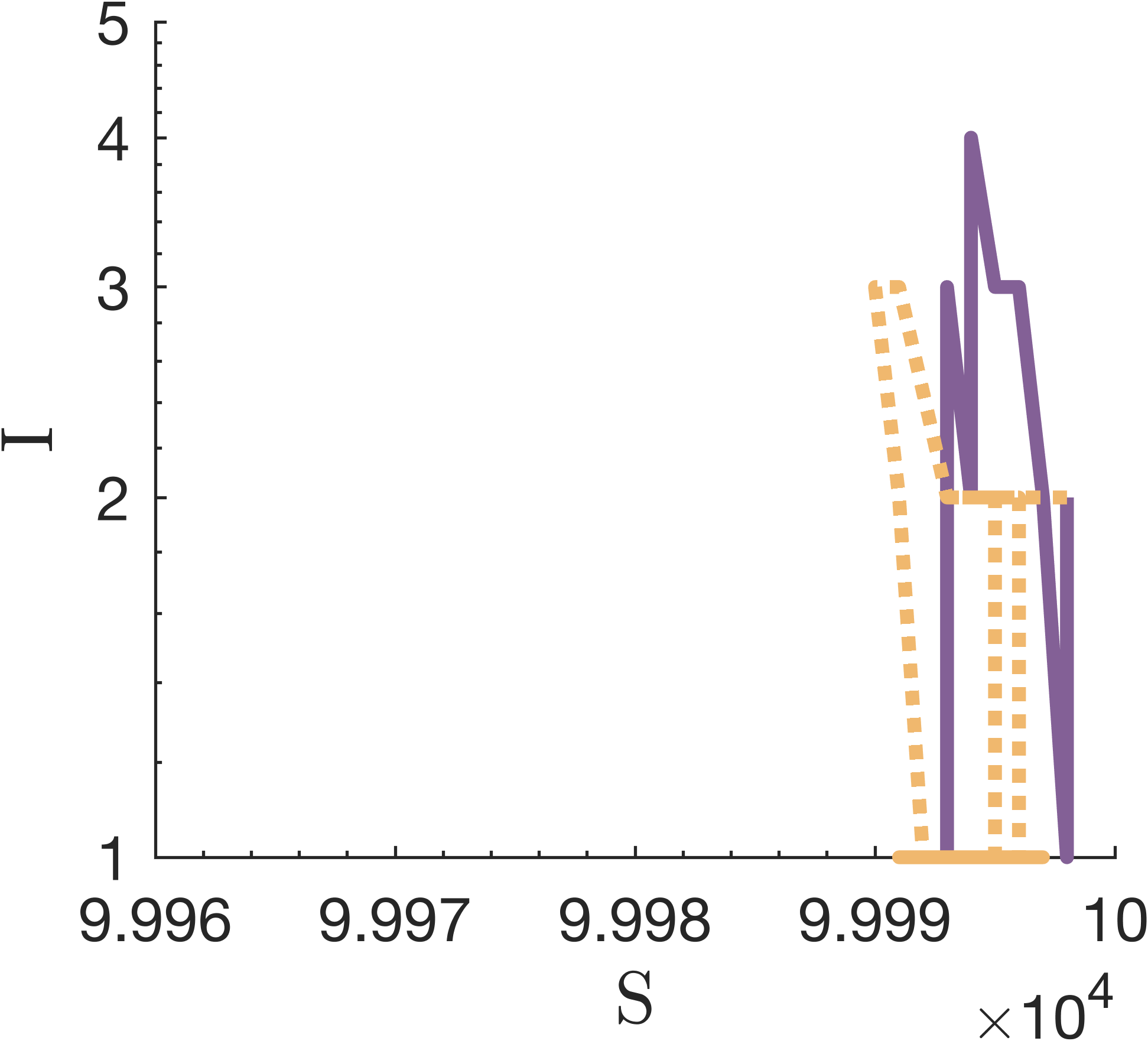}
     \caption{\label{fig:SIRS_Ext_SI}}
     \end{subfigure}
     \begin{subfigure}[b]{0.25\textwidth}
     \centering
     \includegraphics[width=1.0\linewidth]{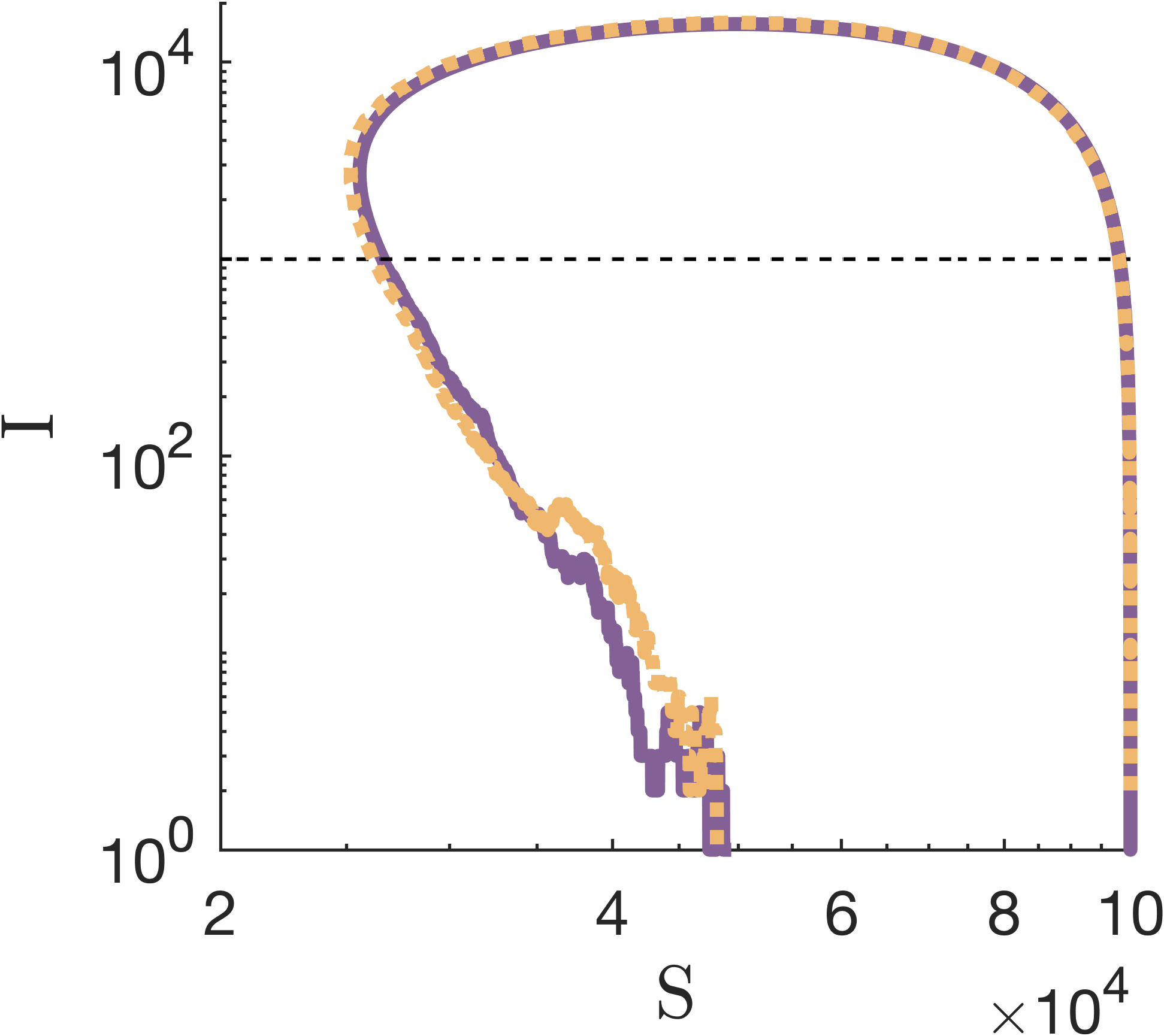}
     \caption{\label{fig:SIRS_Fad_SI}}
     \end{subfigure}
     \begin{subfigure}[b]{0.25\textwidth}
     \centering
     \includegraphics[width=1.0\linewidth]{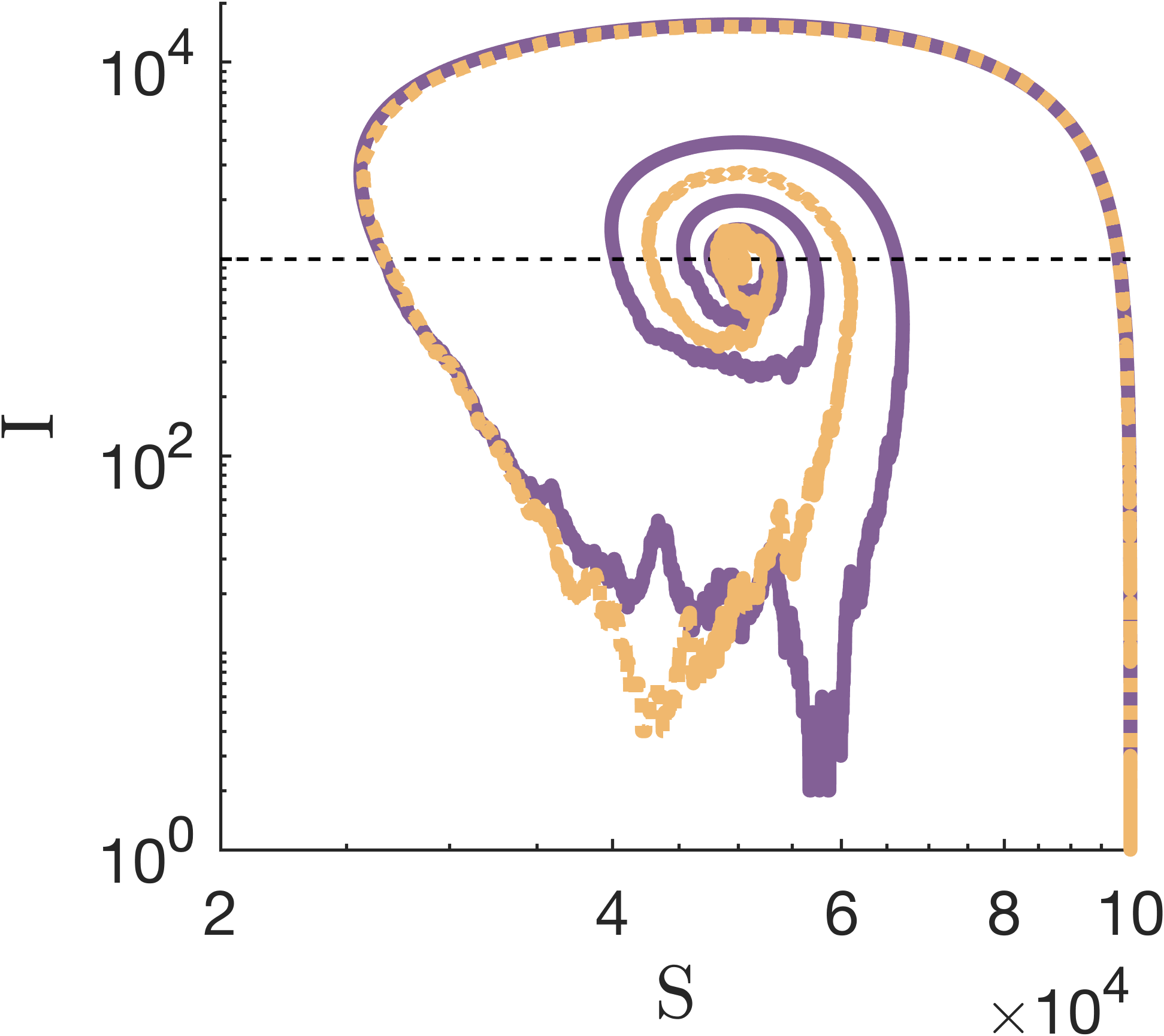}
     \caption{\label{fig:SIRS_End_SI}}
     \end{subfigure}

     \captionsetup{subrefformat=parens}

     \caption{\label{fig:SIRS_Dynamics_of_Interest} Possible simulated trajectories of SIRS model with demography for an initial population size $N(0)=10^5$, from Doob-Gillespie (yellow) and Jump-Switch-Flow (purple, $\Omega=10^3$) methods: \subref{fig:SIRS_Ext_I} two examples of extinction trajectories of the infectious compartment, with associated $S-I$ phase plane in \subref{fig:SIRS_Ext_SI}; \subref{fig:SIRS_Fad_I} two examples of fade-out trajectories of the infectious compartment, with associated $S-I$ phase plane in \subref{fig:SIRS_Fad_SI};  \subref{fig:SIRS_End_I} two examples of endemic trajectories of the infectious compartment, with associated $S-I$ phase plane in \subref{fig:SIRS_End_SI}. }

\end{figure}

As well as being computationally efficient, our Jump-Switch-Flow method is capable of capturing the inherent stochastic nature of the sampled process. For the SIRS model with demography, this is realised by three key different scenarios:
\begin{enumerate}
    \item \textit{Extinction}: all of the individuals recover before the disease manages to spread significantly throughout the population (see Figure \ref{fig:SIRS_Ext_I} and \ref{fig:SIRS_Ext_SI});
    \item \textit{Fade-out}: the disease spreads throughout the population, however it fades-out in the trough proceeding the first wave (see Figure \ref{fig:SIRS_Fad_I} and \ref{fig:SIRS_Fad_SI});
    \item \textit{Endemic}: the disease  persists indefinitely within the population (see Figure \ref{fig:SIRS_End_I} and \ref{fig:SIRS_End_SI}).
\end{enumerate}

\subsection{Simulation experiments: Comparing Jump-Switch-Flow and Doob-Gillespie}
We compare our Jump-Switch-Flow method to the gold standard approach for simulating exact solutions to CTMCs, the Doob-Gillespie method. To do so, we first specified an initial population size of $N(0)=10^5$, and a switching threshold of each compartment of $\Omega=10^3$. We then generated 5,000 simulations using the parameters in Table \ref{tab:SIRS_Params}.

To capture the dynamics of the model, we track the infectious compartment as a key compartment of interest. In doing so, we then obtain distributions for the peak number of infectious individuals and also the time for the infection to peak, for the fade-out scenario. We also track the cumulative number of infected individuals after a certain amount of time for the endemic scenario. These results are presented in Figure \ref{fig:N05_Study}. 

\begin{figure}[h!]
\centering
    \begin{subfigure}[t]{0.9\textwidth}
    \centering {Jump-Switch-Flow}
    \end{subfigure}

    \begin{minipage}[t][0.1cm][t]{0.27\textwidth}
        \centering
        \begin{subfigure}{\linewidth}
            \includegraphics[trim={0 0.0cm 0 0},clip, width=\linewidth]{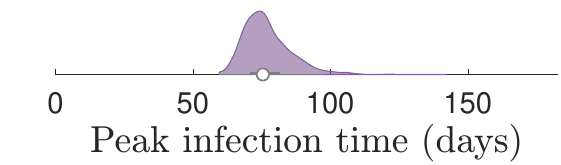}
            \caption{\label{fig:JSF_N05_Fad_peak_5}}
        \end{subfigure}
    \end{minipage}
    \begin{minipage}[t][0.1cm][t]{0.08\textwidth}
        \hspace{0.5cm}
    \end{minipage}
    \begin{minipage}[t][0.1cm][t]{0.3\textwidth}
        \hspace{0.5cm}
    \end{minipage}
    \begin{minipage}[t][0.1cm][t]{0.08\textwidth}
        \hspace{0.5cm}
    \end{minipage}
    
    \begin{minipage}[t][1cm][t]{0.3\textwidth}
        \centering
        \begin{subfigure}{\linewidth}
            \includegraphics[width=\linewidth]{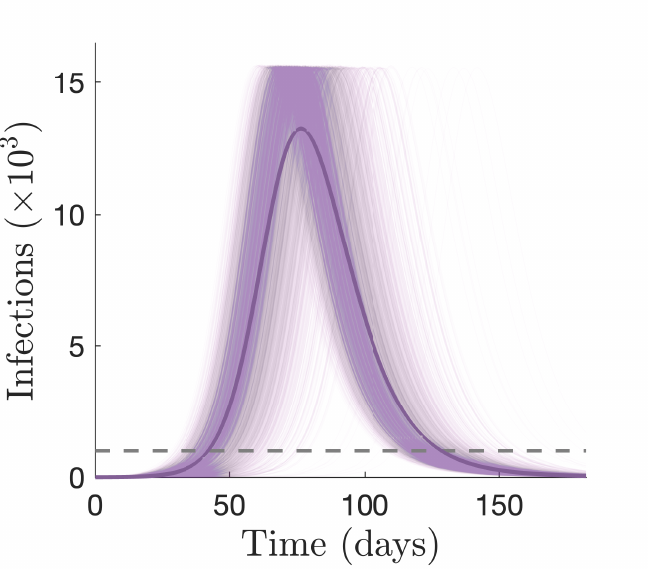}
            \caption{\label{fig:JSF_N05_Fad_5}}
        \end{subfigure}
    \end{minipage}
    \begin{minipage}[t][1cm][t]{0.08\textwidth}
        \centering
        \begin{subfigure}{\linewidth}
            \includegraphics[width=\linewidth]{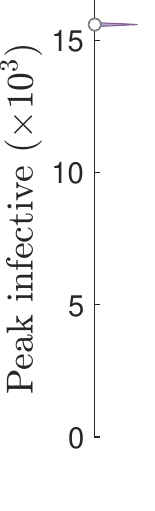}
            \caption{\label{fig:JSF_N05_Fad_peaknumb_5}}
        \end{subfigure}
    \end{minipage}
    \begin{minipage}[t][1cm][t]{0.3\textwidth}
        \centering
        \begin{subfigure}{\linewidth}
            \includegraphics[width=\linewidth]{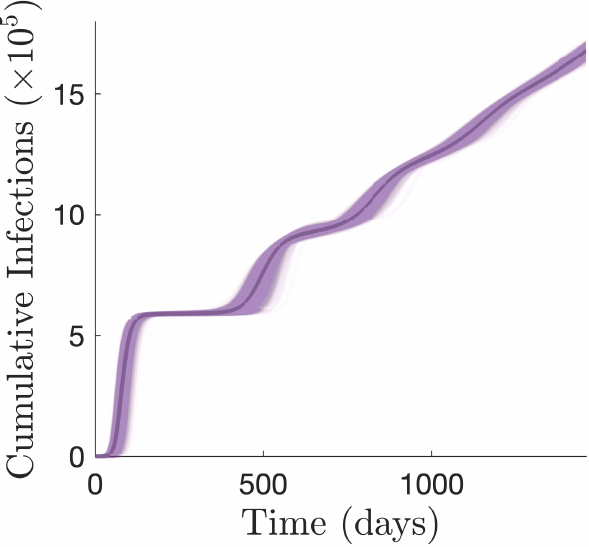}
            \caption{\label{fig:JSF_N05_Cummulative_5}}
        \end{subfigure}
    \end{minipage}
    \begin{minipage}[t][1cm][t]{0.08\textwidth}
        \centering
        \begin{subfigure}{\linewidth}
            \includegraphics[width=\linewidth]{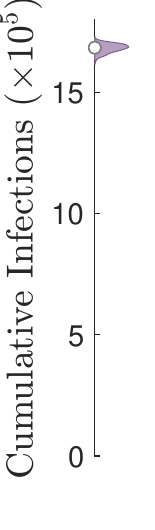}
            \caption{\label{fig:JSF_N05_Cummulative_dist_5}}
        \end{subfigure}
    \end{minipage}

    \begin{subfigure}[t]{0.9\textwidth}
    \centering {Doob-Gillespie}
    \end{subfigure}

    \begin{minipage}[t][0.1cm][t]{0.27\textwidth}
        \centering
        \begin{subfigure}{\linewidth}
            \includegraphics[width=\linewidth]{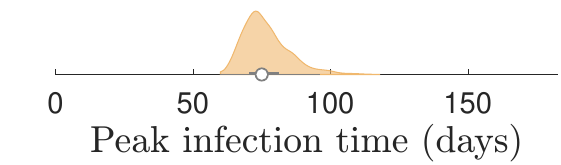}
            \caption{\label{fig:Gil_N05_Fad_peak_5}}
        \end{subfigure}
    \end{minipage}
    \begin{minipage}[t][0.1cm][t]{0.08\textwidth}
        \hspace{0.1cm}
    \end{minipage}
    \begin{minipage}[t][0.1cm][t]{0.3\textwidth}
        \hspace{0.1cm}
    \end{minipage}
    \begin{minipage}[t][0.1cm][t]{0.08\textwidth}
        \hspace{0.1cm}
    \end{minipage}

    \begin{minipage}[t][1cm][t]{0.3\textwidth}
        \centering
        \begin{subfigure}{\linewidth}
            \includegraphics[width=\linewidth]{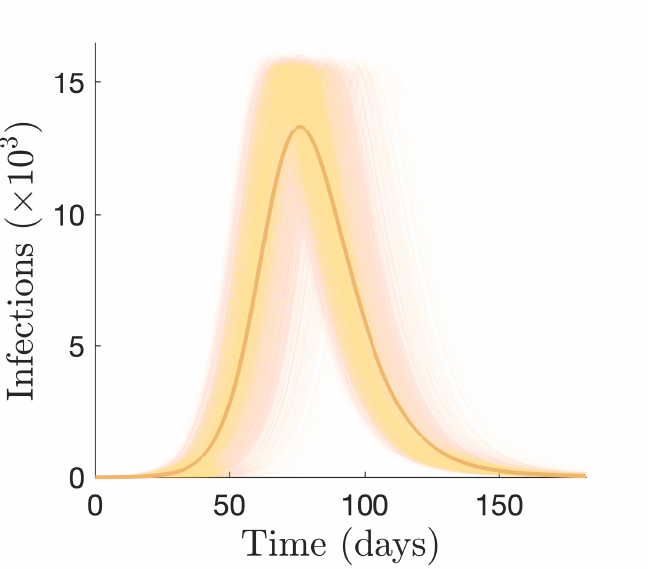}
            \caption{\label{fig:Gil_N05_Fad_5}}
        \end{subfigure}
    \end{minipage}
    \begin{minipage}[t][1cm][t]{0.08\textwidth}
        \centering
        \begin{subfigure}{\linewidth}
            \includegraphics[width=\linewidth]{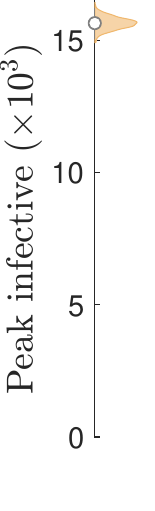}
            \caption{\label{fig:Gil_N05_Fad_peaknumb_5}}
        \end{subfigure}
    \end{minipage}
    \begin{minipage}[t][1cm][t]{0.3\textwidth}
        \centering
        \begin{subfigure}{\linewidth}
            \includegraphics[width=\linewidth]{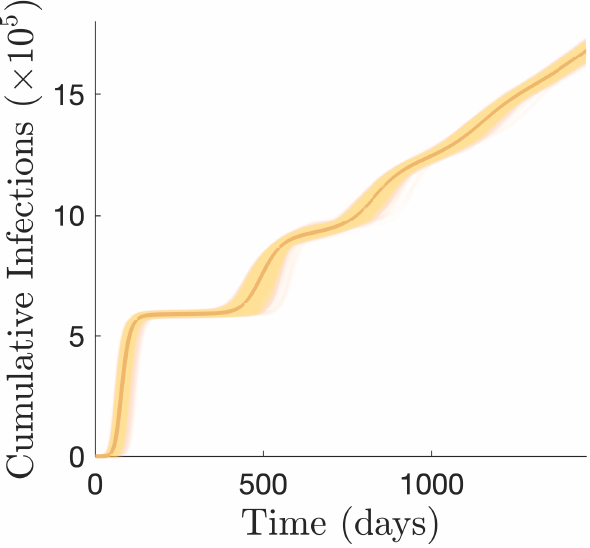}
            \caption{ \label{fig:Gil_N05_Cummulative_5}}
        \end{subfigure}
    \end{minipage}
    \begin{minipage}[t][1cm][t]{0.08\textwidth}
        \centering
        \begin{subfigure}{\linewidth}
            \includegraphics[width=\linewidth]{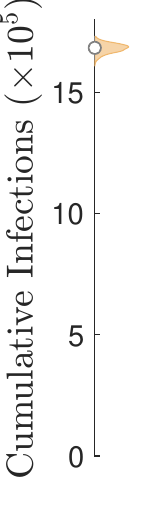}
            \caption{\label{fig:Gil_N05_Cummulative_dist_5}}
        \end{subfigure}
    \end{minipage}

    \captionsetup{subrefformat=parens}
    
     \caption{\label{fig:N05_Study} We compare the infectious compartment from realisations for Jump-Switch-Flow \subref{fig:JSF_N05_Fad_5} and Doob-Gillespie \subref{fig:Gil_N05_Fad_5} solutions, for a population size of $N(0) = 10^5$. We use a switching threshold of $\Omega = 10^3$. We also show the distributions of the time to infectious peak (\subref{fig:JSF_N05_Fad_peak_5} and \subref{fig:Gil_N05_Fad_peak_5}), and infectious peak values (\subref{fig:JSF_N05_Fad_peaknumb_5} and \subref{fig:Gil_N05_Fad_peaknumb_5}). We compare the cumulative infections over the simulations for Jump-Switch-Flow \subref{fig:JSF_N05_Cummulative_5} and Doob-Gillespie \subref{fig:Gil_N05_Cummulative_5} solutions and the final distributions (\subref{fig:JSF_N05_Cummulative_dist_5} and \subref{fig:Gil_N05_Cummulative_dist_5}). Point averages are also presented as a solid line.}
    
\end{figure}

Comparing Figures \ref{fig:JSF_N05_Fad_peak_5} and \ref{fig:Gil_N05_Fad_peak_5}, we can immediately observe that the distributions for time to peak for Jump-Switch-Flow method appears to be representative of that for the Doob-Gillespie method. This is because the delay in the epidemic take off at low population is fully captured with this sufficiently large choice of switching threshold, $\Omega=10^3$, enabling the full stochastic effects to be accurately represented. 

However, if we compare the distributions for the peak number of infectious individuals (\ref{fig:JSF_N05_Fad_peaknumb_5} and \ref{fig:Gil_N05_Fad_peaknumb_5}), we can see that they differ significantly, for this chosen switching threshold. This difference can be understood if we consider how the process is behaving. Once the infectious compartment is sufficiently large for the Jump-Switch-Flow method, the method switches from a stochastic representation to a deterministic representation. Therefore, once the infectious compartment becomes deterministic, given that the susceptible population is also deterministic, its peak value is determined by the number of both infectious and susceptible individuals at this time. Since the number of infectious people at this time will always be the same, $I=\Omega=10^3$ people, any variation in peak value is due to the number of susceptible individuals, $S$, at the time of switching.

If the summary statistic we instead compare is the distribution of the cumulative number of infected individuals after four years of epidemic circulation, we observe that, even for a relatively low switching threshold, the Jump-Switch-Flow method performs well comparatively to the Doob-Gillespie method, see Figures \ref{fig:JSF_N05_Cummulative_dist_5} and \ref{fig:Gil_N05_Cummulative_dist_5}. We also note that for these parameter choices and switching threshold, the infectious compartment switches between jumping and flowing states multiple times, yet the Jump-Switch-Flow method produces a cumulative distribution comparable to that of Doob-Gillespie. This indicates that the Jump-Switch-Flow method is robust for long term simulations.

The final summary statistics we are interested in are the probabilities of extinction, fade-out and endemic scenarios. Table \ref{tab:SIRS_N05_scenario_outcomes_10T3} shows the probability of each scenario computed from 5,000 simulated samples. Here, we see that both methods produce comparatively similar results, which indicates that the Jump-Switch-Flow method is capable of capturing the inherent stochasticity in the exact simulation of the system via the Doob-Gillespie method. 

\begin{table}[!ht]
    \centering
    \begin{tabular}{|c|c|c|c|}
    \hline
        Method & Extinction & Fade-out & Endemic \\
        \hline
        Jump-Switch-Flow & $0.2476 \pm 0.0120$ & $0.4774 \pm 0.0138$ & $0.2750 \pm 0.0124$ \\
        Doob-Gillespie & $0.2520 \pm 0.0120$ & $0.4668 \pm 0.0138$ & $0.2812 \pm 0.0125$ \\
        \hline
    \end{tabular}
    \caption{Probabilities (with associated 95\% confidence interval) of scenario outcomes for Jump-Switch-Flow and Doob-Gillespie, from 5,000 simulations.}
    \label{tab:SIRS_N05_scenario_outcomes_10T3}
\end{table}

We also compare the Jump-Switch-Flow and Doob-Gillespie methods where the switching threshold has been set to the population size, resulting in no compartments switching into the flowing state. These results are further discussed in \ref{sec:N03_Study}, where we show the two methods are indistinguishable.

\subsection{Simulation experiments: Comparing Jump-Switch-Flow (with no switching) and Doob-Gillespie}\label{sec:N03_Study}
We compare the Jump-Switch-Flow method to the Doob-Gillespie method. To implement the latter, we set the switching threshold to be the total population size, to ensure no compartments switch into a flowing regime, {and therefore all compartments are completely stochastic.} We can therefore investigate how our Jumping process compares to exact solutions of the Continuous Time Markov Chains. 

\begin{figure}[H]
\centering
    \begin{subfigure}[t]{0.99\textwidth}
    \centering {Jump-Switch-Flow}
    \end{subfigure}

    \begin{minipage}[t][0.1cm][t]{0.27\textwidth}
        \centering
        \begin{subfigure}{\linewidth}
            \includegraphics[trim={0 0.0cm 0 0},clip, width=\linewidth]{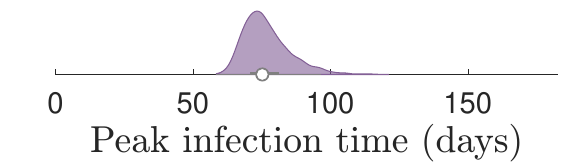}
            \caption{\label{fig:JSF_N05_Fad_peak_exact}}
        \end{subfigure}
    \end{minipage}
    \begin{minipage}[t][0.1cm][t]{0.08\textwidth}
        \hspace{0.1cm}
    \end{minipage}
    \begin{minipage}[t][0.1cm][t]{0.3\textwidth}
        \hspace{0.1cm}
    \end{minipage}
    \begin{minipage}[t][0.1cm][t]{0.08\textwidth}
        \hspace{0.1cm}
    \end{minipage}

    \begin{minipage}[t][1cm][t]{0.3\textwidth}
        \centering
        \begin{subfigure}{\linewidth}
            \includegraphics[width=\linewidth]{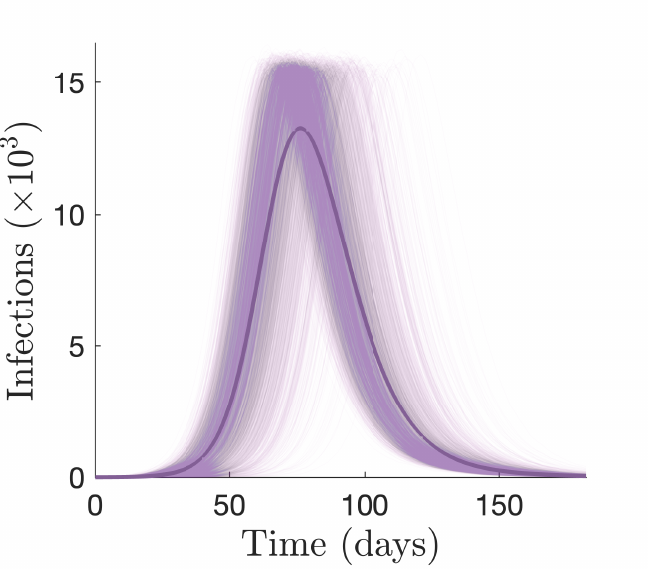}
            \caption{\label{fig:JSF_N05_Fad_exact}}
        \end{subfigure}
    \end{minipage}
    \begin{minipage}[t][1cm][t]{0.08\textwidth}
        \centering
        \begin{subfigure}{\linewidth}
            \includegraphics[width=\linewidth]{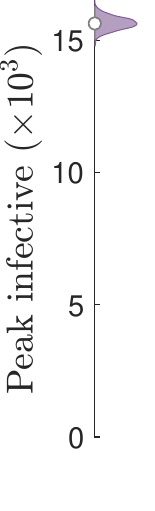}
            \caption{\label{fig:JSF_N05_Fad_peaknumb_exact}}
        \end{subfigure}
    \end{minipage}
    \begin{minipage}[t][1cm][t]{0.3\textwidth}
        \centering
        \begin{subfigure}{\linewidth}
            \includegraphics[width=\linewidth]{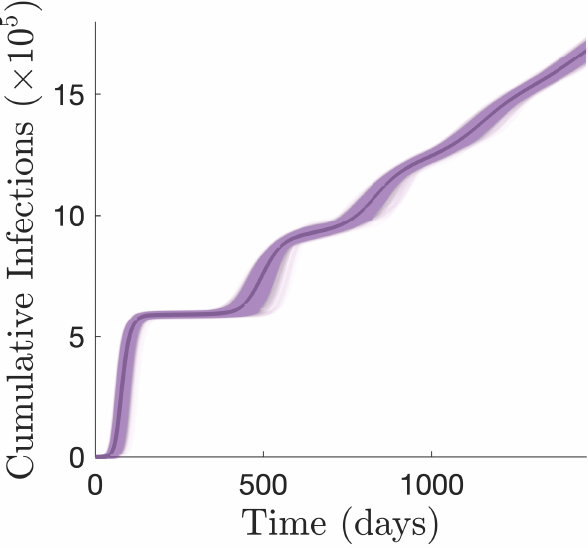}
            \caption{\label{fig:JSF_N05_Cummulative_exact}}
        \end{subfigure}
    \end{minipage}
    \begin{minipage}[t][1cm][t]{0.08\textwidth}
        \centering
        \begin{subfigure}{\linewidth}
            \includegraphics[width=\linewidth]{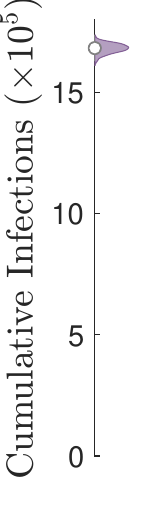}
            \caption{\label{fig:JSF_N05_Cummulative_dist_exact}}
        \end{subfigure}
    \end{minipage}

    \begin{subfigure}[t]{0.99\textwidth}
    \centering {Doob-Gillespie}
    \end{subfigure}

    \begin{minipage}[t][0.1cm][t]{0.27\textwidth}
        \centering
        \begin{subfigure}{\linewidth}
            \includegraphics[width=\linewidth]{figures/SIRSFigures/Gil_N05_Fad_peak.pdf}
            \caption{\label{fig:Gil_N05_Fad_peak}}
        \end{subfigure}
    \end{minipage}
    \begin{minipage}[t][0.1cm][t]{0.08\textwidth}
        \hspace{0.1cm}
    \end{minipage}
    \begin{minipage}[t][0.1cm][t]{0.3\textwidth}
        \hspace{0.1cm}
    \end{minipage}
    \begin{minipage}[t][0.1cm][t]{0.08\textwidth}
        \hspace{0.1cm}
    \end{minipage}

    \begin{minipage}[t][1cm][t]{0.3\textwidth}
        \centering
        \begin{subfigure}{\linewidth}
            \includegraphics[width=\linewidth]{figures/SIRSFigures/Gil_N05_Fad.pdf}
            \caption{\label{fig:Gil_N05_Fad}}
        \end{subfigure}
    \end{minipage}
    \begin{minipage}[t][1cm][t]{0.08\textwidth}
        \centering
        \begin{subfigure}{\linewidth}
            \includegraphics[width=\linewidth]{figures/SIRSFigures/Gil_N05_Fad_peaknumb.pdf}
            \caption{\label{fig:Gil_N05_Fad_peaknumb}}
        \end{subfigure}
    \end{minipage}
    \begin{minipage}[t][1cm][t]{0.3\textwidth}
        \centering
        \begin{subfigure}{\linewidth}
            \includegraphics[width=\linewidth]{figures/SIRSFigures/Gil_N05_Cummulative.pdf}
            \caption{ \label{fig:Gil_N05_Cummulative}}
        \end{subfigure}
    \end{minipage}
    \begin{minipage}[t][1cm][t]{0.08\textwidth}
        \centering
        \begin{subfigure}{\linewidth}
            \includegraphics[width=\linewidth]{figures/SIRSFigures/Gil_N05_Cummulative_dist.pdf}
            \caption{\label{fig:Gil_N05_Cummulative_dist}}
        \end{subfigure}
    \end{minipage}

    \captionsetup{subrefformat=parens}
    
     \caption{\label{fig:N05_Study} We compare the infectious compartment from realisations for Jump-Switch-Flow \subref{fig:JSF_N05_Fad_exact} and Doob-Gillespie \subref{fig:Gil_N05_Fad} solutions, for a population size of $N_0 = 10^5$. We use a switching threshold of $\Omega = 10^3$. We also show the distributions of the time to infectious peak (\subref{fig:JSF_N05_Fad_peak_exact} and \subref{fig:Gil_N05_Fad_peak}), and infectious peak values (\subref{fig:JSF_N05_Fad_peaknumb_exact} and \subref{fig:Gil_N05_Fad_peaknumb}). We compare the cumulative infections over the simulations for Jump-Switch-Flow \subref{fig:JSF_N05_Cummulative_exact} and Doob-Gillespie \subref{fig:Gil_N05_Cummulative} solutions and the final distributions (\subref{fig:JSF_N05_Cummulative_dist_exact} and \subref{fig:Gil_N05_Cummulative_dist}). Point averages are also presented as a solid line.}
    
\end{figure}

Using parameter values in Table 3 of the main text, we generated 5,000 simulations, tracking the same quantities of interest. As with the case with compartment switching, the  peak infection time distribution of JSF (Figure \ref{fig:JSF_N05_Fad_peak_exact}) and Doob-Gillespie (Figure \ref{fig:Gil_N05_Fad_peak}) are comparable. However, this time, as well as a similar median peak infective individuals, the distributions of both JSF (Figure \ref{fig:JSF_N05_Fad_peaknumb_exact}) and Doob-Gillespie (Figure \ref{fig:Gil_N05_Fad_peaknumb}) are also comparable. Lastly, as we saw with compartment switching, the distributions of cumulative infections after four years are comparable for both JSF (Figure \ref{fig:JSF_N05_Cummulative_dist_exact}) and Doob-Gillespie (Figure \ref{fig:Gil_N05_Cummulative_dist}).

Table \ref{tab:SIRS_N05_scenario_outcomes_NoSwitch} shows that the probabilities of scenario outcomes are also comparable for both Jump-Switch-Flow (with no switching) and Doob-Gillespie. 

\begin{table}[!ht]
    \centering
    \begin{tabular}{|c|c|c|c|}
    \hline
        Method & Extinction & Fade-out & Endemic \\
        \hline
        Jump-Switch-Flow & $0.2512 \pm 0.0120$ & $0.4700 \pm 0.0139$ & $0.2788 \pm 0.0123$ \\
        Doob-Gillespie & $0.2520 \pm 0.0120$ & $0.4668 \pm 0.0139$ & $0.2812 \pm 0.0125$ \\
        \hline
    \end{tabular}
    \caption{Probabilities (with associated 95\% confidence interval)  of scenario outcomes for Jump-Switch-Flow (with no switching) and Doob-Gillespie, from 5,000 simulations.}
    \label{tab:SIRS_N05_scenario_outcomes_NoSwitch}
\end{table}

\subsection{Simulation experiments: Sensitivity of Jump-Switch-Flow to switching threshold}

\begin{figure}[H]
  \centering
  \begin{subfigure}[b]{0.49\textwidth}
    \includegraphics[width=0.95\linewidth]{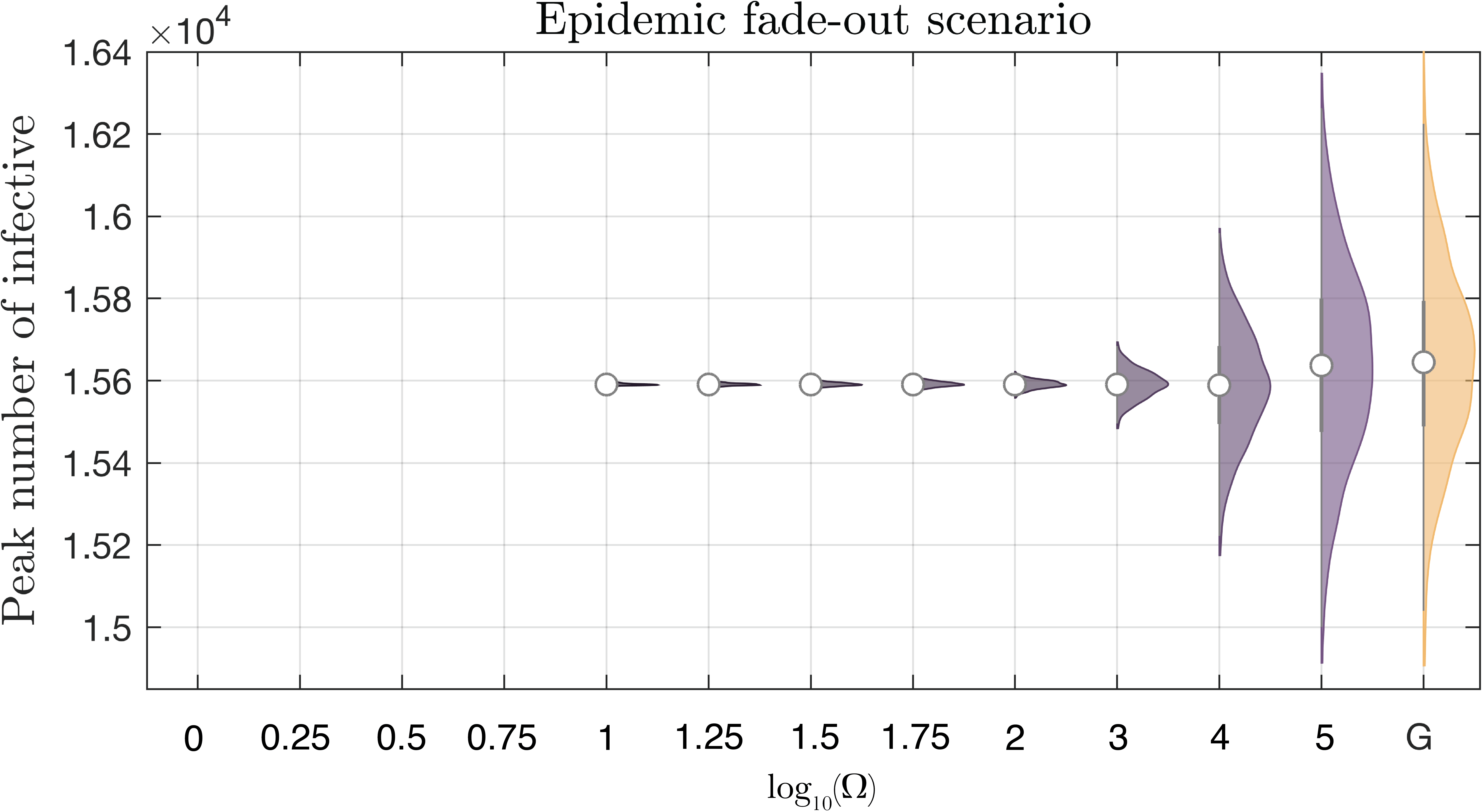}
    \caption{\label{fig:N0_5_FadePeakInf}}
  \end{subfigure}
  \begin{subfigure}[b]{0.49\textwidth}
    \includegraphics[width=0.95\linewidth]{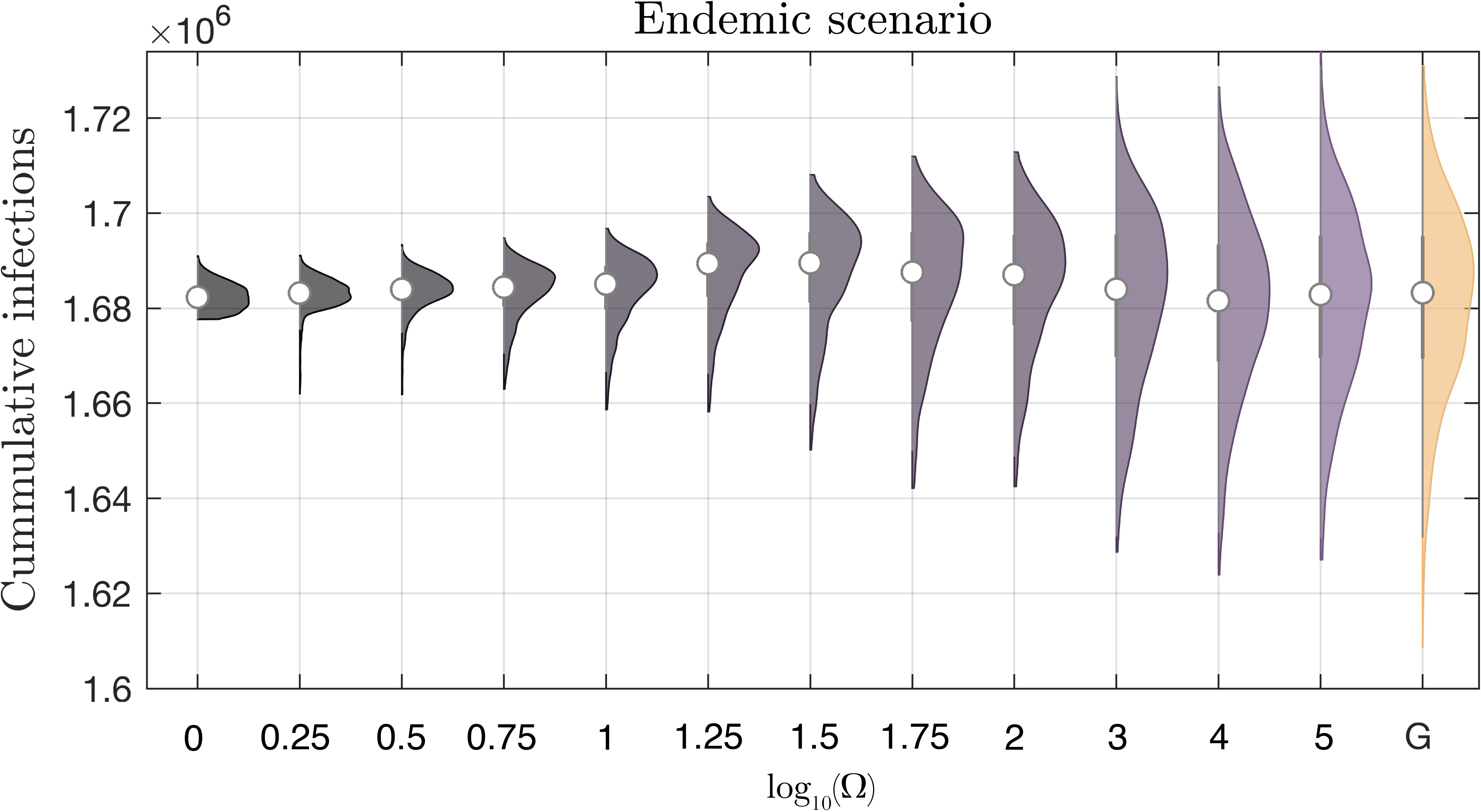}
    \caption{\label{fig:N0_5_EndCumInf}}
  \end{subfigure}
  
  \begin{subfigure}[b]{0.49\textwidth}
    \includegraphics[width=0.95\linewidth]{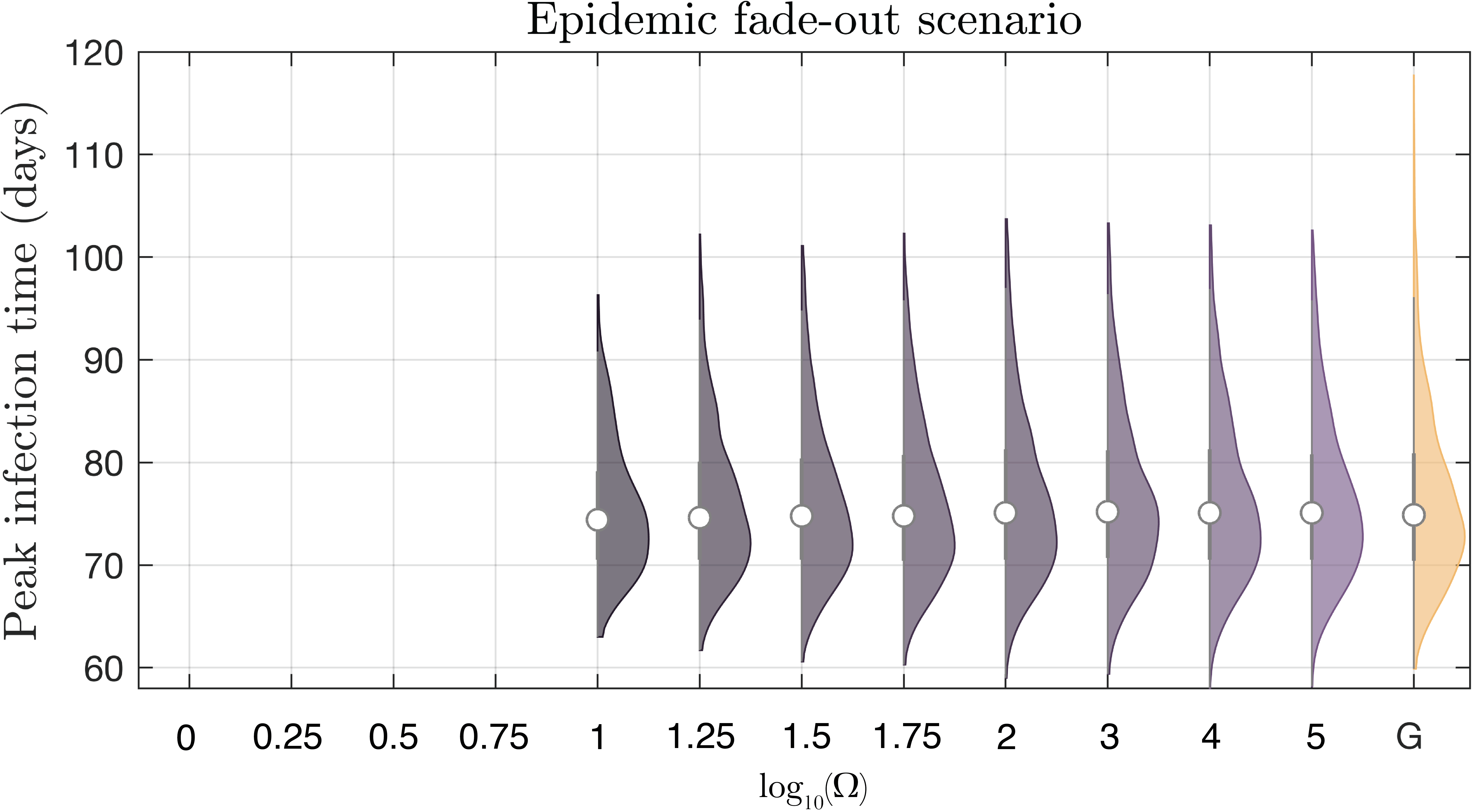}
    \caption{\label{fig:N0_5_FadePeakTime}}
  \end{subfigure}
  \begin{subfigure}[b]{0.49\textwidth}
    \includegraphics[width=0.95\linewidth]{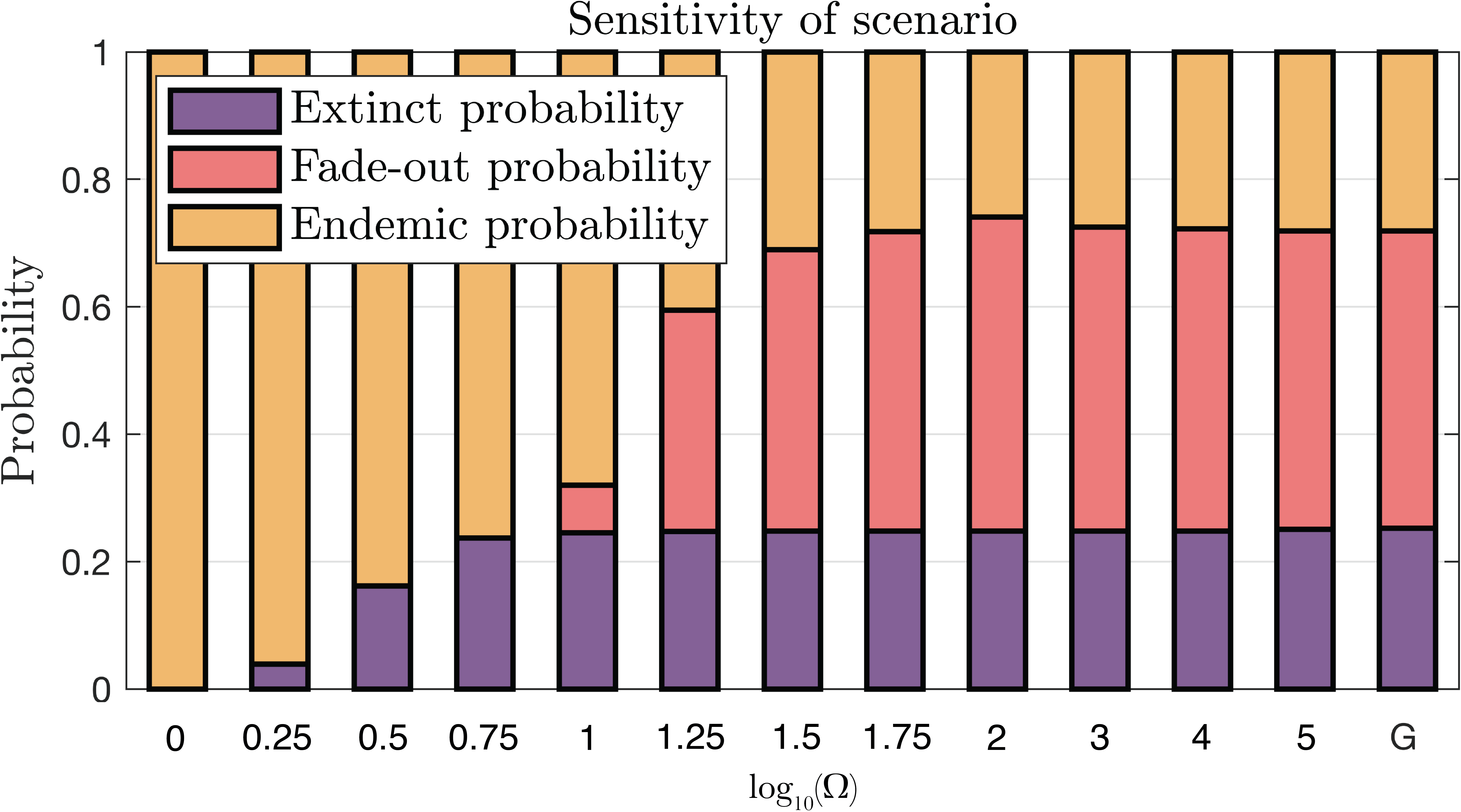}
    \caption{\label{fig:N0_5_Probs}}
  \end{subfigure}

    \captionsetup{subrefformat=parens}

    \caption{Summary statistics of interest to observe the sensitivity of the Jump-Switch-Flow method with respect to the switching threshold. \subref{fig:N0_5_FadePeakInf} Sensitivity to the distribution of peak number of infective individuals for the fade-out scenario. \subref{fig:N0_5_FadePeakTime} Sensitivity to the distribution of peak infection time for the fade-out scenario. \subref{fig:N0_5_EndCumInf} Sensitivity to the distribution of the cumulative number of infected individuals after four years for  the endemic scenario. \subref{fig:N0_5_Probs} Sensitivity to the probability of each scenario occurring. In each of the plots \subref{fig:N0_5_FadePeakInf}-\subref{fig:N0_5_FadePeakTime} grey to purple shaded distributions represent results obtained with the Jump-Switch-Flow method whilst the yellow distributions are the gold standard Doob-Gillespie algorithm (for which the placeholder `G' is used to signify) which exhibit exact stochasticity but at the cost of high computational requirements.}
  \label{fig:N0_5_1000_ALL_summary}
\end{figure}

In this section, we demonstrate how the accuracy of the Jump-Switch-Flow method can exhibit sensitivity with respect to the switching threshold (Figure \ref{fig:N0_5_1000_ALL_summary}). In Figure \ref{fig:N0_5_FadePeakInf}, we can see that the distribution of peak number of infective individuals for the fade-out scenario is highly sensitive to the switching threshold. For a small switching threshold ($\Omega=10^1$), we see that the distribution exhibits low variance, is representative of the deterministic expectation, and is not capturing the distribution obtained with the Doob-Gillespie method. As the switching threshold is increased, we see that the distribution of the peak number of infective individuals increases in variance. However, the Jump-Switch-Flow method is not representative of the Doob-Gillespie method until a large switching threshold ($\Omega=N(0)=10^5$), where all compartments are in the jumping state.

Figure \ref{fig:N0_5_FadePeakTime} presents the distribution of peak infection time for the fade-out scenario for various switching thresholds, and comparing to the Doob-Gillespie method. Here, we see that the Jump-Switch-Flow method with a relatively small switching threshold ($\Omega=10^2$) is capable of capturing a representative distribution of the Doob-Gillespie method.

Similarly, Figure \ref{fig:N0_5_EndCumInf} presents the distribution of the cumulative number of infected individuals after four years for
the endemic scenario. Here, we again see that the Jump-Switch-Flow method with a relatively small switching threshold ($\Omega=10^3$) is also representative of the Doob-Gillespie method.

Lastly, Figure \ref{fig:N0_5_Probs} shows the probability of each scenario occurring (extinction, fade-out, endemic). Here, we see that the Jump-Switch-Flow method with a switching threshold of $\Omega \leq 10^1$ over-represents the endemic scenario. However, increasing the switching threshold, we observe that the probability of each scenario occurring quickly becomes representative of those found by the Doob-Gillespie method.

\subsection{Simulation experiments: Computational efficiency of Jump-Switch-Flow}

We now present the computational efficiency of the Jump-Switch-Flow method, and compare it to the Doob-Gillespie method and Tau-Leaping method (which is regarded as  a computationally efficient, approximate, method), {and the Tau-hybrid method, provided by GillesPy2 \cite{matthew2023gillespy2}. We implement two variants of Tau-leaping with a fixed step size of $\Delta \tau = 0.01$ days, and $\Delta \tau = 0.1$ days respectively, two variants of the C++ implementation of the Tau-hybrid method, with time steps of size $\Delta t = 0.01$ days and $\Delta t = 0.1$ days respectively, one implementation of the Python Tau-hybrid method, with time steps of size $\Delta t = 0.1$ days, and the Jump-Switch-Flow method with a time step size of $\Delta t = 0.01$ days. }

\begin{figure}[H]
  \centering

  \begin{subfigure}[b]{0.45\textwidth}
     \centering
    \includegraphics[width=0.93\linewidth]{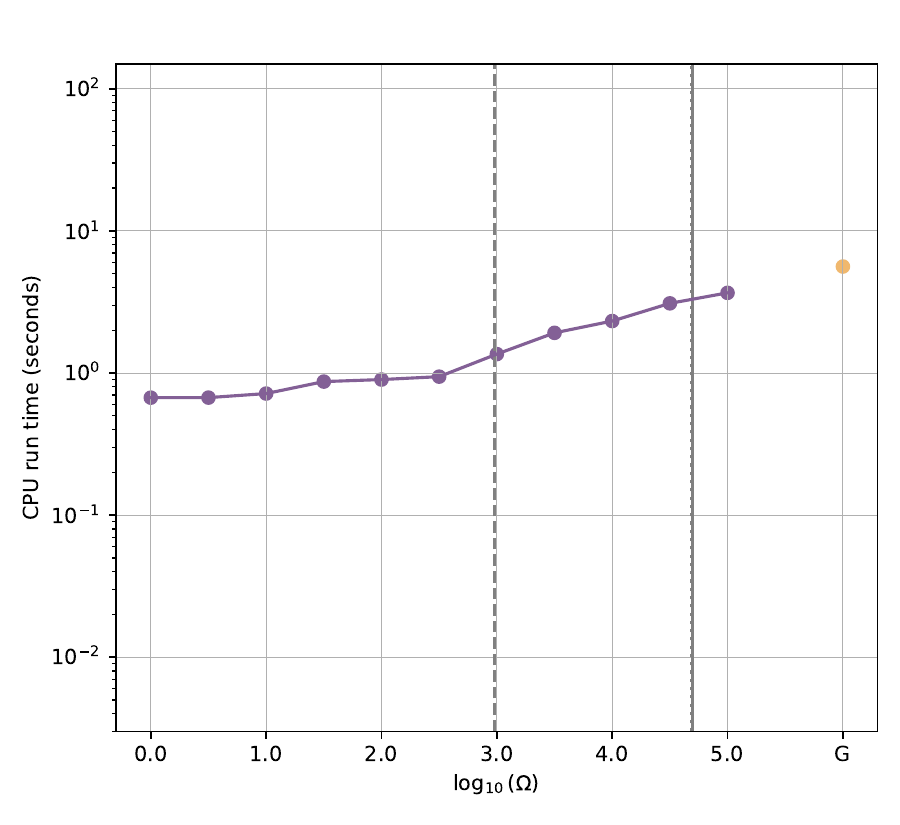}
     \caption{\label{fig:CPU_RunTime_N05}}
    \end{subfigure}
    \begin{subfigure}[b]{0.45\textwidth}
     \centering
    \includegraphics[width=0.98\linewidth]{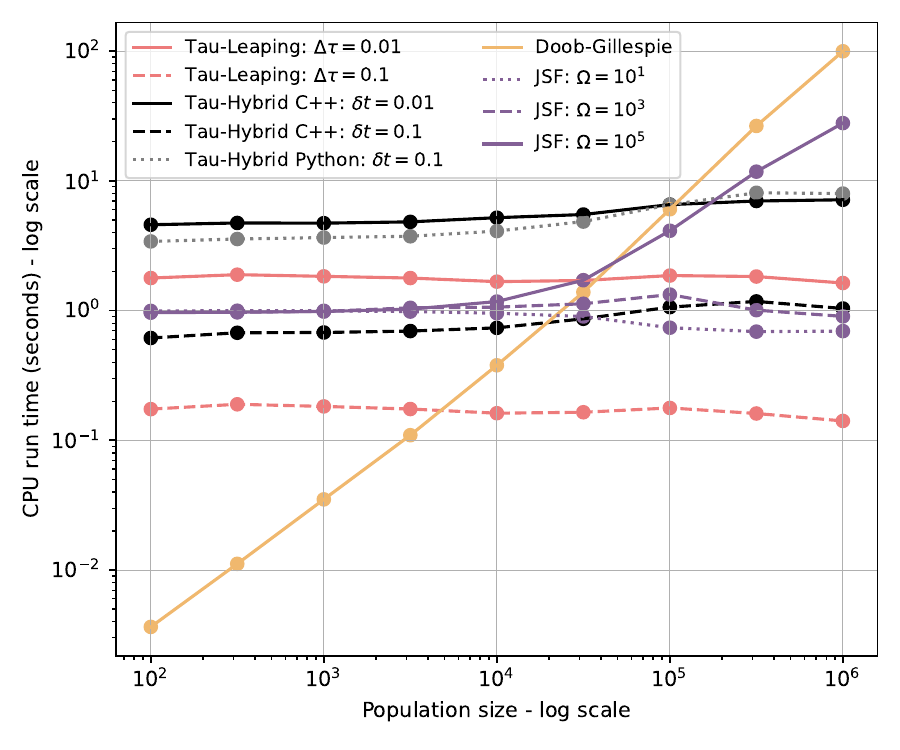}
     \caption{\label{fig:CPU_RunTime_all}}
    \end{subfigure}
    
    \captionsetup{subrefformat=parens}
  
  \caption{ \subref{fig:CPU_RunTime_N05} The running time to simulate the scenario for four years for an initial population size of $N(0) = 10^5$, with varying switching threshold, $\Omega$ (grey to purple), compared with the Doob-Gillespie (`G') method (yellow). The endemic steady state number of infectious individuals (vertical dashed line), susceptible individuals (vertical solid line), and recovered individuals (vertical dotted line) are presented.
  \subref{fig:CPU_RunTime_all} The running time for the Jump-Switch-Flow method for an appropriate threshold choice is
    approximately constant as the population size varies (for example $\Omega=10^3$). If a poor switching threshold is chosen, the running time begins to scale exponentially (for example  $\Omega=10^5$). 
    For the Tau-Leaping method, the running time is constant, {as we use fixed step size of $\Delta \tau = 0.01$ (solid line) and $\Delta \tau = 0.1$ (dashed line).
    We also present the Tau-hybrid method, in both Python and C++, with various different time meshes, C++ with a fine mesh ($\delta t = 0.01$, black solid line) and a corse mesh ($\delta t = 0.1$, black dashed line), and Python with a corse mesh ($\delta t = 0.1$, grey dotted line).}
    \label{fig:CPU_time_Figs}}
\end{figure}

Figure \ref{fig:CPU_RunTime_N05} shows how the computational efficiency varies with the switching thresholds, $\Omega$, for the SIRS model with demography presented in the previous section, and compares it to the Doob-Gillespie method. We see that as we increase the switching threshold, the Jump-Switch-Flow method requires more computational time to simulate the scenario for four years, but is always more efficient than the Doob-Gillespie method. {This increased efficiency is because the Jump-Switch-Flow method utilises the Next Reaction Method to compute the stochastic events, a method known to be more computationally more efficient than the direct, first reaction, Doob-Gillespie method.}

To observe how the computational efficiency of our Jump-Switch-Flow method varies with the total (population) size of the system, we {consider three switching thresholds: $\Omega = 10^1$ (dotted purple), $\Omega = 10^3$ (dashed purple), and $\Omega = 10^5$ (solid purple).} Figure \ref{fig:CPU_RunTime_all} compares the Jump-Switch-Flow method (purple) to the Doob-Gillespie method (yellow), to the computationally more efficient approximation Tau-Leaping method ({$\Delta \tau = 0.01$, dashed red, and $\Delta \tau = 0.1$, solid red}), {and to the Tau-hybrid method ($\Delta t = 0.01$ solid black, and $\Delta t = 0.1$ dashed black) provided by GillesPy2, noting the log-log scale. We simulate the SIRS system described above, for a total time of $4 \times 365$ days.}
We can see that the computational efficiency of the Doob-Gillespie method scales exponentially with system size, as expected. The Tau-Leaping method initially has constant computational time scaling.
In contrast, our Jump-Switch-Flow method is constant in time up to an initial system size of $N(0) = 10^5$, after which it begins to slowly decrease (for $\Omega =10^1$ and $\Omega=10^3$). This decrease is caused by more of the scenario being simulated in the flowing state, as the system becomes highly deterministic with a larger population. {We also observe our Jump-Switch-Flow method increasing exponentially between $N(0)=10^{4}$, for the largest switching threshold of $\Omega=10^5$ (solid purple). We observe our Jump-Switch-Flow method performing better, in comparison to Doob-Gillespie exact method, as we use the Next Reaction Method to sample the Jump times. This is despite our implementation being required to perform extra computation to check the regime of the Jump-Switch-Flow process, and implement any relevant regime changes (even though they are not required in this particular instance). We also observe that for large, realistic population sizes (i.e. $N(0) \geq 10^5$), the Tau-hybrid method with the corse time mesh (black dashed line) performs at best comparable to, or worse than the Jump-Switch-Flow method, if a suitable $\Omega$ is chosen (such as $\Omega = 10^3$. Moreover, as we specify a finer time mesh (solid black line) we observe that the Tau-hybrid method performs worse across all population sizes. We anticipate this is due to the Tau-hybrid method utilising an arbitrary ODE solver, an adaptive $\Delta \tau$ that is calculated at each step, and also due to the dynamic switching behaviour between stochastic and deterministic regimes that requires extra computation of the propensities to determine if a switching event should occur. } 

\subsection{Simulation experiments: Inference with simulated data, parameter estimates}
To test how the Jump-Switch-Flow method behaves when used to perform parameter estimation, we first generated synthetic data of a known SIRS mode with demography. We simulate data for 400 days using the parameter values given in Figure 2C, with the Doob-Gillespie algorithm, and select a trajectory which experiences epidemic fade-out (at day 250), and an initial condition of two infected individuals ($I(0)=2$), no recovered individuals ($R(0)=0$), and a total population of $N(0) = 10^5$ individuals. We track the number of infectious individuals within the population each day, and use a truncated binomial distribution to add noise into the data set, reflecting a realistic measurement process. We then couple our Jump-Switch-Flow method, using a switching threshold of $\Omega = 10^3$, with the open-source particle filter
package pypfilt \citep{Moss2024}, using 2,000 particles. We first fix the demographic parameters to $\kappa = \mu = 1/(85 \times 365)$ (since these are assumed to be standard within a given population). We perform the inference on the data set for the first 100 days of the epidemic to estimate $\beta$, $\gamma$ and $\omega$, and then use our parameter estimates to predict forward in time the possible trajectories based on the obtained predictions. Using these predictions, we also estimate the probability of epidemic fade-out occurring at times $t=100$ to $t=400$ days.

\begin{figure}[H]
    \centering
    \includegraphics[ width=0.75\linewidth]{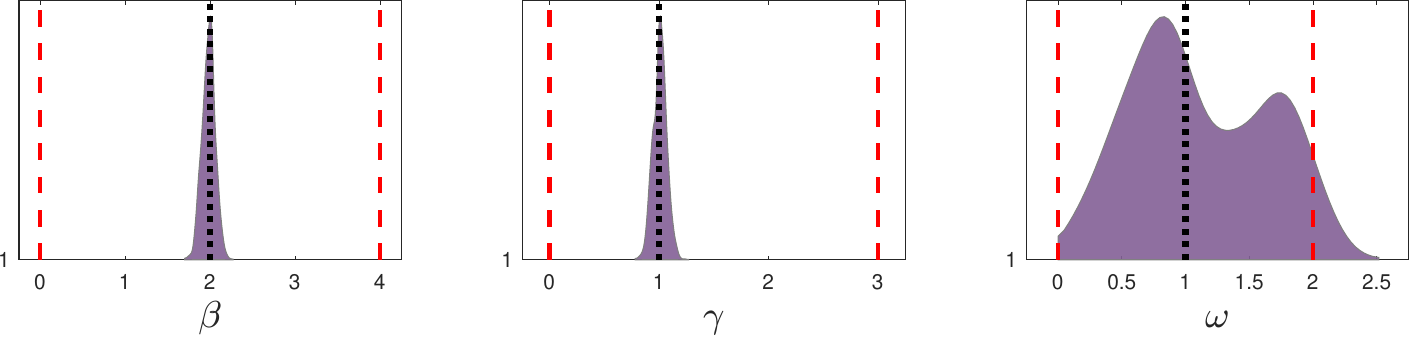}
    \caption{Posterior distributions of parameters for the SIRS model with demography, obtained via a particle filter with 2,000 particles, and the Jump-Switch-Flow sampler. The red dashed lines indicate the initial bounds on prior samples. The vertical dotted black lines indicate the true value used to produce the simulated data.}
    \label{fig:SIRS_Param_Ests}
\end{figure}

Figure \ref{fig:SIRS_Param_Ests} shows the posterior distributions of parameters. The dotted black lines show the true value used to produce the simulated data, and the dashed red lines indicate the initial bounds on the prior estimates. We see that both $\beta$ and $\gamma$ are well estimated, with the posteriors centered around the true values. However $\omega$ not well estimated. This is because $\omega$ acts on the order of $1$ year, while the window of data used to fit the parameters to the data is 100 days ($\approx 0.3$ years).

\section{Inference Study: Estimating TEIRV model parameters using a particle filter}

We implemented the Refractory Cell Model described by the authors for the viral load data obtained from nasal swabs, using our Jump-Switch-Flow method with a switching threshold on all compartments $\Omega = 10^2$. 
We implemented this in particle filter pypfilt \cite{Moss2024} to estimate the model parameters, using 6,000 particles per simulation.
In previous analysis \cite{ke2022daily}, the decay of virion was fixed at $c = 10$ virions/day, and the eclipse of cells to become infectious as $k=4$ cells/day, leaving the remaining parameters to be estimated. 
Therefore, we similarly fix $c=10$ virions/day and $k=4$ cells/day. {The prior distributions used for the analysis with the TEIRV model are given in Table \ref{tab:TIV_model}.}

\begin{table}[]
\centering
\begin{tabular}{lr}
\hline
\multicolumn{1}{c}{\textbf{Parameter}} & \multicolumn{1}{c}{\textbf{Prior}} \\ \hline
$\log_{e}{(V_0)}$   & $\text{Uniform}(0,5)$   \\
$c$   & Fixed at $10.0$   \\
$k$   & Fixed at $4.0$   \\
$\beta \times 10^7$   & $\text{Uniform}(0,20)$ \\
$\Phi \times 10^5$   & $\text{Uniform}(0,15)$ \\
$\rho$   & $\text{Uniform}(0,1)$ \\
$\delta$ & $\text{Uniform}(1,10)$ \\
$\pi$   & $\text{Uniform}(200,400)$   \\ \hline
\end{tabular}
\caption{The prior distributions used for the analysis with the TEIRV model. \label{tab:TIV_model}}
\end{table}

\subsection{Parameter Estimation and Viral Clearance Prediction}

Figure 9 of the main text highlights how the estimated viral loads closely resemble the data collected from the 6 patients. We also show how the estimated $R_0$ values for the patients vary. These $R_0$ estimates are consistent with the original study by \cite{ke2022daily}. However, understanding how the parameter estimates vary between patients assists in understanding how the fitting process behaves.

\begin{figure}[H]

  \centering
  \begin{subfigure}[b]{\textwidth}
    \centering
    \includegraphics[width=0.8\linewidth]{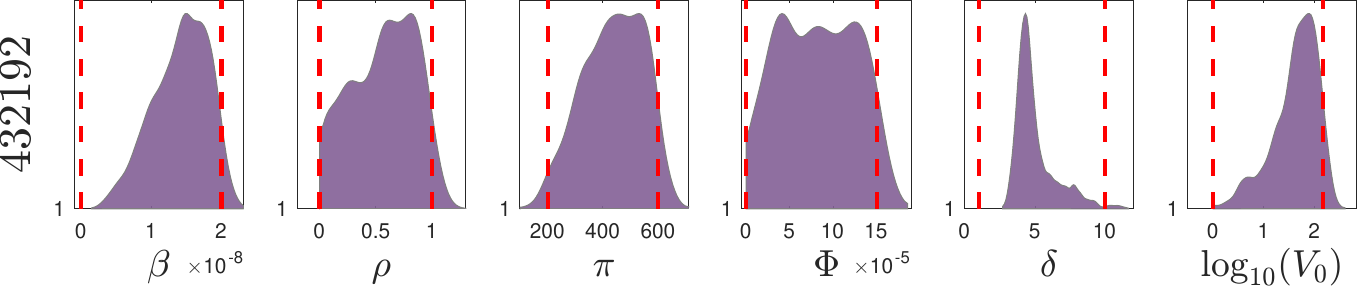}
    \caption{\label{fig:covid_p1_params}}
  \end{subfigure}

  \centering
  \begin{subfigure}[b]{\textwidth}
    \centering
    \includegraphics[width=0.8\linewidth]{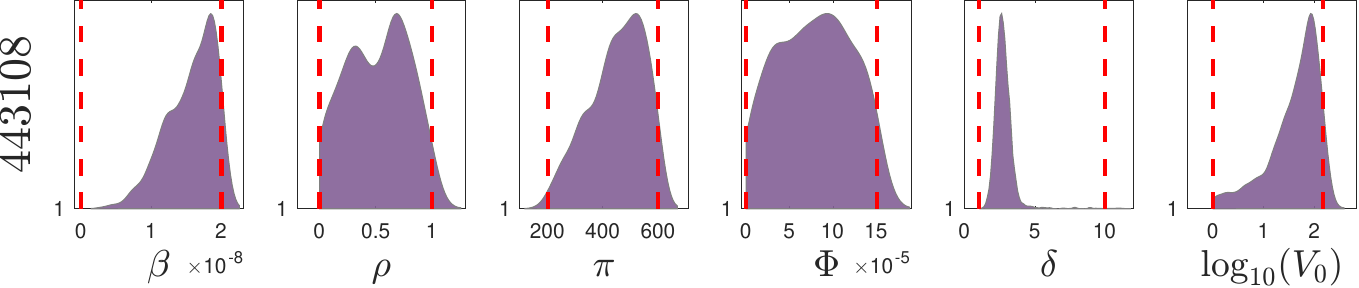}
    \caption{\label{fig:covid_p2_params}}
  \end{subfigure}

  \centering
  \begin{subfigure}[b]{\textwidth}
    \centering
    \includegraphics[width=0.8\linewidth]{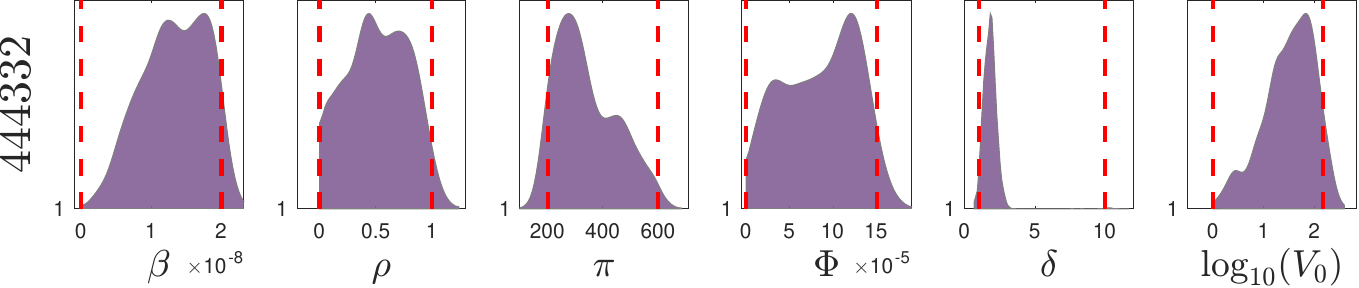}
    \caption{\label{fig:covid_p3_params}}
  \end{subfigure}

  \centering
  \begin{subfigure}[b]{\textwidth}
    \centering
    \includegraphics[width=0.8\linewidth]{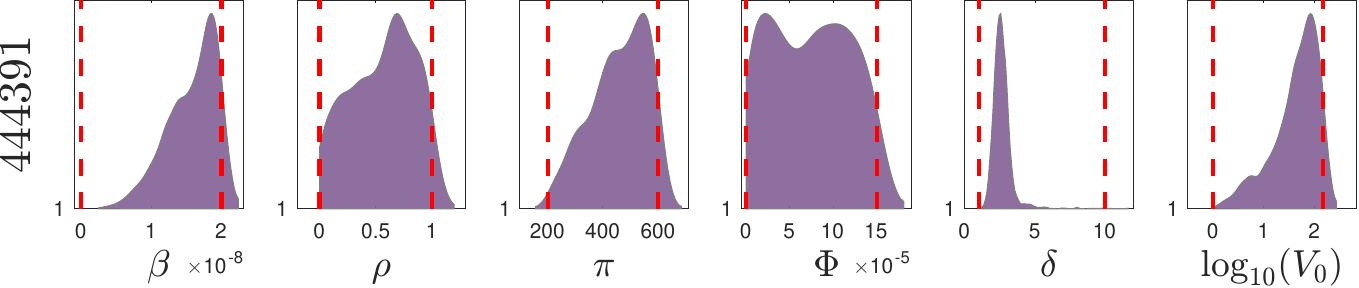}
    \caption{\label{fig:covid_p4_params}}
  \end{subfigure}

  \centering
  \begin{subfigure}[b]{\textwidth}
    \centering
    \includegraphics[width=0.8\linewidth]{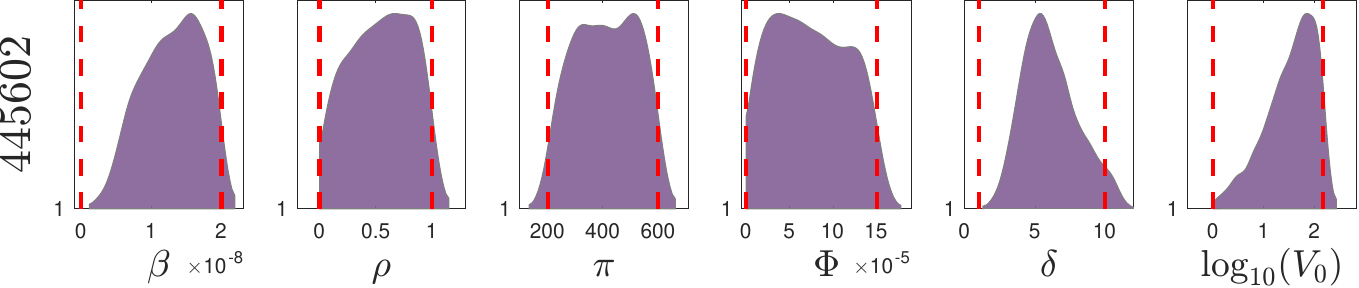}
    \caption{\label{fig:covid_p5_params}}
  \end{subfigure}

  \centering
  \begin{subfigure}[b]{\textwidth}
    \centering
    \includegraphics[width=0.8\linewidth]{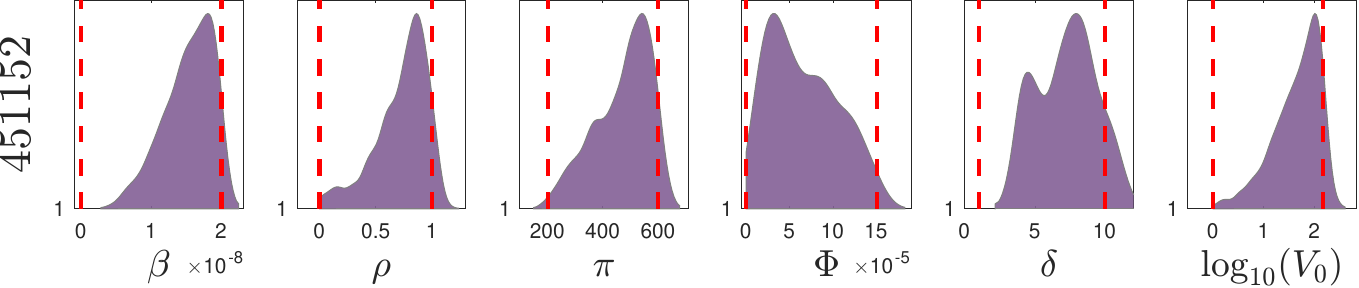}
    \caption{\label{fig:covid_p6_params}}
  \end{subfigure}

    \captionsetup{subrefformat=parens}

    \caption{Posterior distributions of the model parameters and initial viral load for each of the selected SARS-CoV-2 infections. The red dashed lines indicate the initial (uniform) bounds on prior samples.}
  \label{fig:Covid_inference_params}
\end{figure}

The posterior distributions for each of the 6 patient's parameter estimates are provided in Figure \ref{fig:Covid_inference_params}. These posterior distributions are found by fitting a violin envelope from the 6,000 estimates obtained from the particle filtering process. We also show the prior distributions, as red dashed lines, which show the lower and upper bounds for the uniform initial guesses.

We can see that, often, parameters $\rho$ and $\Phi$ are not identified here, as the posterior distributions closely resemble the priors (excluding patient 451152). We also observed that for all patients, $\beta$ appears to be well estimated. The parameter $\pi$ is also well estimated for patients 443108, 444332, 444391 and 451152, however patients 432192 and 445602 provide less insight. The parameter $\delta$ is very well estimated by all patients, excluding patient 451152. Lastly, the initial viral loads, $\text{log}_{10}(V_0)$, are all estimated well, and also estimated to take a large value.

\end{document}